\documentclass[twocolumn,10pt,aps,nofootinbib]{revtex4-1}
\usepackage{amsmath}
\usepackage{amsfonts}
\usepackage{amssymb}
\usepackage{graphicx}
\usepackage{caption}
\usepackage{subcaption}
\usepackage{float}
\usepackage{caption}
\usepackage{braket}
\allowdisplaybreaks
\newcommand {\bp}{\begin{pmatrix}}
\newcommand {\ep}{\end{pmatrix}}
\newcommand{\be}{\begin{equation}} \newcommand{\ee}{\end{equation}}
\newcommand{\bea}{\begin{eqnarray}}\newcommand{\eea}{\end{eqnarray}}

\DeclareMathOperator{\sgn}{sgn}

\begin{document}
\title{Edge states and persistent current in a $\mathcal{PT}$-symmetric
extended Su-Schrieffer-Heeger model with generic boundary conditions}

\author{Supriyo Ghosh}
\email[]{supriyoghosh711@gmail.com}
\author{Pijush K. Ghosh}
\email[]{pijushkanti.ghosh@visva-bharati.ac.in}
\author{Shreekantha Sil}
\email[]{shreekantha.sil@visva-bharati.ac.in}
\affiliation{ Department of Physics, Siksha-Bhavana, Visva-Bharati,
Santiniketan, PIN 731 235, India.}
\date{\today}

\begin{abstract}

We consider a generalization of the  Su-Schrieffer-Heeger(SSH) model by including
next-nearest-neighbour(NNN) interaction and balanced loss-gain(BLG), and subjecting the
whole system to an external uniform magnetic field. We study the band-structure, edge
states and persistent current in this extended SSH model under General Boundary
Condition(GBC) of which the periodic, anti-periodic and open boundary conditions appear as special
cases. It is shown that the point band gap decreases with the increasing value of the
strength of the NNN interaction and vanish beyond a critical value for both topologically
trivial and non-trivial phases. Further, the line gap exhibits closed-loop like structures
for non-vanishing NNN interaction under the Periodic Boundary Condition(PBC). 
The Zak phase receives no contribution from the NNN interaction under the PBC.
We show that the NNN interaction has no effect on the persistent
current in the half-filled limit for the case of PBC, while for other fillings less than the
half-filling, it  enhances the magnitude of the current significantly. We numerically study
the variation of the persistent current with respect to the system parameters under the GBC,
for the case the Hamiltonian admits entirely real spectra, and show that its magnitude increases
with the increasing strength of the BLG. We show that the model without the NNN interaction is exactly 
solvable for a class of GBC of which PBC, anti-periodic boundary condition(APBC) and anti-hermitian boundary
condition(AHBC) arise as special cases.
We obtain analytic expressions for the edge states in the case of Open Boundary
Condition(OBC) and AHBC for vanishing NNN interaction. 
We show numerically for OBC that edge states in the
topologically trivial phase appear for non-vanishing NNN interaction only when the strength
of the loss-gain term is greater than the modulus of the difference between
the intercell and intracell hopping strengths. In the topologically non-trivial phase,
the edge states under OBC exists only up to a critical value of the NNN strength
and vanishes beyond this critical value. 
The bulk-boundary correspondence(BBC) for unbroken $\mathcal{PT}$-phase is similar to hermitian SSH model, 
while non-Hermitian skin effect(NHSE) is observed for broken $\mathcal{PT}$-phase. 
\end{abstract}

\maketitle
\tableofcontents{}

\section{Introduction}
One of the recurring themes in contemporary research
in physics is non-hermitian systems with parity-time($\mathcal{PT}$)
symmetry\cite{bender1} and/or pseudo-hermiticity\cite{ali}. The primary motivation
stems from the fact that non-hermitian Hamiltonians having unbroken
$\mathcal{PT}$ symmetry admits entirely real spectra, and a consistent
quantum description is allowed with a modified norm in the Hilbert space\cite{bender1}.
The same is true for a pseudo-hermitian system provided a positive-definite metric
exists in the Hilbert space\cite{ali}. The studies on non-hermitian systems within the ambit of
$\mathcal{PT}$ symmetry and pseudo-hermiticity have provided plethora of interesting and new results
in the realm of quantum field theories\cite{Bender2004prd,Jones2006JPhysA}, open quantum
systems\cite{Rotter2009JPhysA}, optical systems with complex refractive indices\cite{Klaiman2008Prl,
Kivshar2010Pra, Ramezani2012Prl, Longhi2009Prl, Musslimani2008Prl, Luo2013Prl}, Jaynes-Cummings and
Tavis-Cummings models\cite{Pkg2005Jpa}, the Anderson models for disordered systems\cite{Goldsheid1998Prl,
Heinrichs2001Prb,Molinari2009JPhysA}, the Dirac Hamiltonians of topological insulators\cite{Pkg2012JPhys},
quantum phase transitions\cite{Deguchi2009Pre}, level statistics\cite{PKG2009Pre}, transverse Ising
model\cite{Deguchi2009Jpa}, quantum many-particle solvable models\cite{Pkg2010Jpa,Pkg2011Ijtp,DSJPA2019},
quantum chaos\cite{SG2024Chaos} and lattice models\cite{Bendix2009Prl,JinSong2009Pra,joglekar,pla1,physicaE,Meden2023Rep}.

The studies on non-hermitian lattice models deserve a special attention due to its possible
applicability in condensed matter and optical systems. The earlier investigations in this context
were mostly on one dimensional tight-binding chain with balanced loss-gain.
The phase-transition in a $\mathcal{PT}$-symmetric tight-binding chain with balanced loss-gain terms
at two arbitrary sites have been studied\cite{joglekar}. Further, exact solvability
of a tight-binding chain with conjugated imaginary potentials at two edge has been
shown\cite{JinSong2009Pra}. The studies on spectral and transport properties in a tight-binding
lattice with loss or gain in alternate lattice sites reveal some intriguing physical
properties\cite{pla1,physicaE}.

The SSH model, a tight-binding chain with dimerized hopping amplitude,
was originally devised to describe one dimensional polyacetylene polymer chain\cite{Su1979Prl}.
One of the interesting aspects of the SSH model is that it exhibits topological phases,
in particular, topological insulators\cite{Fradkin1983Prb,Kivelson1982Prb,Li2014prb}. 
In the last few years, extensive research have focused on the modification of the SSH model by
incorporating non-hermitian terms in the system which provide a simple framework to study the
interplay of the $\mathcal{PT}$-symmetry and topology in condensed matter physics
\cite{Lieu2018Prb,Klett2017Pra,pra2022,JPhysCon2020,pra2014,Xu2020Pra,JinSong,pra2022,pre2020,Kawabata2019Prx}.
Such modified non-hermitian SSH models have been investigated from the
viewpoint of $\mathcal{PT}$-symmetric phase transition\cite{Klett2017Pra,pra2022,pra2014},
real and complex eigenspectra\cite{Klett2017Pra,JPhysCon2020}, NHSE
BBC\cite{Zhang2022AdvPhysX,Kawabata2019Prx,Halder2023JPhys}, 
localization and transport properties\cite{pra2016}, quantum chaos\cite{pre2020}. 

A unique phenomena in non-Hermitian lattice models under the OBC is the tendency of a large number 
of eigenstates to localize at one of the boundaries which is known as 
NHSE\cite{Yao2018Prl,Zhang2022AdvPhysX}. In general, the NHSE is linked to the
non-reciprocity of the hopping strengths \textemdash the unequal forward and backward hopping
cause the localization of the eigenstates at one of the edges \cite{Zeng2022Prb,Yao2018Prl,Halder2023JPhys}.
The BBC for the hermitian SSH model correctly predicts the parametric regions for zero-mode
edge states under the OBC from the proper identification of the topological phases under the PBC
or the vice versa. The BBC breaks down or gets modified\cite{Yao2018Prl} in presence of NHSE, and the probability densities
of the most of the eigenstates accumulate at edges. The breakdown of the conventional bulk-boundary correspondence(BBC) is 
one of the key features of the NHSE \textemdash the eigenspectra
under OBC and PBC differ significantly. The NHSE exhibits a strong sensitivity 
to boundary conditions. One of the key motivations for investigating lattice models with GBC is to explore the 
dependence of the NHSE on GBC. From this perspective, a recent study has examined the SSH model under GBC
and has shown that NHSE appear even when the boundary conditions 
deviate from the OBC\cite{Guo2021Prl}. 

In this article, we consider a generalized non-hermitian SSH model. The bulk Hamiltonian
consists of the standard SSH model with additional next nearest neighbor(NNN) interaction and
onsite imaginary potentials such that the loss-gain is identically balanced. Further,
the whole system is subjected to an external uniform magnetic field in order to study variation
of the induced persistent current in the system in presence of the NNN interaction and loss-gain
terms. The onsite imaginary potential is the only source of
non-hermiticity in the bulk Hamiltonian. The nonhermiticity is also introduced through GBC for which
the boundary terms are non-hermitian. We denote such boundary condition as Non-hermitian GBC(NGBC),
while boundary condition associated with hermitian boundary terms is denoted as Hermitian GBC(HGBC).
The PBC and OBC are special cases of HGBC, and studied separately. The anti-hermitian boundary
term is a special case of NGBC and denoted as AHBC.
The system as a whole is $\mathcal{PT}$-symmetric for hermitian boundary terms corresponding
to HGBC.

The purpose of this article is to study the combined effect of the NNN interaction,
the BLG terms and the external uniform magnetic field on the properties of SSH model
under the GBC. In particular, we investigate the spectra, persistent current and edge states
in the system: 
\begin{itemize}
\item {\bf Spectra \& Band Gap}: The Hamiltonian is exactly solvable under the PBC,
and analytic expressions for eigenvalues and eigenstates can be obtained in a closed form. The role
of the NNN interaction is to convert a direct band gap into an indirect one. The system
admits point as well as line gaps, and the line gap exhibits closed-loop like structures for
non-vanishing NNN interaction.
The Hamiltonian is shown to be pseudo-hermitian under the PBC, and we obtain analytic expression
for the associated positive-definite metric. The pseudo-hermiticity is used to obtain the equivalent
hermitian Hamiltonian. We show that the Zak phase receives no contribution from the NNN interaction
under PBC, thereby, the classification of topological phases remain the same. In particular, 
the topologically non-trivial phase is obtained whenever the intercell hopping strength is
greater than the intracell hopping strength. 
It is shown that the point band gap decreases with the increasing value of the strength of the
NNN interaction and vanish beyond a critical value for both topologically trivial and non-trivial phases.
For the case of other boundary conditions, the sensitive dependence of the spectra on system parameters
is seen.

\item {\bf Persistent Current}: The persistent current is studied for HGBC as well as NGBC,
except OBC for which the concept of a persistent current is not meaningful. We restrict our
investigation to that region in the parameter-space for which the non-hermitian Hamiltonian
admits entirely real spectra.  We show analytically that the NNN interaction has no effect on
the persistent current in the half-filled limit for the case of PBC. On the other hand, for
other fillings less than the half-filling, the NNN interaction enhances the magnitude of the
current significantly under the PBC. We numerically study the
variation of the persistent current with respect to the system parameters, and show that its magnitude
increases with the increasing strength of the BLG terms for all allowed boundary conditions.

\item {\bf Exact Solvability}: We analyze the extended SSH model with GBC from the viewpoint of exact 
solvability. The Hamiltonian without the NNN interaction is shown to be exactly solvable for a class of
GBC. We obtain exact analytic expressions of bulk as well as edge states along with their eigenvalues.
The parametric regions in which edgestates appear are identified. 
The APBC and AHBC appear as special cases of this GBC, and to the best of our knowledge, no exact analytic
expressions of the eigenstates for these cases have been obtained earlier.

\item {\bf Edge States}: The edge states are studied for HGBC and NGBC, except for the PBC.
We obtain analytic expressions for the edge states in the case of OBC for vanishing NNN interaction.
The study  on edge states for non-vanishing NNN interaction involves roots of a quartic equation which
leads to cumbersome expressions, and evades a simple closed-form analytic expression. We study edge states
numerically for non-vanishing NNN interaction, and show that edge states in the topologically trivial
region appear only when the strength of the loss-gain term is greater than the modulus of the difference
between the intercell and intracell hopping strengths. The edge states in the topologically nontrivial
region appear up to a critical value of the strength of the NNN interaction and vanishes beyond this
critical value. In general, the NNN interaction, the loss-gain and the boundary terms play crucial
role in the creation and destruction of edge states for the GBC. The appearance of the eigenstates
with pair of imaginary eigenvalues is a signature of edge states, which is independently confirmed
by computing the IPR. The modulus of the wave-function of the edge states is also plotted as a
function of cite index to see whether or not they are symmetrically localized between the two
edges.

\end{itemize}

The paper is organized as follows. The generalized Hamiltonian is introduced in Sec. II.
The boundary conditions and symmetry of the system are discussed in this section.
The Sec. III deals with the system with PBC \textemdash the spectra, the band-structure 
the and persistent current are discussed in sections III.A, III.B and III.C, respectively.
The study of the Hamiltonian under GBC is contained in Sec. IV. A class of exactly solvable
SSH models with vanishing NNN interaction and under the GBC are presented in Sec. IV.A and IV.B.
The analytic and numerical
results concerning edge states and spectra for OBC are presented in Sec. IV.C. The numerical
results on the eigenspectra, edge states and persistent current
for the case of HGBC and NGBC are discussed in sections IV.D.1 and IV.D.2, respectively. Finally, 
the obtained results are summarized with future outlook in Sec. V. The Appendix A contains consistency 
condition for the existence of eigenstates under the GBC for the case when intracell hopping strength vanishes. 
In Appendix B, the nature of some trivial solutions which do not span the Hilbert space under a class
of GBC are discussed.
The derivation of exact eigenstates for the AHBC are given in Appendix C. 
\section{The Model}

\begin{figure}[H]
	\centering
	\begin{subfigure}{0.7\linewidth}
		\centering
		\includegraphics[scale=0.35]{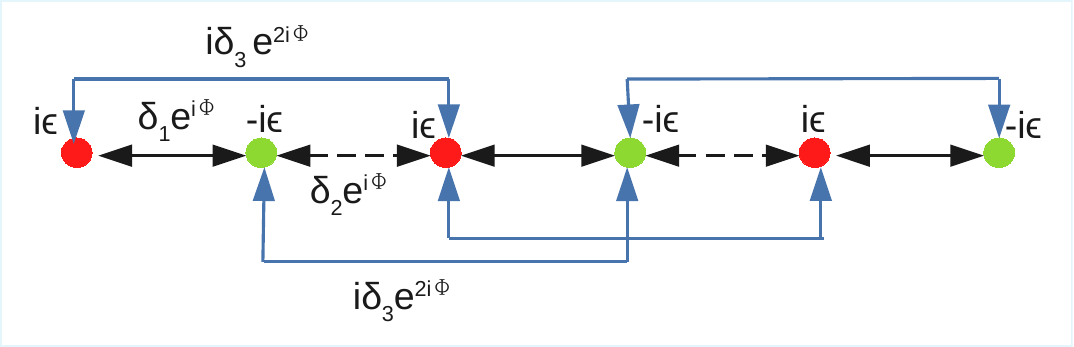}
		\caption{Single chain representation of the SSH model with NNN hopping}
		\label{schematic1}
	\end{subfigure}
	\vfill
	\begin{subfigure}{0.7\linewidth}
                \centering
                \includegraphics[scale=0.35]{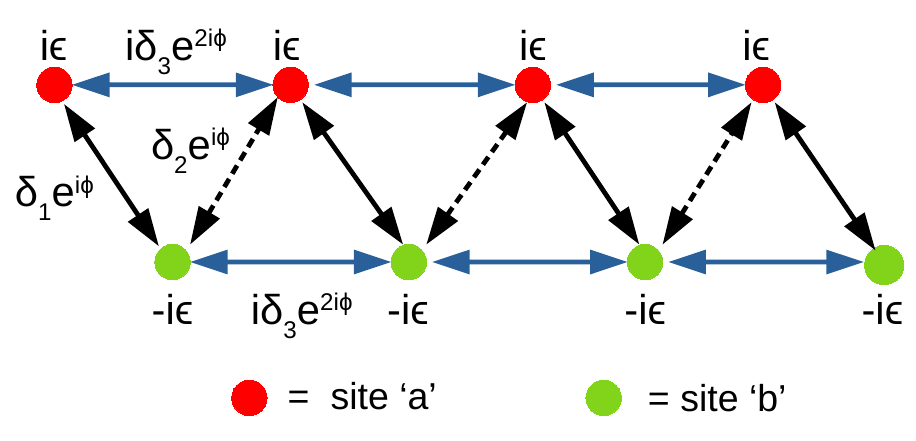}
                \caption{Two-chain representation of the same model with two sublattices shown separately}
                \label{schematic2}
        \end{subfigure}
	\caption{(Color online) Schematic diagram of the lattice model}
	\label{schematic}
\end{figure}

We consider an extended SSH(SSH) model on $N = 2m$ lattice sites described by the bulk Hamiltonian, 
\bea
&& H_{bulk}  =  H_{0} + H_{NNN} + H_{BLG}\nonumber \\
&& H_{0}  =  e^{i \phi} \left ( \delta_1 \sum_{l = 1}^{m}  \ a^{\dagger}_{l} b_l
+ \delta_2 \sum_{l = 1}^{m-1} \ b^{\dagger}_{l} a_{l+1} \right) + h.c. \nonumber \\
&& H_{NNN} = i \delta_3 e^{-2 i \phi} \ \sum_{l=1}^{m-1} \left (  a^{\dagger}_{l+1} a_{l} 
 +  b^{\dagger}_{l+1} b_{l} \right ) + h. c.\nonumber \\
&& H_{BLG} = i \epsilon \sum_{l=1}^{m} \left( a^{\dagger}_{l} a_l - b^{\dagger}_{l} b_{l} \right),
\ \left ( \delta_1, \delta_2, \delta_3, \epsilon,\phi \right ) \in \mathbb{R}
\label{ham_gbc}
\eea
\noindent The schematic diagram of the lattice model is shown in the Fig.~\ref{schematic}. The sub-lattices
`a' and `b' are denoted by the red and green circles, respectively.
\noindent The standard SSH model with OBC is given by $H_0$. The parameters $\delta_1$
and $\delta_2$ denote the intra and inter cell hopping strengths, respectively. The topological phases
are obtained in the system for $\delta_1 < \delta_2$. The SSH model is subjected to an external constant
magnetic field with the total Aharonov–Bohm(AB) flux\cite{Aharonov1959PhysRev} $\Phi$. The
phase factor $\phi$ which modifies the hopping strengths $\delta_1, \delta_2$ is related to the AB-flux
as $\phi=\frac{2 \pi \Phi}{N}$. The term $H_{NNN}$ incorporates next nearest-neighbour interaction for the whole one dimensional system.
It may be recalled that if a closed loop tight binding chain is subjected to an external magnetic field,  the hopping amplitudes
are modified as $t_{ij} \rightarrow t_{ij} exp\left(i\frac{e}{\hbar} \int_{\mathcal{C}} \vec{A} \cdot d\vec{l} \right)$, where
$t_{ij}$ denotes the hopping amplitudes from the site $i$ to site $j$. The integral is taken along the path $\mathcal{C}$ from site $i$ to site $j$.
In our model, intercell and intracell hoppings span one unit distance, while NNN interaction span two
unit distance. Consequently, intercell and intracell hopping amplitudes acquire a phase $e^{i\phi}$, whereas the NNN
hopping amplitudes acquire a phase $e^{2i\phi}$. 
The standard SSH model described by $H_0$ reduces to $m$ disconnected dimers in the flat-band limit, i.e.
either the intercell or the intracell hopping strength vanishes. Such a reduction for $H_0+H_{NNN}$ leads to
a two-leg tight-binding ladder with NNN interaction acting along each leg and $\delta_1/\delta_2$
corresponds to interaction strength across each rung connecting the legs.
The onsite imaginary potential with strength $\epsilon$ is introduced via the non-hermitian term
$H_{BLG}$. The relative sign change between the onsite potentials at the sublattices `a' and 'b' ensures
that loss-gain is balanced.

The bulk Hamiltonian $H_{bulk}$ is subjected to a variety of boundary conditions as 
encoded in $H_{boundary}$,
\bea
&& H_{boundary} = \alpha_{L} \ e^{i\phi} \ b^{\dagger}_{m}a_1 + \alpha_{R} \ e^{-i\phi} \ a^{\dagger}_{1}b_{m}
+ \beta_{L} \ e^{2i\phi} \ a^{\dagger}_{m}a_1\nonumber \\
&& + \beta_{R} \ e^{-2i\phi} \ a^{\dagger}_{1}a_{m} 
+ \beta_{L} \ e^{2i\phi} \ b^{\dagger}_{m}b_1 + \beta_{R} \ e^{-2i\phi} \ b^{\dagger}_{1}b_{m}.
\eea
\noindent The Hamiltonian $H=H_{bulk} + H_{boundary}$ can  be studied with the OBC, PBC, AHBC and GBC
depending on appropriate  choice of the parameters $(\alpha_L, \alpha_R, \beta_L, \beta_R )$. In particular,
\begin{itemize}
\item OBC: $\alpha_L=\alpha_R=\beta_L=\beta_R=0$
\item PBC: $\alpha_L=\alpha_R =\delta_2, \beta_R=-\beta_L = i \delta_3$
\item GBC: There are three distinct possibilities depending on the boundary term being hermitian,
anti-hermitian and non-hermitian. We denote the boundary conditions associated with these four
cases as APBC, HGBC, AHBC and NGBC, respectively. 
\begin{enumerate}
\item APBC: $\alpha_L = \alpha_R = - \delta_2, \beta_R = {\beta}^{*}_L = - i\delta_3$
\item HGBC: $\alpha_L = \alpha_R \neq \delta_2, \beta_R =  {\beta}^{*}_L \neq i \delta_3$
\item AHBC: $\alpha_L=-\alpha_R, \beta_L=-{\beta}^{*}_R$
\item NGBC: $\alpha_L \neq \alpha_R, \beta_L \neq {\beta}^{*}_R$
\end{enumerate}
\end{itemize}
\noindent 
The Hamiltonian $H$ is necessarily non-hermitian due to the balanced loss-gain term, irrespective of the boundary
conditions imposed on the system. There is an additional source of non-hermiticiy for the choice of AHBC and NGBC.
The phase $\phi$ can be completely gauged away from $H$ for the OBC.
In particular, the transformation 
\bea 
a_l \rightarrow e^{-2i(l-1)\phi} a_l, b_l \rightarrow e^{-i(2l-1) \phi} b_l,
\eea
\noindent completely removes $\phi$ from $H$ for $H_{boundary}=0$. However, the boundary terms
$H_{boundary}$ are not invariant under this transformation. Thus, the phase $\phi$ will be considered
as zero for OBC without any loss of generality, and kept for all other choices of boundary conditions. 
It is worth mentioning that $H_{bulk}$ is not invariant under phase rotation if the phase coefficient of 
NNN hopping amplitudes is taken to be $e^{\pm i\phi}$ instead of $e^{\pm 2i\phi}$.
The Hamiltonian $H$ has been analyzed in different limiting cases in earlier studies. The standard SSH model corresponds 
to $\delta_3 = 0$, $\epsilon = 0$ and $\phi = 0$\cite{Su1979Prl}. In the specific limit where $\epsilon = 0$, $\phi = 0$, 
the Hamiltonian represents a Hermitian SSH model incorporating NNN interactions\cite{Li2014prb}. Furthermore, 
when $\delta_3 = 0$ the Hamiltonian describes a $\mathcal{PT}$-symmetric SSH model with BLG, which has also 
been examined in earlier works\cite{Klett2017Pra,Xu2020Pra}.

The extended Hamiltonian $H$ possesses some of the symmetries of the standard SSH Hamiltonian. In particular,
the system is translation invariant for the PBC despite having the loss-gain and NNN term. The Hamiltonian
$H_{bulk}$ is $\mathcal{PT}$ symmetric, where ${\cal{T}}: i \rightarrow - i$ and ${\cal{P}}$ is an
anti-diagonal identity matrix
of order $2m\times 2m$. However,  $H_{boundary}$ is  $\mathcal{PT}$ symmetric only when it is hermitian.
Consequently, the Hamiltonian $H$ loses ${\cal{PT}}$ symmetry for the cases when AHBC and NGBC are used.
The chiral symmetry of the standard SSH model is lost due to the balanced loss-gain terms.
However,in the case of non-Hermitian Hamiltonians, the time-reversal, particle-hole 
and chiral symmetries ramify into two branches, namely $AZ$ and $AZ^{\dagger}$ symmetries\cite{Kawabata2019Prx}. 
The system has the $AZ^{\dagger}$ class particle hole symmetry for $\phi = 0$,
$\beta_{L}^{*} = -\beta_{L}$,$\alpha_{L}^{*} = \alpha_{L}$,$\beta_{R}^{*} = -\beta_{R}$,
$\alpha_{R}^{*} = \alpha_{R}$. The system has particle-hole symmetry  $\mathcal{C}H^{*}\mathcal{C}^{-1} = - H$
with the  unitary matrix $\mathcal{C} = I_{m} \otimes \sigma_{3}$.
In this paper we have restricted the parametric regions $(\delta_1, \delta_2, \delta_3) \geq 0$. 
\section{Periodic Boundary Condition}

In this section, we consider the PBC, i.e. $\alpha_{L} = \alpha_{R} = \delta_2$ and 
$\beta_{R} = -\beta_{L} = i\delta_3$. The Hamiltonian $H$ under PBC with $\phi=\frac{\pi}{2},
\delta_3=\delta_1 \delta_2, \epsilon=\delta_2^2 - \delta_1^2$ was introduced in Ref. \cite{piju-phh}
as a many-particle pseudo-hermitian system with the associated metric $\eta_+$ in the Hilbert
space given by\footnote{The state vector notation
used in Ref. \cite{piju-phh} for $\eta_+$ has been appropriately modified to annihilation and
creation operators.},
\bea
\eta_+= \sum_{i=1}^m \left [ a_i^{\dagger} a_i + b_i^{\dagger} b_i 
+ \left ( \delta_1 a_i^{\dagger} b_i + \delta_2  a_{i+1}^{\dagger} b_i +
h.c. \right ) \right ],\nonumber
\eea
\noindent where positivity of $\eta_+$ is ensured by the sufficient condition
$m \left ( \delta_1^2 + \delta_2^2 \right ) < 1$. In this article, we consider $H$
under PBC for generic values of the parameters $\phi, \delta_1, \delta_2, \delta_3,
\epsilon$, and discuss its pseudo-hermiticity in the momentum space. 

The Hamiltonian $H$  can be diagonalized
exactly in the momentum space. In particular, with the introduction of annihilation operators
$a_k$ and $b_k$ in the momentum-space,
\bea
a_{l} = \frac{1}{\sqrt{m}} \sum_{k} e^{ilk} \ a_{k},
\ b_{l} = \frac{1}{\sqrt{m}} \sum_{k} e^{ilk} \ b_{k}; \ \left(l = 0, 1 \dots ,m\right) \nonumber
\eea
\noindent the Hamiltonian $H$ can be written as, 
\bea
H = \sum_{k} \bp a_{k}^{\dagger} \ b_{k}^{\dagger} \ep H_{k} \bp a_{k} \\ b_{k} \ep \nonumber
\eea
\noindent The Bloch Hamiltonian $H_k$ takes the form,
\bea
H_{k} & = & d_0 I_{2} + \vec{d} \cdot \vec{\sigma},
\eea
\noindent where $\sigma_i's$ are the three Pauli matrices, $I_2$ is the $2 \times 2 $ identity matrix,
and the components of the vector $\vec{d}$ and $d_0$ are given by,
\bea
d_0 & = & 2 \delta_3 \sin(k + 2 \phi) \nonumber \\
d_1 & = & \delta_1 \cos(\phi) + \delta_2 \cos(k + \phi) \nonumber \\
d_2 & = & \delta_2 \sin(k+\phi) - \delta_1 \sin(\phi) \nonumber \\
d_3 & = & i\epsilon \nonumber
\eea
\noindent With the introduction of a complex parameter $d \equiv d_1 + i d_2$ and
the matrices $\sigma_{\pm}:=\frac{1}{2} ( \sigma_1 + i \sigma_2 )$, $H_k$
can be equivalently written as $ H_k=d_0 I_2 + d^* \sigma_+ + d \sigma_- + i \epsilon
\sigma_3$, where a $^*$ denotes complex conjugation.

The Bloch Hamiltonian is ${\mathcal{PT}}$-symmetric with the
identification of ${\mathcal{P}}:=\sigma_1$ and ${\mathcal{T}}: i \rightarrow
- i$. A complementary approach for studying non-hermitian systems is the
pseudo-hermiticity, i. e.  $H_k^{\dagger}=\eta H_k \eta^{-1}$. We provide in this article
a comprehensive discussion on pseudo-hermiticity of $H_k$ for completeness. The matrix $\eta$
\bea
\eta=\frac{{\vert \alpha \vert}{\vert d \vert}}{\epsilon} \sin(\theta_{\alpha}-
\theta_d ) I_2 + \alpha^* \sigma_+ + \alpha \sigma_-, \
\eea
\noindent satisfies the condition of pseudo-hermiticity\cite{pkg-nlse}, where
$\alpha={\vert \alpha \vert} e^{i \theta_{\alpha}},\
d ={\vert d \vert} e^{i \theta_{d}}$. The matrix $\eta$ depends on a
complex parameter $\alpha$, which can be chosen independently of the system parameters,
and is not unique. The operator ${\cal{O}}:= \eta_2^{-1} \eta_1$ is a symmetry
generator of $H_k$, i.e.  $[H_k,{\cal{O}}]=0 $, where  $\eta_1$ and $\eta_2$ correspond
to $\eta$ at $\alpha=\alpha_1$ and $\alpha=\alpha_2$, respectively, with $\theta_{\alpha_1}
\neq \theta_{\alpha_2}$.  The eigenvalues of $\eta$,
$\lambda_{\pm}= \frac{{\vert \alpha \vert}{\vert d \vert}}{\epsilon} \sin(\theta_{\alpha}-
\theta_d ) \pm {\vert \alpha \vert}$,
are positive-definite for $\frac{\vert d \vert}{\epsilon} \sin(\theta_{\alpha}-
\theta_d ) > 1$ with the constraint $0 < \theta_{\alpha} - \theta_d < \pi$ for $\epsilon > 0$ and
$\pi < \theta_{\alpha} - \theta_d < 2 \pi$ for $\epsilon < 0$.
The positive definite metric $\eta_+$ in the Hilbert space is fixed as,
\bea
\eta_+= \frac{{\vert d \vert}}{\vert \epsilon \vert} I_2 +
\alpha^* \sigma_+ + \alpha \sigma_-,  \ \alpha=e^{i (\theta_d +\frac{\pi}{2})}, \
{{\vert d \vert}} > {\vert \epsilon \vert}.\nonumber 
\eea
\noindent for which $\lambda_{\pm}=\frac{{\vert d \vert}}{\vert \epsilon \vert} \pm 1$.
We introduce an operator $\rho:=\sqrt{\eta_+}$ which can be expressed as,
\bea
\rho= C_+ \ I_2 + C_- ( \alpha^* \sigma_+ + \alpha \sigma_-),
\eea
\noindent where $C_{\pm} = \frac{1}{2} \left ( \sqrt{\lambda_+} \pm \sqrt{\lambda_-} \right)$.
The Hamiltonian $h=\rho H_k \rho^{-1}$ is hermitian:
\bea
h = d_0 I_2 + \sqrt{{\vert d \vert}^2 - {\vert \epsilon \vert}^2} \left ( e^{-i \theta_d} \sigma_+ +
e^{i \theta_d}  \sigma_- \right )
\eea
\noindent The eigenvalues $E_{\pm}$ and the corresponding eigenfunctions $\phi_{\pm}$ of $h$ are,
\bea
E_{\pm}(k) = d_0 \pm \sqrt{{\vert d \vert}^2 - {\vert \epsilon \vert}^2},\
\phi_{\pm}= \frac{1}{\sqrt{2}} \bp 1\\ \pm e^{i \theta_d} \ep.
\label{ev-ef}
\eea
\noindent The Hamiltonian $H_k$ has the eigenvalues $E_{\pm}(k)$ with the corresponding eigenfunction
$\psi_{\pm}=\rho^{-1} \phi_{\pm}$. A modified norm $\langle\cdot | \cdot \rangle_{\eta_+} =
\langle \cdot |\eta_+ \cdot \rangle$ is used in the Hilbert space of a pseudo-hermitian system,
and $\langle \psi_{\pm}|\eta_+ \psi_{\mp} \rangle=\langle \phi_{\pm} | \phi_{\mp}\rangle=0,
\langle \psi_{\pm}|\eta_+ \psi_{\pm} \rangle=\langle \phi_{\pm} | \phi_{\pm}\rangle=1$. 
The Hamiltonian $h$ with the standard norm or the Hamiltonian $H_{k}$ with the modified norm 
describe the same physical scenario. \\  
{\bf \underline{Zak Phase} :}  
The Berry curvature and the associated gauge potential can be defined in terms of the modified
norm in the Hilbert space of $H_k$ or equivalently with the standard norm in the Hilbert space of $h$\cite{Pkg2012JPhys}.
The eigen functions $\psi_{\pm}$ and $\phi_{\pm}$ do not depend on the
strength of the NNN interaction $\delta_3$, since $\delta_3$ appears in $d_0$ and 
$\left [d_0 I_2, \vec{d} \cdot \vec{\sigma} \right ] =0 $. 
The global Berry phase for $\delta_3=0$ has been calculated in Ref. \cite{Liang2013Pra,Lieu2018Prb} which
is equally valid for $\delta_3 \neq 0$. Thus, {\it the NNN interaction has
no contribution to the Zak phase}. Further, the classification of topologically  trivial and
non-trivial phases is independent of $\delta_3$ \textemdash the topologically  trivial and
non-trivial phases are characterized by $\delta_1 > \delta_2$ and $\delta_1 < \delta_2$,
respectively.

\subsection{Spectrum \& Eigenstate}

We remove the restriction $\vert d \vert > \vert \epsilon \vert $ henceforth,
and allow the eigenvalues of $H_k$ to be real as well as complex. The energy
eigenvalues of $H_k$ are still given by Eq. (\ref{ev-ef}) despite the operator
$\eta_+$ becoming non-positive-definite and $\rho, h$ becoming non-hermitian.
We use explicit forms of $d_0$ and $R=\sqrt{{\vert d \vert}^2 - {\vert \epsilon \vert}^2}$
to express $E_{\pm}$ as,
\bea
E_{\pm} =   2\delta_3 \sin(k+2\phi) \pm 
\sqrt{ \delta_1^2 + \delta_2^2 + 2 \delta_1 \delta_2 \cos(k+2\phi) - \epsilon^2}\nonumber
\eea
\noindent where $k$ is quantized as $k = \frac{2s\pi}{m}$, \ $s = 0,1,2....(m-1)$
due to the PBC. The eigenvalues can be real or complex depending on $R$ being real
or purely imaginary, respectively. The entire spectra
is real for the condition $\left( \frac{\delta_1}{\epsilon}  -
 \frac{\delta_2}{\epsilon} \right)^{2} \geq 1$, while complex eigenvalues appear for
$\left( \frac{\delta_1}{\epsilon}  -  \frac{\delta_2}{\epsilon} \right)^{2} < 1 $.
The condition for reality of the spectra does not depend at all on the hopping strength of the NNN
interaction $\delta_3$. However, the nature of band gap crucially depends on $\delta_3$ for entirely
real spectra as well as for complex eigenvalues. The phase $\phi$ shifts the  values of sine
and cosine functions for each $k$ and reality condition of the spectra is independent of it. The gradient
of $E$ as a function of $\phi$ changes and is responsible for persistent current in the system which
will be discussed later. We discuss below the band structure of the system in the thermodynamic limit
$N \rightarrow \infty$ for which the momentum $k$ may be treated as a continuous variable and $\phi \rightarrow 0$.

\subsection{Band structure}

The SSH model may admit point gap for entirely real spectra and line gap for complex
eigenvalues. The Hamiltonian $\tilde{H}_k=H_k - d_0 I_2$ has the following chiral symmetry
$\sigma_3 \tilde{H}_k^* \sigma_3 = - \tilde{H_k}$, where a $*$ denotes complex conjugation.
This symmetry implies that if $E$ is an eigenvalue of $\tilde{H}_k$, then $-E^*$ is also
its eigenvalue. For the case of unbroken ${\mathcal{PT}}$ symmetry, $E$ is real, and both
$E$ and $-E$ are eigenvalues of $\tilde{H}_k$. This implies that the $2 \delta_3 \sin(k)
\pm E$ are eigenvalues of $H_k$.

\begin{figure}
\centering
        \begin{subfigure}{0.32\linewidth}
                \centering
                \includegraphics[width=\textwidth]{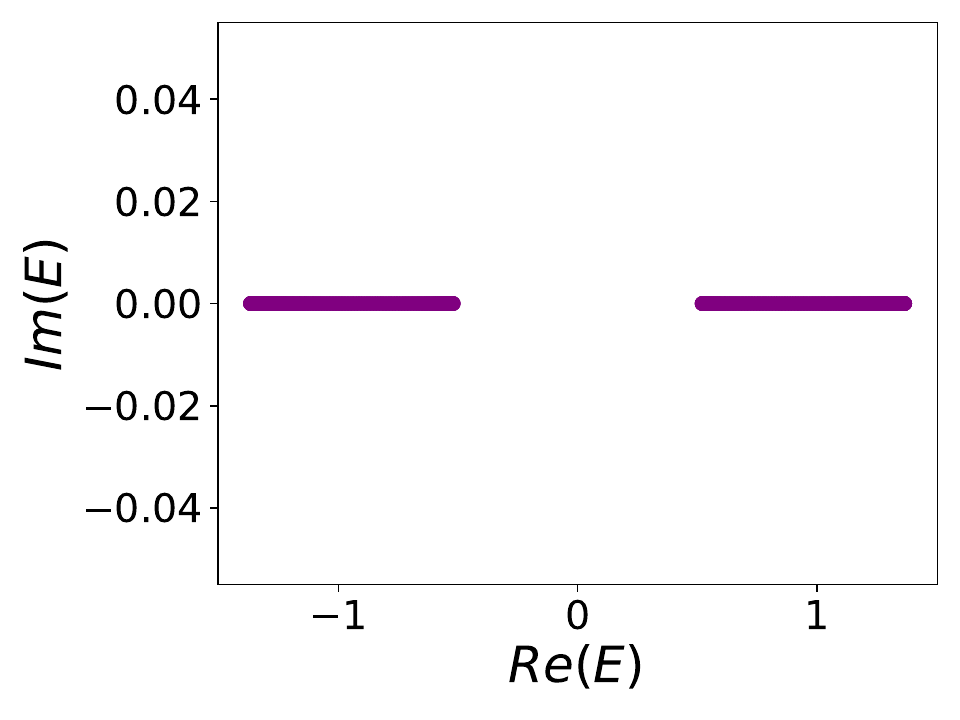}
                \caption{}
                \label{}
        \end{subfigure}
        \hfill
        \begin{subfigure}{0.32\linewidth}
                \centering
                \includegraphics[width=\textwidth]{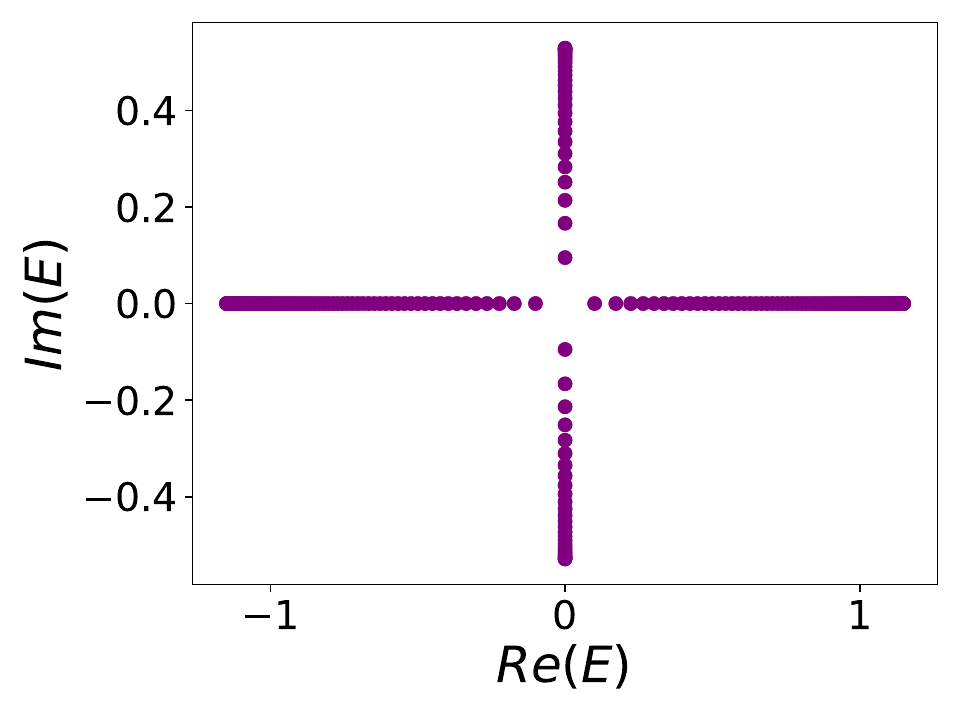}
                \caption{}
                \label{}
        \end{subfigure}
	\hfill
	\begin{subfigure}{0.32\linewidth}
                \centering
                \includegraphics[width=\textwidth]{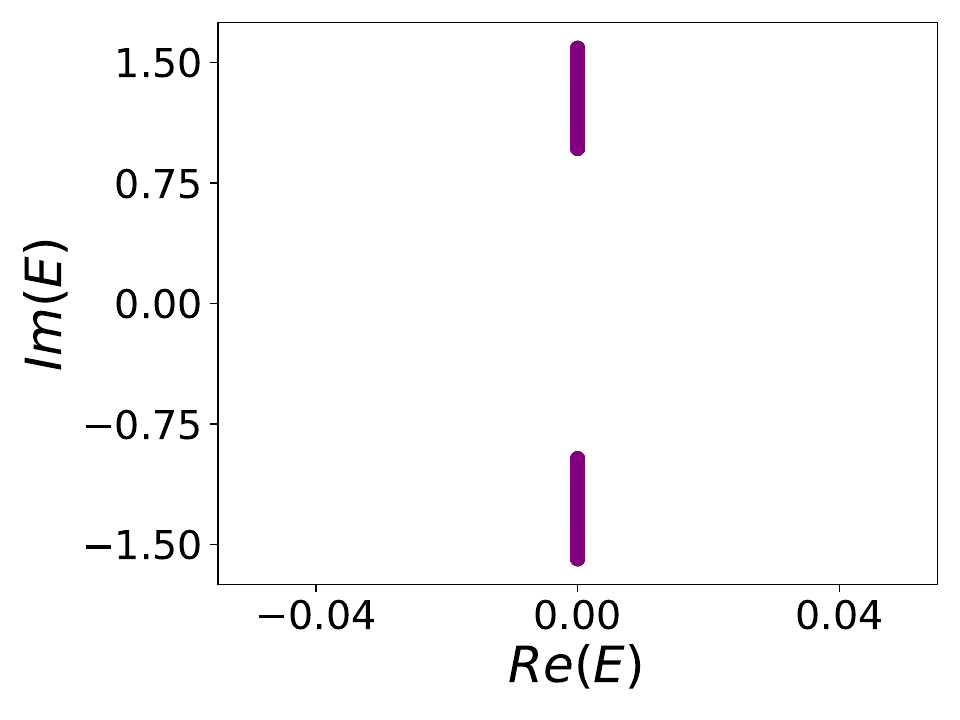}
                \caption{}
                \label{}
        \end{subfigure}
        \vfill
        \begin{subfigure}{0.32\linewidth}
                \centering
                \includegraphics[width=\textwidth]{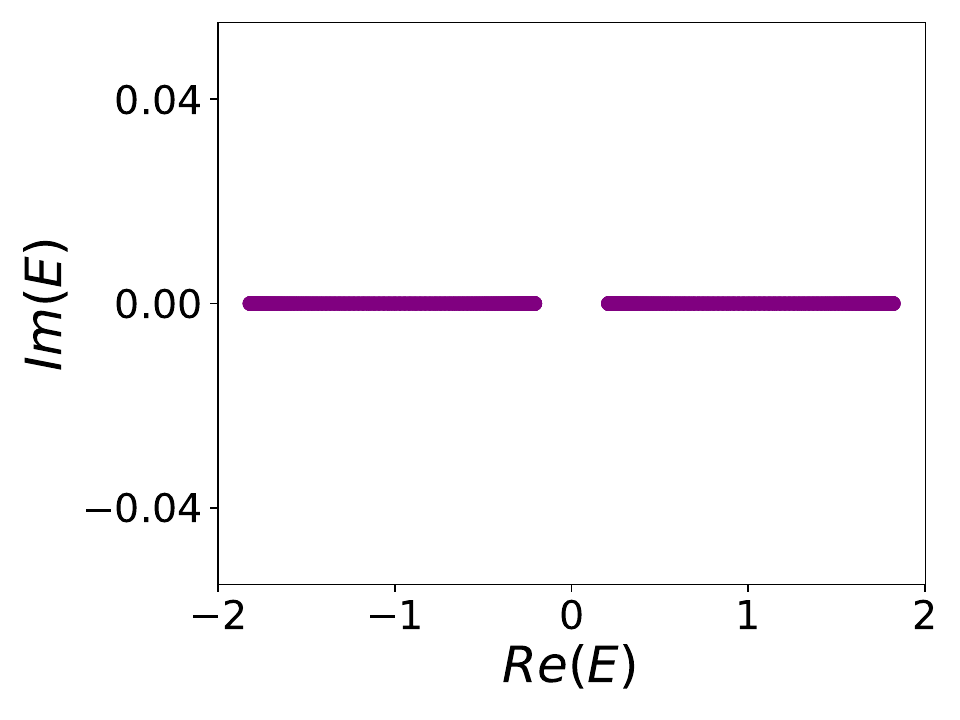}
                \caption{}
                \label{}
        \end{subfigure}
	\hfill
	\begin{subfigure}{0.32\linewidth}
                \centering
                \includegraphics[width=\textwidth]{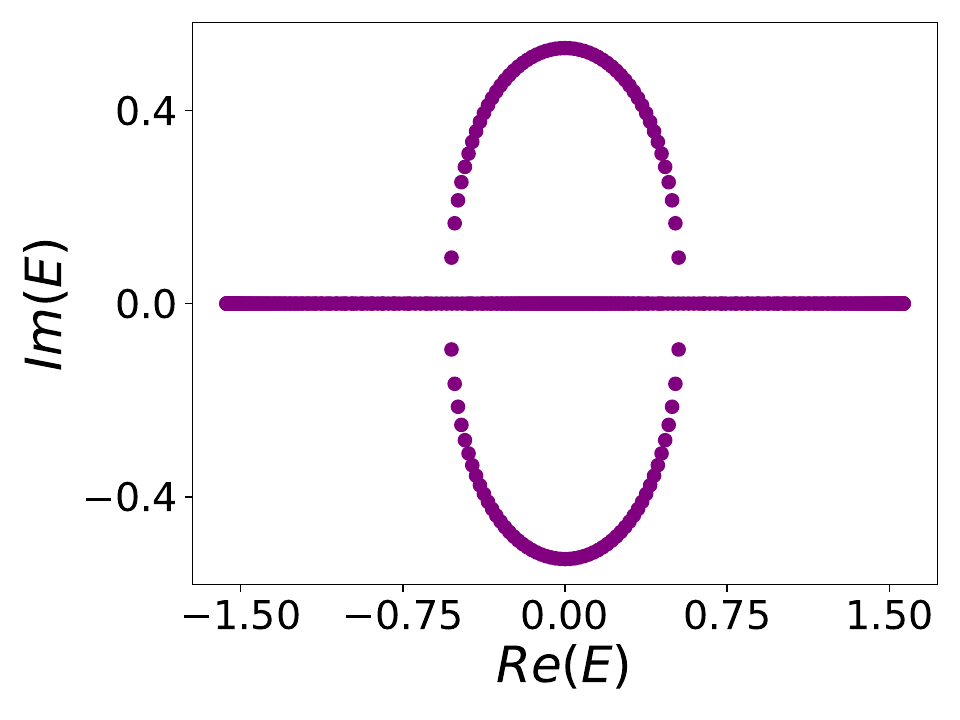}
                \caption{}
                \label{}
        \end{subfigure}
	\begin{subfigure}{0.32\linewidth}
                \centering
                \includegraphics[width=\textwidth]{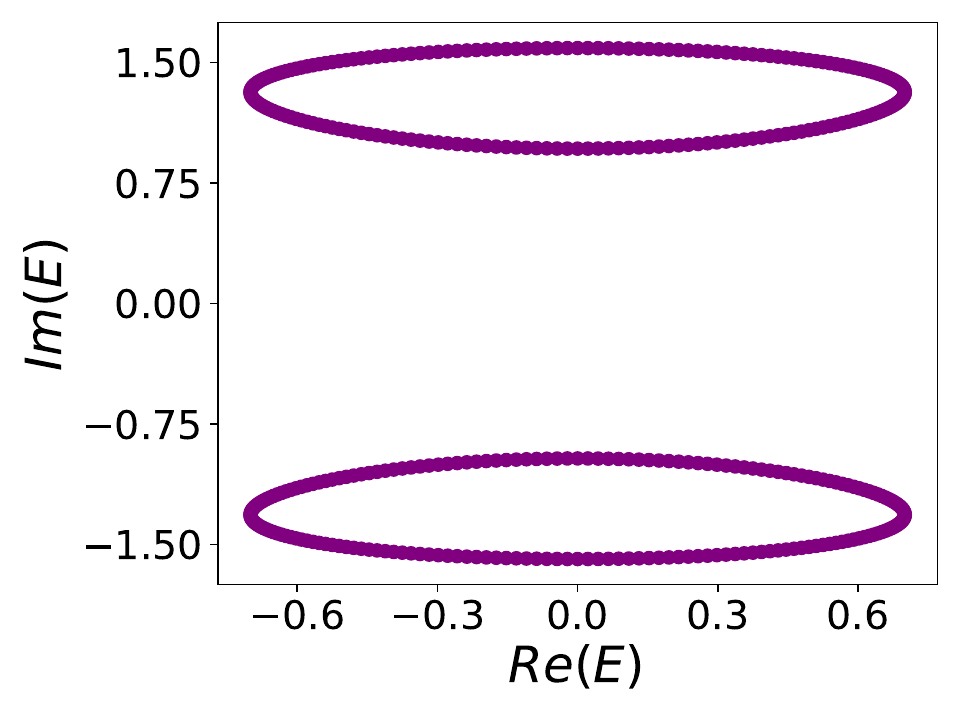}
                \caption{}
                \label{}
        \end{subfigure}

\caption{(Color online) Plot of the real vs. imaginary part of the eigenspectra under
the PBC in topologically trivial region; 
Parameter Values : $\phi = 0$, $\delta_1 = 1$, $\delta_2 = 0.4$, $N = 400$; 
Fig.~a : $\epsilon = 0.3$,$\delta_3 = 0$; Fig.~b: $\epsilon = 0.8$, $\delta_3 = 0$;
Fig.~c : $\epsilon = 1.7$,$\delta_3 = 0$; Fig.~d: $\epsilon = 0.3$, $\delta_3 = 0.35$;
Fig.~e : $\epsilon = 0.8$,$\delta_3 = 0.35$; Fig.~f: $\epsilon = 1.7$, $\delta_3 = 0.35$.}
\label{re_vs_imag_pbc1}
\end{figure}

\begin{figure}
\centering
        \begin{subfigure}{0.48\linewidth}
                \centering
                \includegraphics[width=\textwidth]{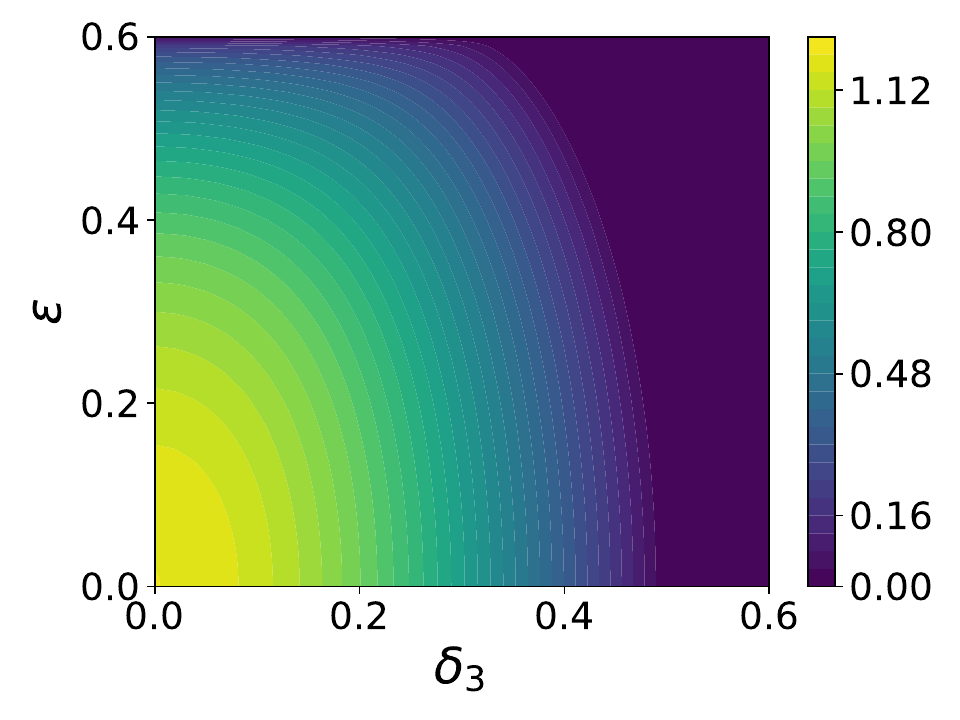}
                \caption{}
                \label{bg1_pbc}
        \end{subfigure}
        \hfill
        \begin{subfigure}{0.48\linewidth}
                \centering
                \includegraphics[width=\textwidth]{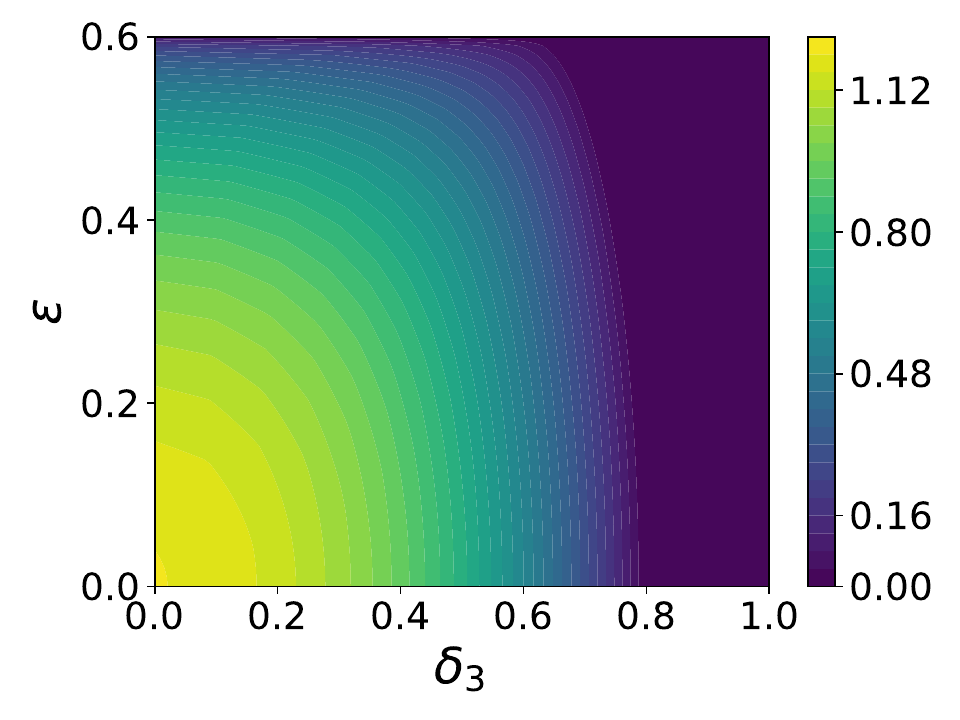}
                \caption{}
                \label{}
        \end{subfigure}
	\caption{(Color online) Contour plot of the band gap in PBC with the variation of the $\delta_3$ and 
	$\epsilon$. Parametric values : $\phi = 0$, (a) : $\delta_1 = 1$, $\delta_2 = 0.4$ ; 
	(b): $\delta_1 = 1$, $\delta_2 = 1.6$.}
	\label{bandgap}
\end{figure}

The  gapless mode exists  if $Det(H_k)=0$ for at least one $k$, i. e. $d_0 = \pm R$. 
The condition for the existence of band gap is $f(k) \equiv Det(H_k) \neq 0 \ \forall \ k$.
The function $f(k)$ has the form,
\bea
f(k) = 4 \delta_3^2 \sin^2 k - \left ( \delta_1^2 +\delta_2^2 + 
2 \delta_1 \delta_2 \cos k-\epsilon^2 \right ).
\eea
\noindent The values of $f(k)$ at its extrema are given by,
\bea
f(0) & = & f(2 \pi)=- \left [ \left \{ \left ( \delta_1 - \delta_2 \right )^2 - \epsilon^2 \right \}
+ 4 \delta_1 \delta_2 \right ],\nonumber \\
f(\pi) & = & - \left [ \left (\delta_1-\delta_2 \right )^2 - \epsilon^2 \right ]\nonumber \\
f(\tilde{k}_{\pm}) & = & \underbrace{4 \delta_3^2 + \frac{\delta_1^2 \delta_2^2}{4 \delta_3^2} +
\epsilon^2}_{t_1} -\underbrace{(\delta_1^2+\delta_2^2)}_{t_2} \equiv t_1-t_2,
\eea
\noindent where $\tilde{k}_{\pm} \equiv \pm \cos^{-1} \left ( -
\frac{ \delta_1 \delta_2 }{4 \delta_3^2} \right )$. The extrema at
$k=\tilde{k}_{\pm}$ exists only if $\delta_1 \delta_2 \leq 4 \delta_3^2$ and $\delta_3 \neq 0$.
The parameter space can be divided into two regions based on the
number of extrema of $f(k)$ \textemdash (i) Region-I: 
$\delta_1 \delta_2 > 4 \delta_3^2$ and (ii) Region-II: $\delta_1 \delta_2 < 4 \delta_3^2$.
The Region-II does not exist in the SSH model without the NNN interaction.
We denote the minimum and maximum of $f(k)$ as $f_{min}$ and $f_{max}$, respectively.
The condition $f(k) \neq 0 \ \forall \ k$ can be satisfied if either $f_{min} > 0 \ \forall
\ k$ or $f_{max} <0 \ \forall \ k$. We now discuss band gaps for three cases characterized
by (i) entirely real energy eigenvalues, (ii) real plus complex energy eigenvalues,
and (iii) entirely complex energy eigenvalues:
\begin{itemize}
\item $\epsilon < {\vert  \delta_1  -  \delta_2  \vert}$: The spectra
is entirely real. The function $f(k))$ has local minima at $k=0, 2 \pi$ and
a local maximum at $k=\pi$ in Region-I. The band gap exists, since both $f_{min}=f(0)=f(2 \pi)$
and $f_{max}=f(\pi)$ are negative definite. In Region-II, the function $f(k))$ has local
minima at $k=0, \pi, 2 \pi$ and maxima at $k=\tilde{k}_{\pm}$. The band gap exists for
$t_1 < t_2$ for which $f_{min}=f(0)=f(2\pi)$ and $f_{max}=f(\tilde{k}_{\pm})$ are
negative definite.

\item ${\vert  \delta_1 - \delta_2  \vert} < \epsilon < (\delta_1 + \delta_2)  $: The energy spectra
consists of real as well as complex eigenvalues. There are no band gaps in Region-I,
since $f_{min}=f(0)=f(2\pi)$ is negative definite, while $f_{max}=f(\pi) > 0$. There
are no band gaps in Region-II either, since $f(k)=0$ for at least one value of $k$
irrespective of the value of $f(\tilde{k}_{\pm})$.

\item $\epsilon > (\delta_1  +  \delta_2) $:  The
energy spectra is entirely complex. The real part of the energy is continuous and there
is no real energy gap. However, there exists line gap in Region-I, since both $f_{min}=f(0)$
and $f_{max}=f(\pi)$ are positive definite. In Region-II, the line band gap exists for
$t_1 > t_2$. The bands above and below the line $Im(E)=0$ do not meet.  
\end{itemize}
\noindent We plot real part vs imaginary part of the energy eigenvalues in Fig. \ref{re_vs_imag_pbc1}.
The first and the second rows in the panel consist of plots corresponding to the  Region-I and 
Region-II, respectively. Further, figures depicting the three cases discussed above, corresponding to allowed ranges of $\epsilon$, are included
along each row. The band structure representing two closed loops symmetrically placed around the $Im(E)$ axis
arises solely due to the NNN interaction. The contour plot of band gap as a function
of $\epsilon$ and $\delta_3$ is given in Fig. \ref{bandgap}.

\subsection{Persistent Current}

\begin{figure}
\centering
        \begin{subfigure}{0.48\linewidth}
                \centering
                \includegraphics[width=\textwidth]{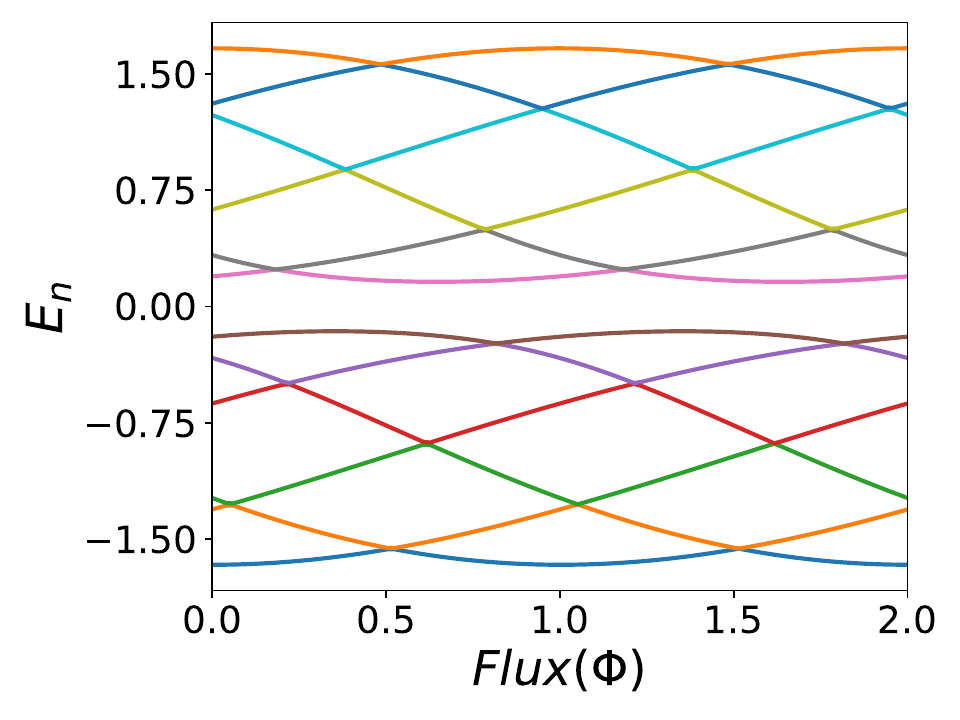}
                \caption{}
                \label{}
        \end{subfigure}
        \hfill
        \begin{subfigure}{0.48\linewidth}
                \centering
                \includegraphics[width=0.9\linewidth]{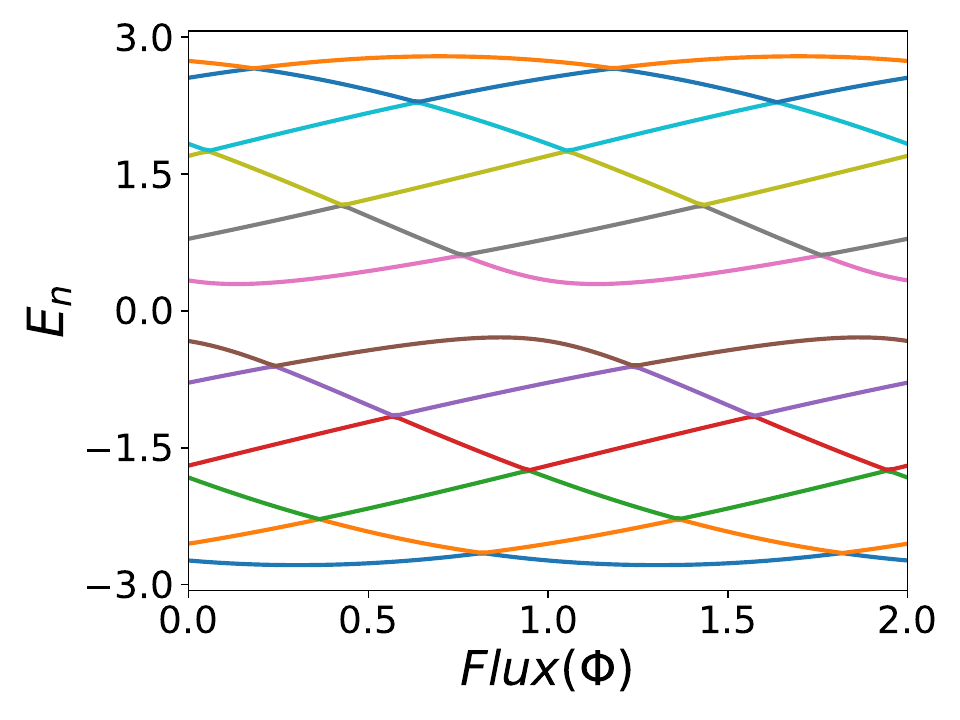}
                \caption{}
                \label{}
        \end{subfigure}
	\vfill
	\begin{subfigure}{0.48\linewidth}
                \centering
                \includegraphics[width=\textwidth]{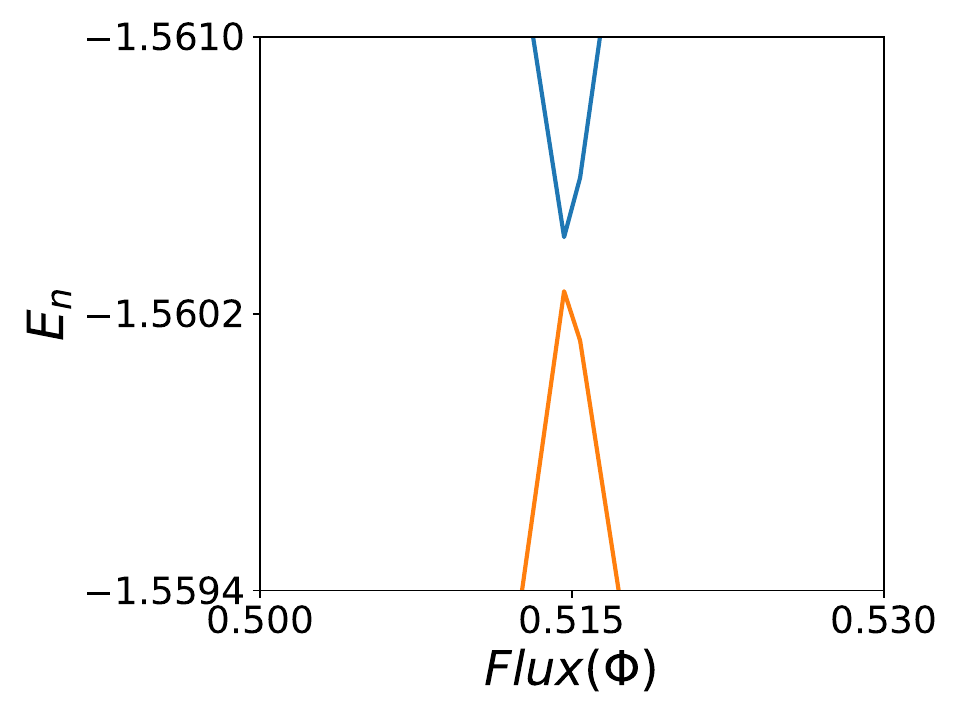}
                \caption{}
                \label{elong1}
        \end{subfigure}
        \hfill
        \begin{subfigure}{0.48\linewidth}
                \centering
                \includegraphics[width=0.9\linewidth]{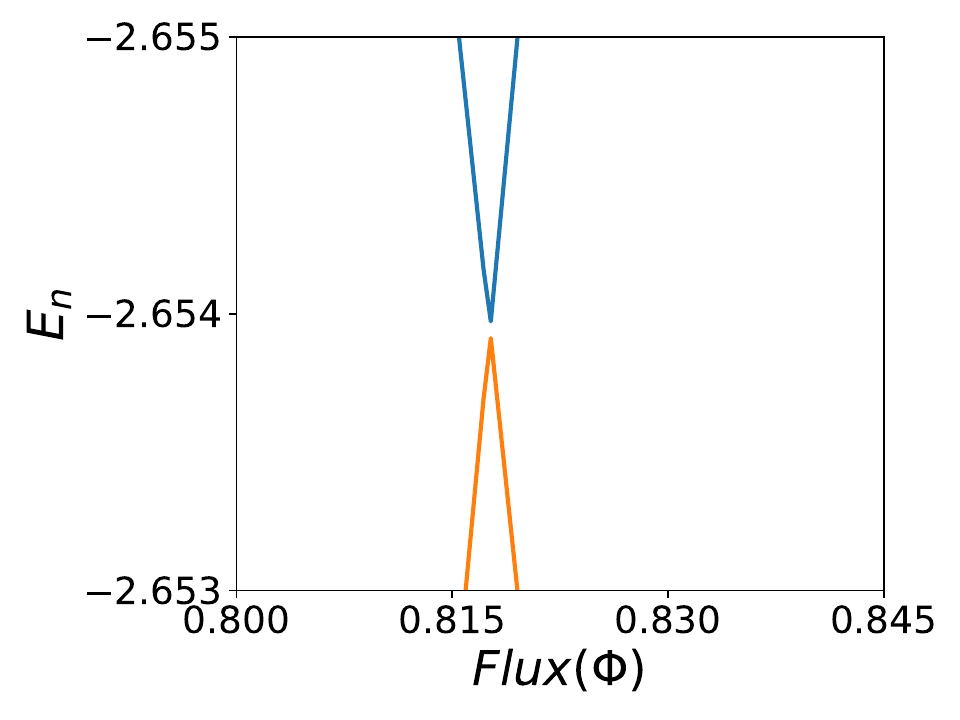}
                \caption{}
                \label{elong2}
        \end{subfigure}
	\caption{(Color online) Plot of eigenvalues with respect to $\Phi$ under PBC. In second row
	a portion of the plots in the first row has been elongated. In 
	Fig.~(a) we have considered $\delta_1 > \delta_2$ and in Fig.~(b) $\delta_1 < \delta_2$.
	Parameter Values : $\epsilon = 0.5$, $\delta_3 = 0.3$, $N = 12$, $\delta_1 = 1$,  
	Fig(a),Fig(c) : $\delta_2 = 0.4$; Fig(b),Fig(d) : $\delta_2 = 1.6$.}
        \label{energy_phi_pbc}
\end{figure}

The band spectra have been discussed in the thermodynamic limit $N \rightarrow \infty$ for which $\phi \rightarrow 0$.
The eigenspectra of the Hamiltonian (\ref{ham_gbc}) is shown in Fig.~\ref{energy_phi_pbc} as a function of $\Phi$
for $N=12$. The size of the system is taken to be small for a better presentation of the results \textemdash appearance of many
states for large $N$ makes the figure clumsy. The level repulsion with the separation of the
neighbouring points being of the order of $10^{-3}$ can be seen in an amplified version shown in
Figs. (\ref{elong1}) and (\ref{elong2}). This is partly due to numerical approximations, and partly due to
the finite size effect. The apparent level repulsion disappears for large $N$. This is consistent with the fact that
the system is integrable. The variation in  the energy with respect to the flux is a signature of persistent current
in the system. We discuss below persistent current in the parametric region $\epsilon < \vert \delta_1 - \delta_2 \vert$
for which real band gap exists.

The study of the persistent current is significant in the context of quantum phenomena in the condensed 
matter physics. The possibility of the persistent current in a non-superconducting metal ring due to an 
external magnetic field was first theoretically proposed by B\"uttiker,Imry and Landauer\cite{Buttiker1983Pla}. Recently, study
of the persistent current in the SSH model has drawn attention of the researchers\cite{Maity2023ScRep}.
We consider that $N_{e}$ number of electron populate the lowest $N_{e}$ number of energy states 
such that the ground state energy $E_{g} = \sum_{j=0}^{N_{e} - 1} E_{j}$. 
The first order derivative of the ground state energy $E_g$ with respect to the flux $\Phi$ is
a measure of the circular current in the system,
\bea
I & = & -\frac{\partial E_{g}}{\partial \Phi}. \nonumber 
\eea
\noindent We refer to a few earlier works where persistent current in a non-hermitian system is obtained 
from the derivative of complex eigenvalues\cite{Cayao2024Prb,Li2024Prb} and the expectation values are obtained by 
considering left-right or right-right eigenvectors\cite{Kornich2023Prl}. However, such approaches show anomalies 
at exceptional points\cite{Shen2024Prl}. 
We restrict our discussions in this article to the case of entirely real spectra of the system
such that $I$ is always real, and there is no ambiguity in sorting out the energy eigen values of
a single-particle in ascending order while filling up the Fermi level. It may be checked that an expression of $I$
defined in terms of left and right eigenvectors of the associated non-hermitian Hamiltonian exactly
reduces to the above expression for the special case when the system admits entirely real spectra.
The transition from entirely real spectra to that of complex-conjugate pairs occurs at the exceptional
points/lines/surfaces. We have restricted our discussion to the entirely real spectra, and there is
no scope of encountering any exceptional points.

\begin{figure}
\centering
        \begin{subfigure}{0.48\linewidth}
                \centering
                \includegraphics[width=\textwidth]{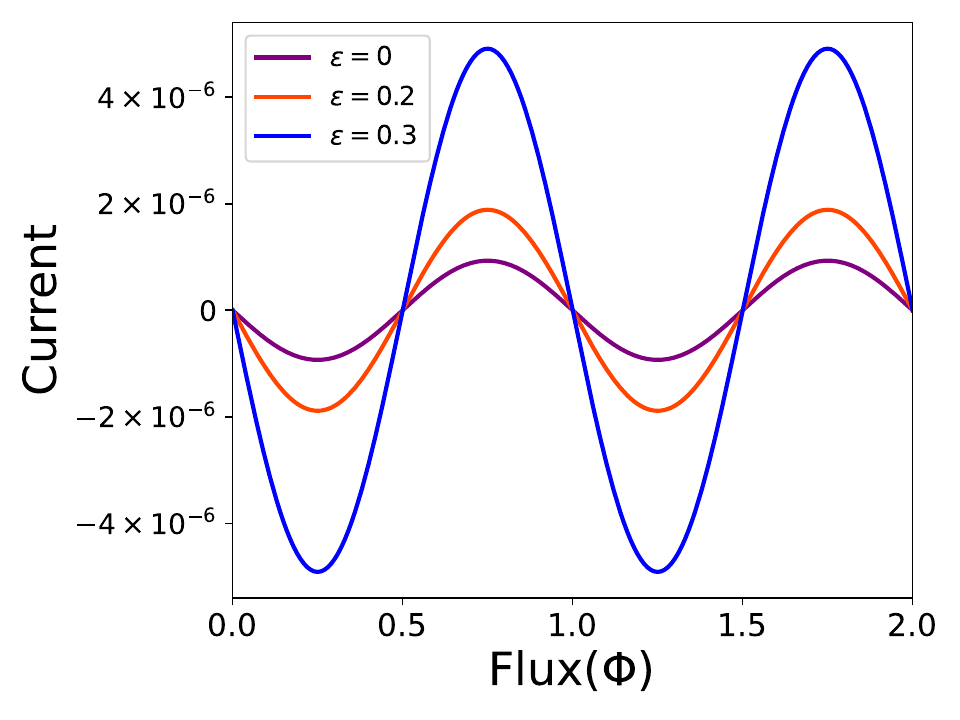}
                \caption{}
                \label{chf1}
        \end{subfigure}
        \hfill
        \begin{subfigure}{0.48\linewidth}
                \centering
                \includegraphics[width=\textwidth]{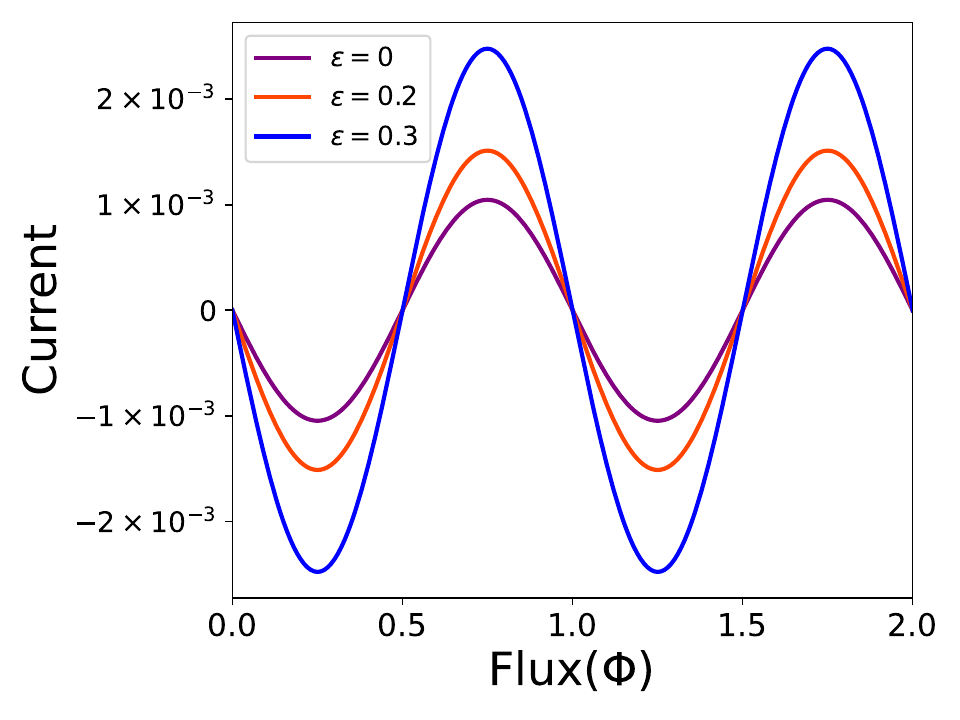}
                \caption{}
                \label{chf2}
        \end{subfigure}
	\caption{(Color online) Plot of the persistent current with the flux in the half filled limit under PBC. 
	Parametric value : $N = 30$, $\delta_3 = 0.3$;  Fig(a): $\delta_1 = 1$, $\delta_2 = 0.4$; 
	Fig(b): $\delta_1 = 1$, $\delta_2 = 1.6$}
	\label{current_hf}
\end{figure}

\begin{figure}
\centering
        \begin{subfigure}{0.49\linewidth}
                \centering
                \includegraphics[width=\textwidth]{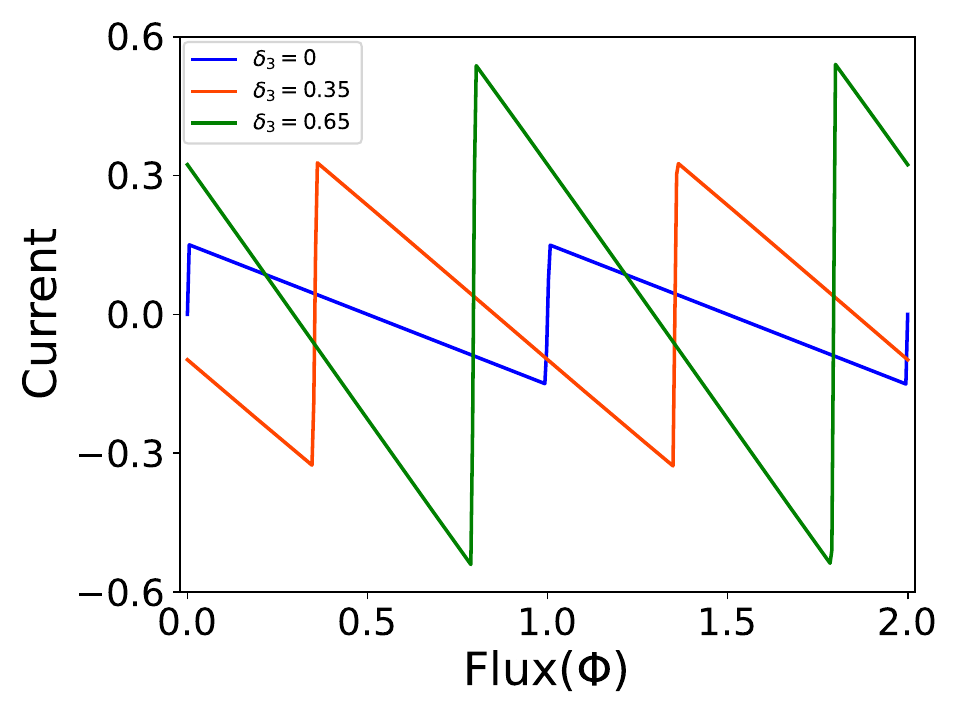}
                \caption{}
                \label{cqf1}
        \end{subfigure}
        \hfill
        \begin{subfigure}{0.49\linewidth}
                \centering
                \includegraphics[width=\textwidth]{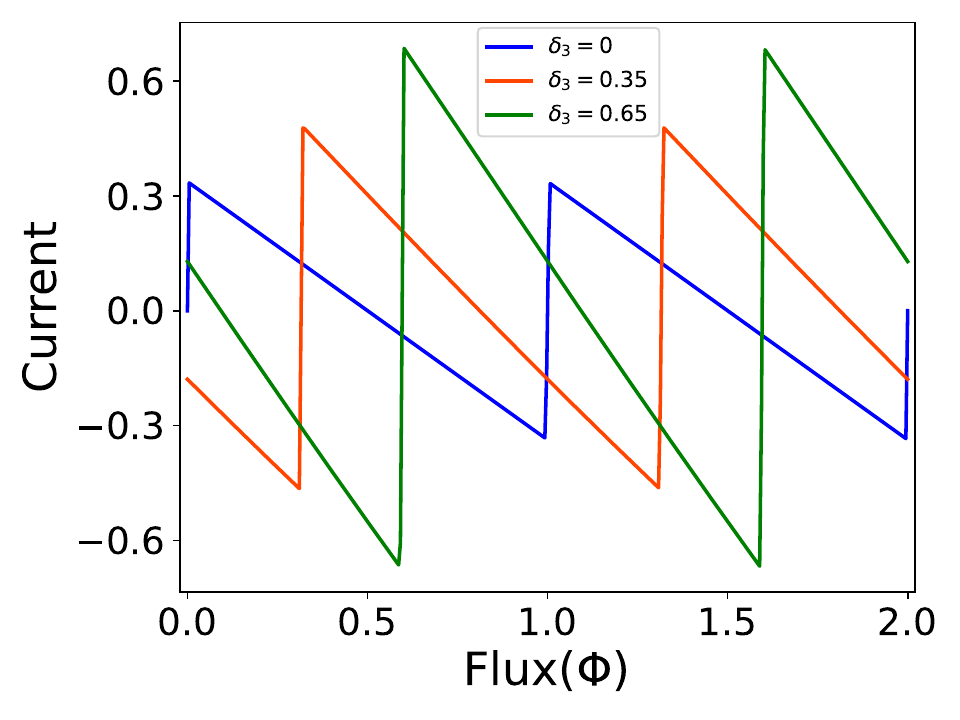}
                \caption{}
                \label{cqf2}
        \end{subfigure}
	\caption{(Color online) Plot of the persistent current with the flux in the quarter filled limit under PBC.
        Parametric value : $N = 32$, $\epsilon = 0.3$, $\delta_1 = 1$;  Fig(a): $\delta_2 = 0.4$;
        Fig(b): $\delta_2 = 1.6$}
        \label{current_qf}
\end{figure}

\begin{figure}
\centering
        \begin{subfigure}{0.46\linewidth}
                \centering
		\includegraphics[width=\textwidth]{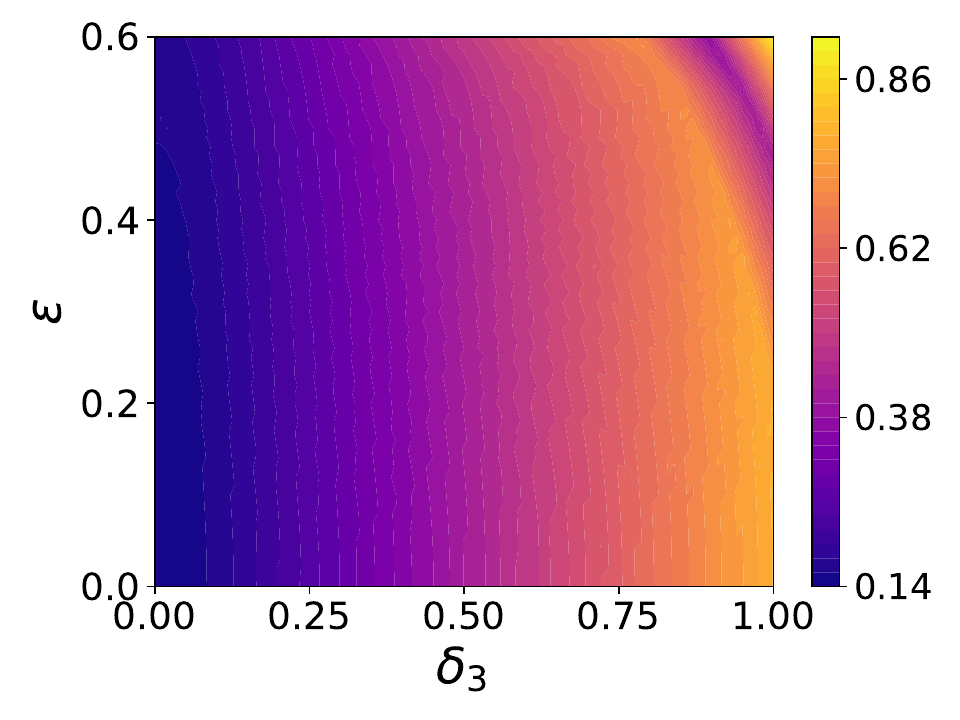}
                \caption{}
                \label{cur1_qf}
        \end{subfigure}
        \hfill
        \begin{subfigure}{0.46\linewidth}
                \centering
                \includegraphics[width=\textwidth]{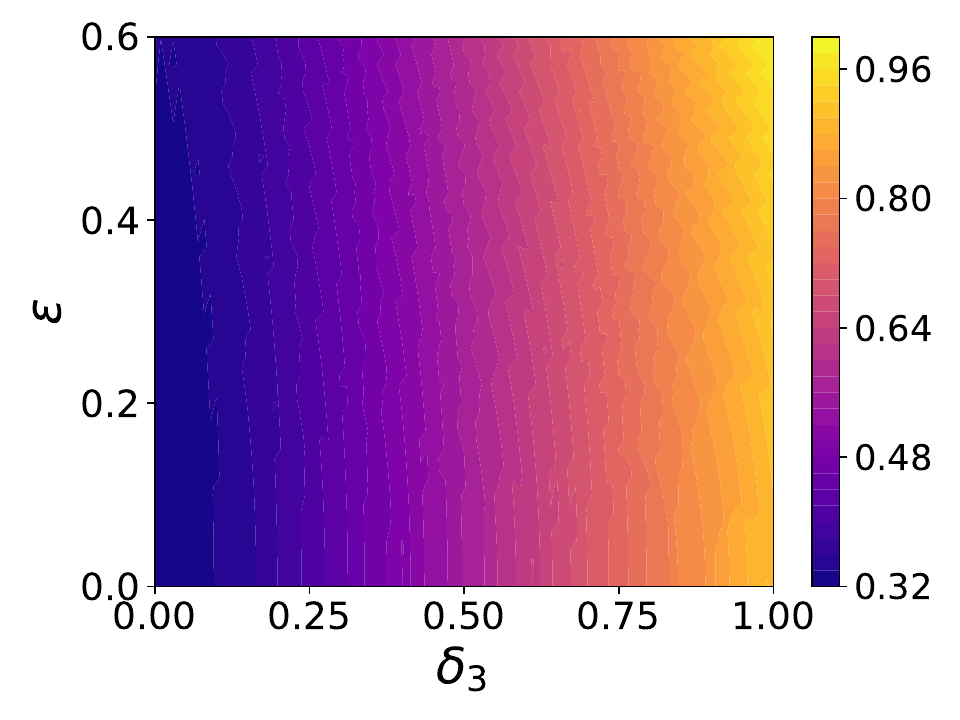}
                \caption{}
                \label{cur2_qf}
        \end{subfigure}
	\vfill
	\begin{subfigure}{0.46\linewidth}
                \centering
                \includegraphics[width=\textwidth]{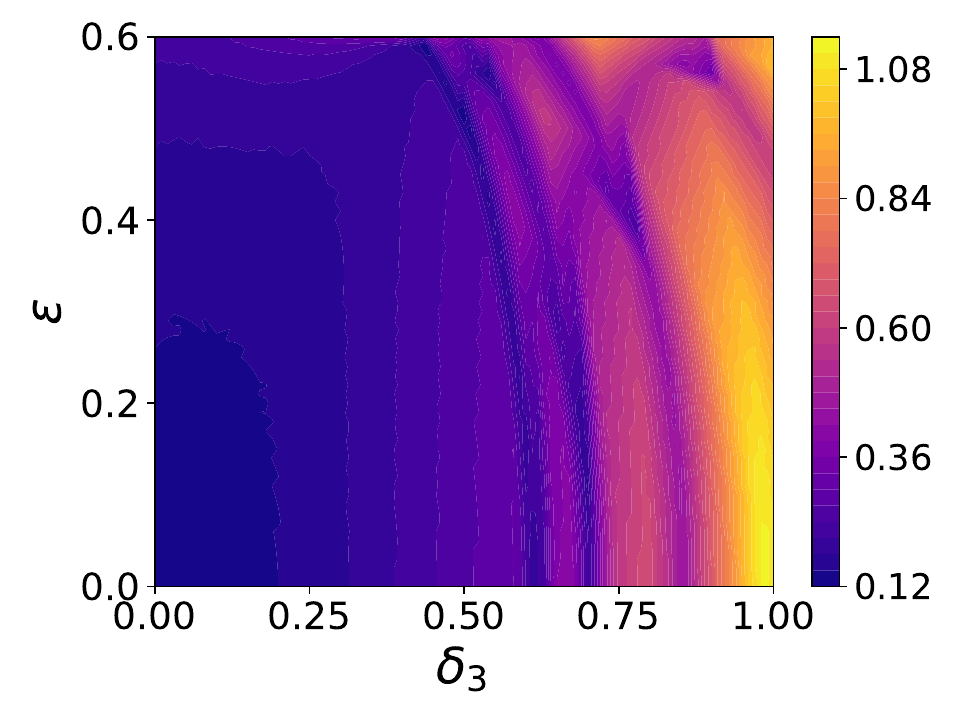}
                \caption{}
                \label{cur1_3b8f}
        \end{subfigure}
        \hfill
        \begin{subfigure}{0.46\linewidth}
                \centering
                \includegraphics[width=\textwidth]{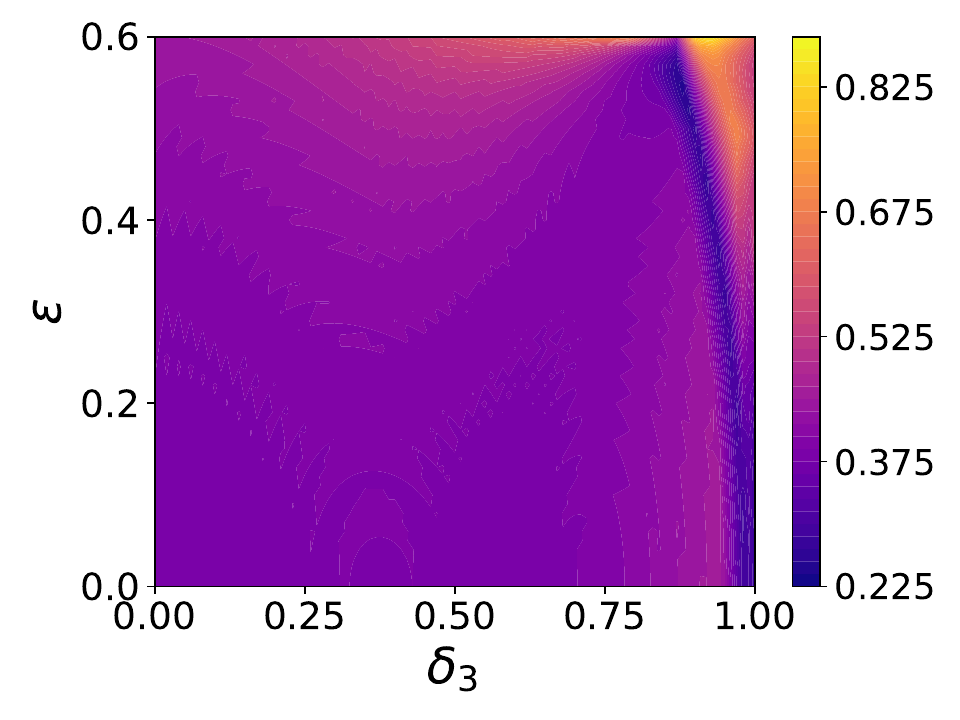}
                \caption{}
                \label{cur2_3b8f}
        \end{subfigure}
	\caption{(Color online) Contour plot of the peak values of the persistent current under PBC. 
	Current has been calculated in the flux range $0\leq \Phi \leq 1$. 
        Parametric values : $N = 32$, $\delta_1 = 1$;  Fig(a): $\delta_2 = 0.4$, $N_e = 8$;
	Fig(b): $\delta_2 = 1.6$, $N_e = 8$; Fig(c): $\delta_2 = 0.4$,$N_e = 12$; 
	Fig(d) : $\delta_2 = 1.6$,$N_e = 12$.}
        \label{current_peak}
\end{figure}

The expression of persistent current is obtained as 
\bea
I    & = & I_1 + I_2 \\ 
I_1  & = & - \frac{4 \pi}{N} \sum_{s_1} \Big[ 2\delta_3 \cos(\frac{2 \pi s_1}{m}+2\phi) +
\frac{\delta_1 \delta_2}{R_{s_1}} \sin(\frac{2 \pi s_1}{m} + 2\phi) \Big]\nonumber \\ 
I_2 & = & - \frac{4 \pi}{N} \sum_{s_2} \Big[ 2\delta_3 \cos(\frac{2 \pi s_2}{m}+2\phi) -
\frac{\delta_1 \delta_2}{R_{s_2}} \sin(\frac{2 \pi s_2}{m} + 2\phi) \Big]\nonumber 
\eea
\noindent where $R_{s_{i}} = \sqrt{\delta^2_{1} + \delta_{2}^2 + 2\delta_1\delta_2 \cos\left(\frac{2\pi s_{i}}{m} 
+ 2\phi\right) - \epsilon^2}$. The summation runs over the possible set of values of $s_1,s_2$ corresponding to 
the lowest $N_{e}$ number of energy states, where the indices $s_1$ and $s_2$ correspond to the lower and upper bands, 
respectively. In the presence of a bandgap, the half-filled limit corresponds to fully occupied  
lower band, i.e., $s_1 = 0, 1, \dots ,(m-1)$, while the upper band remains unoccupied, implying $I_2 = 0$ i.e., $I = I_1$. 
The summation in the first term of $I_1$ can be computed analytically by using the identity,
\bea
\sum_{s=0}^{m-1} \cos(s y+x) = \cos(x + \frac{m-1}{2} y) \sin(\frac{m y}{2}) \ cosec(\frac{y}{2}), \nonumber 
\eea
\noindent while we can not find any analytic expression for the second summation due to the $\frac{1}{R}$ factor. 
Using the above identity, it can be shown that, due to the term $sin(\frac{m y}{2})$ with $y = \frac{2\pi}{m}$, 
$\delta_3$ dependent part vanishes. So, in the presence of a bandgap, the expression for the persistent current 
in the half filled limit is, 
\bea
I & = & \sum_{s=0}^{m-1} \frac{-4 \pi \delta_1 \delta_2 \sin(\frac{2 \pi s_1}{m} + 2\phi)}{ N
\sqrt{ \delta_1^2 + \delta_2^2 - \epsilon^2 + 2 \delta_1 \delta_2 \cos \left ( \frac{2 \pi s_1}{m} + 2 \phi \right )
} }
\eea
\noindent The following important results follow from the above expression of $I$:
\begin{itemize}
\item The persistent current is independent of the strength of the NNN interaction  $\delta_3$ in the limit  
of half filled band, $N_{e} = \frac{N}{2} = m$ and before the onset of band overlap. 
\item The magnitude of the persistent current increases with the increased strength of the loss-gain strength
$\epsilon$ within its allowed range $0 \leq \epsilon < \vert \delta_1 - \delta_2\vert$ for fixed
$\delta_1, \delta_2, \phi, N$.
\end{itemize}
\noindent The persistent current $I$ for the half-filled case is plotted as a function
of the flux $\Phi$ for $\delta_1 > \delta_2$ and $\delta_1 < \delta_2$ in Figs. \ref{chf1}
and \ref{chf2}, respectively. The current increases with the BLG parameter $\epsilon$ for
both the cases. Further, the current is enhanced significantly for $\delta_1 < \delta_2$
compared to the case $\delta_1 > \delta_2$. The persistent current for the case of quarter
filling ($\frac{N_e}{N}=\frac{1}{4}$) is plotted in Figs. \ref{cqf1} and \ref{cqf2} for various
values of $\delta_3$. The magnitude of the current is expectedly
high compared to the half filled case. Finally, the contour plot in Fig. \ref{current_peak} depicts
maximum magnitude of the current as a function of $\delta_3$ and $\epsilon$. The magnitude increases with
the increasing values of both $\delta_3$ and $\epsilon$ prior to band overlap. In the region where two 
bands overlap, the persistent current exhibits non-uniform behavior due to the different responses of the 
two bands to the external magnetic field. In particular, the fractured valleys appearing in Figs.~\ref{cur1_qf}, \ref{cur1_3b8f}
and \ref{cur2_3b8f} at intermediate values of the parameters $(\epsilon, \delta_3)$ are manifestation of band crossing. 

\section{General Boundary Condition}

The bulk SSH Hamiltonian in real space is described in terms of a $2 m \times 2 m$
pentadiagonal matrix with imaginary elements along the main diagonal. The first sub-diagonal
and super-diagonal contain intercell and intracell hoping strength, while the NNN interaction
strength is contained in the second sub-diagonal and super-diagonal. The eigenvalues of this
matrix can not be found analytically for generic values of the parameters or for physically
interesting cases like $\delta_1 \neq \delta_2, \epsilon \neq 0$.
The SSH model with asymmetric hopping strengths reveal some interesting features
under GBC\cite{Guo2021Prl}. This Hamiltonian is non-hermitian due to the asymmetric
hopping strengths and a new phase diagram exists which is different from the
cases with PBC and OBC. We investigate the Hamiltonian $H$ under GBC
\textemdash the PBC appears as limiting cases and has been discussed
in the previous section. As discussed in Sec. II, the GBC can be broadly categorized
into two cases \textemdash (i) HGBC  and (iii) NGBC. The boundary
terms are hermitian for HGBC, while boundary terms are additional source 
of non-hermiticity for NGBC. 

We choose the ansatz for the eigenfunction $\Psi$ of $H$ with eigenvalue $E$ as,
\bea
\Psi = \sum_{n=1}^m \left(\psi_{n,a} \ a^{\dagger}_{n} + \psi_{n,b} \ b^{\dagger}_{n}\right) 
\vert 0 \rangle,
\eea
\noindent where the coefficients $\psi_{n,a}$ and $\psi_{n,b}$ are to be determined.
The eigenvalue equation $H \Psi = E \Psi$ gives a series of equations consisting of bulk 
equations as,
\bea
\left(E-U\right) \Psi_{n} = J \Psi_{n+1} + J^{\dagger} \Psi_{n-1},
\label{bulk_eqn_gbc}
\eea
\noindent with $n = 2, 3, \dots, m-1$ and the boundary equations, 
\bea
\left(E-U\right) \Psi_{1} & = & J \Psi_{2} + J_R \Psi_{m} \nonumber \\
\left(E-U\right) \Psi_{m} & = & J_L \Psi_{1} + J^{\dagger} \Psi_{m-1}, \
\label{boundary_eqn_gbc}
\eea
\noindent where $\Psi_{n} = \left( \psi_{n,a} \ \psi_{n,b} \right)^{T}$ and, 
\bea
U & = & \left ( \delta_1 e^{i\phi} \sigma_{+} + h. c. \right )
+ i\epsilon \sigma_{3}, \
J  =  -i\delta_3 e^{2i\phi}\sigma_0 + \delta_2 e^{i\phi} \sigma_{-},\nonumber \\ 
J_R & = & \beta_R e^{-2i\phi} \sigma_0 + \alpha_R e^{-i\phi} \sigma_{+},
J_L  = \beta_L e^{2i\phi} \sigma_{0} + \alpha_{L} e^{i\phi} \sigma_{-}.\nonumber
\eea 
We choose the ansatz $\Psi_{n} = \bp A \\ B \ep z^{n}$ where $z$ are
complex numbers. The motivation for such a choice is that for PBC $z \sim e^{ik},
\ k \in \mathbb{R}$. The specific form of the complex numbers $z$ for the GBC
will be determined from the consistency conditions arising out of the bulk and
boundary equations. The bulk equation (\ref{bulk_eqn_gbc}) reduces
to the following form, 
\bea
\left[ (E - U) - J z - J^{\dagger} \frac{1}{z} \right] \bp A \\ B \ep = 0 . 
\label{gbc_block_eq}
\eea
after the substitution of $\Psi_{n}$ into it. The condition for the existence of nontrivial
solution is that the determinant of the matrix $M=E -U - J z - J^{\dagger} {z^{-1}}$
vanishes, i.e. $Det(M)=0$. The vanishing determinant condition fixes the energy,
\bea
E & =   & i\delta_3 \left(e^{-2i\phi} \frac{1}{z} - e^{2i\phi}  z \right) \nonumber \\
  & \pm & \sqrt{\delta_1^2 + \delta_2^2 + \delta_1 \delta_2 \left(z e^{2i\phi} 
  + \frac{1}{z} e^{-2i\phi}\right) - \epsilon^2}
\eea
in terms of system parameters and yet to be determined complex numbers $z$. The complex
numbers $z$ are completely determined for fixed $E$ by employing the boundary conditions.
In particular, for a fixed value of energy $E$, $\tilde{z} = z \ e^{2i\phi}$ satisfy the
following quartic equation, 
\bea
&& \tilde{z}^4 + \left(\frac{\delta_1 \delta_2}{\delta_3^2} - \frac{2iE}{\delta_3} \right) 
 \tilde{z}^3 + \left(\frac{\delta_1^2 + \delta_2^2 - \epsilon^2 - E^2}{\delta_3^2} 
- 2 \right) \tilde{z}^2 \nonumber \\
&&  + \left(\frac{\delta_1 \delta_2}{\delta_3^2} 
+ \frac{2iE}{\delta_3} \right) \tilde{z} + 1 = 0, 
\label{gbc_root}
\eea
\noindent which can be solved analytically leading to four roots.
The analytic expressions of these roots for generic values
of the parameters prove to be  cumbersome for further analysis, and we do not pursue the same here.
We look for specific solutions such that roots are expressed in a simple closed form.
We note in this regard that $z, z^{-1}$ are solutions of Eq. (\ref{gbc_root})
either for (i) $\delta_1 \delta_2 = 0$ or (ii) $\delta_3 = 0$. Both of these limiting cases 
correspond to physically interesting situations \textemdash the system corresponds to tight-binding
ladder for the case (i), while NNN interaction vanishes for the case (ii). The quartic equation takes
a bi-quadratic form for $\delta_1 \delta_2 = 0$ and all the roots can be expressed in a simple closed form
as given in the Appendix A. However, the consistency condition of the wave-functions at the boundary
leads to a highly nonlinear equation, and appears to evade any analytic solution which can be
utilized for extracting physical information. We thus present the
relevant calculations in the Appendix A instead of presenting it in the main text. 
The results for the case $\delta_3 = 0$ corresponding to the SSH model with BLG under GBC are
presented below. 

The quartic equation reduces to a quadratic equation for $\delta_3 = 0$, 
\bea
\tilde{z}^{2} + \left( \frac{\delta_1^2 + \delta_2^2 - \epsilon^2 - E^2}{\delta_1 \delta_2} \right) 
 \tilde{z} + 1 = 0,
\label{quad-exact}
\eea
\noindent and the two roots $\tilde{z}_{1}, \tilde{z}_{2}$ satisfy the condition
$\tilde{z}_1\tilde{z}_2 =  1$. We choose $\tilde{z}_1 = e^{i\theta}$ and $\tilde{z}_2 =  e^{-i\theta}$,
where $\theta$ is to be determined from the consistency condition at the boundary.
It may be noted that the exact analytic solutions of Eq. (\ref{quad-exact}) do not specify $z$
completely due to their dependence on $E$. The boundary condition is used to find $E$ and $z$.
The solution of Eq. (\ref{bulk_eqn_gbc}) is the superposition of two solutions,
\bea
\Psi_{n} & = & \bp c_1 A_1 z_{1}^{n} + c_2 A_2 z_{2}^{n} \\
 c_1 B_1 z_{1}^{n}  + c_2 B_2 z_{2}^{n} \ep,
\eea
which must satisfy the boundary equation, 
$ H_{B} \bp c_1 \\ c_2 \ep = 0
$ where,
\bea
H_{B} = \bp \left(\delta_2 - \alpha_R z_{1}^{m}\right) B_1 e^{-i\phi}  \ 
\left(\delta_2 - \alpha_R z_{2}^{m}\right) B_2 e^{-i\phi} \\ 
\left(\delta_{2}z_{1}^{m} - \alpha_L \right) e^{i\phi} A_1 z_1 \ 
\left(\delta_{2}z_{2}^{m} - \alpha_L \right) e^{i\phi} A_2 z_2 \ep. \nonumber 
\eea
The condition for the non-trivial solution is $Det[H_{B}] = 0$, which 
when expressed in terms of the $n^{th}$ order Chebyshev polynomial of the second kind $U_{n}(\cos\theta)
\equiv \frac{\sin (n+1) \theta}{\sin \theta}$, is given by the equation,
\bea
(2x + \eta_1) U_{m-1}(x)  -(1+\eta_2) U_{m-2}(x) = \Delta,
\label{boundary_theta}
\eea
\noindent where $\eta_1 = \frac{\delta_2}{\delta_1} - \frac{\alpha_L \alpha_R}{\delta_1\delta_2}$, 
$\eta_2 = \frac{\alpha_R \alpha_L}{\delta_{2}^2}$, $x = \cos(\theta)$ and
$\Delta = \frac{\alpha_L}{\delta_2} e^{2im\phi} 
+ \frac{\alpha_R}{\delta_2} e^{-2im\phi}$. We have used the recurrence relation $U_{n+1}(x)=2 x U_n(x) - U_{n-1}(x)$
while deriving the above equation.
The energy expression in terms of $\theta$ for $\delta_3 = 0$ is 
\bea
E_{\pm} = \pm \sqrt{\delta_1^2 + \delta_2^2 + 2 \delta_1 \delta_2 \cos(\theta) - \epsilon^2}. 
\label{energy_it_theta}
\eea
\noindent The analytic expressions of the eigenvalues and eigenstates under PBC and OBC can
be obtained after 
obtaining the quantization condition on $\theta$ from Eq.~(\ref{boundary_theta}). 

It appears that Eq.~(\ref{boundary_theta}) is not analytically solvable in it's generic form. We discuss
below a few limiting cases for which Eq.~(\ref{boundary_theta}) admits exact, analytical solutions,
leading to exact solvability of the Hamiltonian:  
\begin{enumerate}
\item $\alpha_L \alpha_R = \delta_2^2$ : The PBC and APBC arise as special cases for $\alpha_L=\alpha_R=\delta_2$
and $\alpha_R=\alpha_L=-\delta_2$, respectively. The Hamiltonian is exactly solvable for generic values of the
parameters satisfying $\alpha_L\alpha_R = \delta_2^2$.
\item AHBC : The Hamiltonian is exactly solvable for $\alpha_L=-\alpha_R=\delta_2$ such that $\alpha_L \alpha_R=
-\delta_2^2$ which corresponds to AHBC. 
\item $\alpha_L = \alpha_R = 0$ :  This corresponds to OBC. The Hamiltonian is not exactly solvable. However,
closed-form expressions for edge states and the associated energy can be found analytically. 
\end{enumerate}

\subsection{$\alpha_R \alpha_L = \delta_2^2$}

We introduce a parameter $\nu = \sgn(\frac{\alpha_L}{\delta_2}) = \sgn(\frac{\alpha_R}{\delta_2})$
such that $\nu=\pm 1$ depending on the signs of $\alpha_L, \alpha_R$ and $\delta_2$. The PBC and APBC
correspond to $\nu=1$ and $\nu=-1$, respectively. However, the converse is not true, i.e. $\nu=\pm 1$ does
not necessarily correspond to PBC or APBC.
The parameters take the values $\eta_1 = 0$ and $\eta_2 = 1$  leading to a reduction of 
Eq.~(\ref{boundary_theta}) to the following form, 
\bea
&& \left[ \cos(m\theta) -\nu \cosh(u) \right] \sin(\theta)  =  0,\nonumber \\
&& u = ln\vert\frac{\alpha_L}{\delta_2}\vert + 2im \phi
\label{boundary_eq_con1}
\eea
The solutions of Eq. (\ref{boundary_eq_con1}) are,
\bea
\theta_{\pm} & = & \frac{2s\pi}{m} \pm \frac{1}{m} \arccos\left(\nu\cosh(u)\right)\nonumber \\
& = & \frac{2s\pi}{m} \pm \left(\frac{\pi}{2m} \left(1 - \nu\right) + 2\phi\right) 
\mp \frac{i}{m} ln\vert\frac{\alpha_L}{\delta_2}\vert, \ 
\nonumber
\eea
\noindent where $s = 0, 1 \dots (m-1)$. The solutions corresponding to $\sin\theta = 0$ are
neglected, since they lead to trivial solution $\Psi=0$.
The relevant calculations are presented in the Appendix B ensuring that there are total $m$ number
of solutions.
We consider only $\theta=\theta_+$ henceforth since both $\theta_{\pm}$ give the same set of eigenvalues.
In order to find the wavefunctions $\Psi_n$, we substitute the ratio $\frac{A}{B}$ obtained
from Eq. (\ref{gbc_block_eq}) into the matrix equation $ H_{B} \bp c_1 \\ c_2 \ep = 0$ leading to the following
expressions,
\bea
c_2 A_2 & = & - c_1 A_1 \left(\frac{\delta_2 z_1^{m+1} - \alpha_L z_1}{\delta_2 z_2^{m+1} - \alpha_L z_2}\right)  \nonumber \\
c_2 B_2 & = & - c_1 B_1 \left(\frac{\delta_2 - \alpha_R z_1^m}{\delta_2 - \alpha_R z_2^m}\right) \nonumber \\
	& = & - c_1 A_1 \left(\frac{\delta_2 - \alpha_R z_1^m}{\delta_2 - \alpha_R z_2^m}\right) 
	\left(\frac{(E-i\epsilon)z_1}{\delta_1 e^{i\phi} z_1 + \delta_2 e^{-i\phi}}\right)
\eea
The expression of $\psi_{n,a}$,$\psi_{n,b}$ for eigenstates is obtained as,
\bea
\psi_{n,a} & =      & \frac{2ic_1 A_1 e^{i(\theta-2n\phi)}}{\delta_2 e^{-i m(\theta +2\phi)} - \alpha_L} \left\{\delta_2 
\sin(n-m-1)\theta \ e^{-2im\phi} \right. \nonumber \\ 
           & +      & \left. \alpha_L \sin(1-n)\theta \right\} \nonumber \\
\psi_{n,b} & =      & \frac{2i c_1 A_1 (E -i\epsilon) e^{-i(2n+1)\phi}}{\left(\delta_2 - \alpha_R e^{-im(\theta + 2\phi)} \right)
\left(\delta_1 + \delta_2 e^{-i\theta}\right)} \nonumber \\
	   & \times & \left\{\delta_2 \sin(n\theta) + \alpha_R \sin(m-n)\theta \ e^{-2im\phi}\right\}
\eea
The PBC corresponds to $u = 2im\phi$ and $\nu = 1$. The expression of $\theta$, as given in
Sec. III.A, is reproduced $\theta = \frac{2s\pi}{m} + 2\phi$ for this choice of parameters.
The Hamiltonian $H$ under the PBC has been discussed in detail in Sec. III and we do not repeat
the same. All the values of $\theta$ are also real for APBC and the related results are discussed
in the next section. The values of $\theta$ are complex except for PBC and APBC.

\subsubsection{Anti-Periodic Boundary Condition}

The eigenvalues of H with vanishing NNN interaction under the APBC is given by,
\bea
E_{\pm} = \pm \sqrt{\delta_1^2 + \delta_2^2 + 2\delta_1\delta_2 \cos\left(\frac{\pi}{m}(2 s +1) + 2 \phi \right) -
\epsilon^2}. \nonumber 
\eea
\noindent The energy eigenvalues are entirely real for $\epsilon \leq \vert \delta_1 - \delta_2 \vert$ which
is also the condition for reality of the entire spectra under the PBC.
The band structure of the system in the thermodynamic limit $N \rightarrow \infty$, for which the momentum
$k=\frac{\pi}{m}(2 s +1)$ may be treated as a continuous variable and $\phi \rightarrow 0$, is identical to
the case of PBC with $\delta_3=0$. There are point band gaps for $\epsilon < \vert \delta_1 - \delta_2 \vert$ and
line band gap for $\epsilon > \delta_1 + \delta_2 $. There are no band gaps in the intermediate region
$\vert \delta_1 - \delta_2 \vert < \epsilon < \delta_1 + \delta_2 $. The values of $\theta$ being entirely real for APBC, 
possibility of edgestates are ruled out. The persistent current in the half-filled
limit has the expression,
\bea
I & = & \sum_{s=0}^{m-1} \frac{-4 \pi \delta_1 \delta_2 \sin(\frac{\pi }{m}(2s+1) + 2\phi)}{ N
\sqrt{ \delta_1^2 + \delta_2^2 - \epsilon^2 + 2 \delta_1 \delta_2 \cos \left ( \frac{\pi}{m}(2s+1)
+ 2 \phi \right ) } }\nonumber
\eea
\noindent The relevant results for APBC with $\delta_3 \neq 0$ APBC are obtained numerically.

\subsection{Anti-Hermitian Boundary Condition}

The condition $\alpha_L \alpha_R=-\delta_2^2$ gives $\eta_1 = 2 \frac{\delta_2}{\delta_1}$, $\eta_2 = -1$, and 
Eq.~(\ref{boundary_theta}) reduces to
\bea
\sin(m\theta) \left\{\cos(\theta) + \alpha\right\} = \tilde{\nu} \sinh(u)  \sin(\theta), \
\alpha \equiv \frac{\delta_2}{\delta_1},
\label{eebb}
\eea
\noindent where $\tilde{\nu} = \sgn(\frac{\alpha_L}{\delta_2}) = - \sgn(\frac{\alpha_R}{\delta_2})$. 
It appears that exact solutions of the above equation is possible only for $\sinh(u) = 0$, i.e.
$\alpha_R=-\alpha_L=\delta_2, \phi=\frac{l\pi}{2m}, l \in \mathbb{Z}$ which corresponds to
AHBC. In case of AHBC, Eq. (\ref{eebb}) reduces to,
\bea
\sin(m\theta) \left\{ \cos\theta + \alpha \right\} = 0,
\eea
\noindent with $m$ allowed solutions of $\theta$ as
$\theta_s = \frac{s \pi}{m}, \ s = 1,2 \dots (m-1)$ and $\theta_m = \pi + \arccos(\alpha)$.
The solutions $\theta_m$ is real for $\alpha \leq 1$, while it is complex for $\alpha > 1$.
Note that $\theta = 0$ is not included in the set of solutions since it leads to $\Psi=0$.
The analysis for showing that $\theta = 0$ indeed leads to $\Psi=0$ is similar to the arguments
presented in the Appendix B for the case of Eq. (\ref{boundary_eq_con1}).

The expression of the bulk state energy is 
\bea
E_{\pm} = \pm \sqrt{\delta_1^2 + \delta_2^2 + 2\delta_1\delta_2 \cos(\theta_s) - \epsilon^2}
\eea
\noindent The energy eigenvalues are entirely real for $\alpha \leq 1$ and
$\epsilon < \vert \delta_1 - \delta_2 \vert$. The qualitative nature of band spectra
is similar to that of PBC with $\delta_3=0$ for $\alpha \leq 1$. However,
due to complex values of $\theta_m$ for $\alpha > 1$, 
purely imaginary eigenvalues as a conjugate pair appears which we denote as $E_{edge}$,
\bea
E_{edge} = \pm i \delta_1 \sqrt{\alpha^2 + (\frac{\epsilon}{\delta_1})^2  - 1}.
\eea
The expression of $\psi_{n,a}$,$\psi_{n,b}$ for the eigenstate corresponding to the eigenvalue 
$E_{edge}$ is,
\bea
\psi_{n,a} & = & -(-1)^{n}\frac{2 c_1 A_1 e^{-\zeta}}{1 + (-1)^{m} e^{m\zeta}} \left\{ (-1)^{m} sinh(n-m-1)\zeta \right. \nonumber \\
           & + & \left. sinh(n-1)\zeta\right\} \nonumber \\
\psi_{n,b} & = & -(-1)^{n}\frac{2 c_1 A_1}{1 - \alpha e^{\zeta}} \frac{E - i\epsilon}{\delta_1\left(1 - (-1)^{m} e^{m\zeta}\right)} \left\{ sinh(n\zeta) \right. \nonumber \\
           & - & \left. (-1)^{m} sinh(n-m)\zeta\right\} \nonumber
\eea
This state is identified as the edgestate. Due to the hyperbolic function, the 
probability density of this state will be maximum at the edges. 
The detail calculation for the expression of $\psi_{n,a}$,$\psi_{n,b}$ corresponding to 
bulk and edgestate is shown in Appendix C.

\subsection{Open Boundary Condition}

The study of the system under OBC is mainly based on numerical analysis except for $\delta_3=0$
for which analytical results may be obtained by using the method described at the beginning of this section. 
The phase $\phi$ can be gauged away for OBC and is taken to be zero. 
The OBC corresponds to $\alpha_{L} = \alpha_{R} = 0$ which implies
$\eta_1 = \frac{\delta_2}{\delta_1} = \alpha$, $\eta_2=\Delta=0$. In this limit, Eq.~(\ref{boundary_theta}) 
reduces to, 
\bea
\left(2x + \alpha\right) U_{m-1}(x) - U_{m-2}(x) = 0 
\label{boundary_eq_obc}
\eea
where $x = \cos\theta$. Eq.~(\ref{boundary_eq_obc}) admits
$m$ real solutions for $\alpha < \alpha_c = 1 + \frac{1}{m}$, while for $\alpha > \alpha_c$,
there are $m-1$ real solutions and one complex solution associated with the edge state. 
The critical value of $\theta_c$ is obtained from the relation 
$f_1^{'}(\pi) = f_2^{'}(\pi)$, where $f_1 (\theta)= \sin(m+1)\theta$,
$f_2(\theta) = -\alpha_c \sin(m\theta)$, and $f_{i}^{'}(\pi)=\frac{d f_i}{d \theta}|_{\theta=\pi}$.
Eq. (\ref{boundary_eq_obc}) for complex $\theta = \pi + i \zeta$ reads,
\bea
2m\zeta = ln(\frac{e^{-\zeta} - \alpha}{e^{\zeta} - \alpha}) 
\label{zeta_eq}
\eea
which has solution only when $e^{\zeta} \approx \alpha$. We set 
$e^{\zeta} = \alpha + \mu_{\zeta} (\mu_{\zeta} \rightarrow 0)$, and
obtain $\mu_{\zeta}$ as 
\bea
\mu_{\zeta} = \frac{\frac{1}{\alpha} - \alpha}{\alpha^{2m} + \frac{1}{\alpha^2}}
\eea 
by considering the approximation $(\alpha + \mu_{\zeta})^{2m} \approx \alpha^{2m}$ and
$\frac{1}{\alpha + \mu_{\zeta}} \approx \frac{1}{\alpha} - \frac{\mu_{\zeta}}{\alpha^2}$.
The expression for $\zeta$, 
\bea
\zeta = ln\left[ \alpha \left( 1 + \frac{1 - \alpha^2}{1 + \alpha^{2m+2}} \right)\right],
\eea
and the corresponding energy becomes 
\bea
E_{edge} = \pm \sqrt{ \delta_1^2 + \delta_2^2 - 2\delta_1 \delta_2 \cosh(\zeta) - \epsilon^2}.
\eea
In the thermodynamic limit, $e^{\zeta} = \alpha$ and $E_{edge} = \pm i \epsilon$. 
From the boundary Eq. (\ref{boundary_eqn_gbc}) and from the ratio $\frac{A}{B}$, we get 
\bea
c_2 A_2 & = & - c_1 A_1  \frac{1 + \alpha e^{i\theta}}{1 + \alpha e^{-i\theta}} \nonumber \\
c_2 B_2 & = & - c_1 B_1 = -c_1 A_1 \frac{E - i\epsilon}{\delta_1(1 + \alpha e^{-i\theta})}
\eea 
The expressions of $\psi_{n,a}$,$\psi_{n,b}$ corresponding to the bulk states are as follows,
\bea
\psi_{n,a} & = & c_1 A_1 z_1^n + c_2 A_2 z_2^n \nonumber \\
           & = & \frac{2ic_1 A_1}{1 + \alpha e^{-i\theta}} \left( \sin(n\theta) + \alpha \sin((n-1)\theta) \right)  \\
\psi_{n,b} & = & c_1 B_1 z_1^n + c_2 B_2 z_2^n \nonumber \\
           & = & \frac{2ic_1 A_1}{1 + \alpha e^{-i\theta}} \frac{E-i\epsilon}{\delta_1} \sin(n\theta)
\eea
As discussed earlier, we will get one complex $\theta$ while $\alpha > \alpha_{c}$. The complex $\theta$ takes the 
form $\theta = \pi + i \zeta$ where $\zeta$ satisfy Eq.~(\ref{zeta_eq}). The eigenstate corresponding to the 
$\theta = \pi + i \zeta$ is identified as edgestate. The expression of $\psi_{n,a}$,$\psi_{n,b}$ for the edge state is
\bea
\psi_{n,a} & = & (-1)^{n+1} \frac{2c_1 A_1}{1 - \alpha e^{\zeta}} \left( \sinh(n\zeta) - \alpha \sinh(n-1)\zeta \right)\nonumber \\
\psi_{n,b} & = & (-1)^{n+1} \frac{2c_1 A_1}{1 - \alpha e^{\zeta}} \frac{E - i\epsilon}{\delta_1} \sinh(n\zeta)\nonumber
\eea 
\noindent Due to the sine hyperbolic function, the probability density of the eigenstates will be maximum 
at the edges and in the interior portion the probability density will be zero. This kind of the localized 
behavior is the signature of the edge state.

\begin{figure}
\centering
        \begin{subfigure}{0.46\linewidth}
                \centering
                \includegraphics[width=\textwidth]{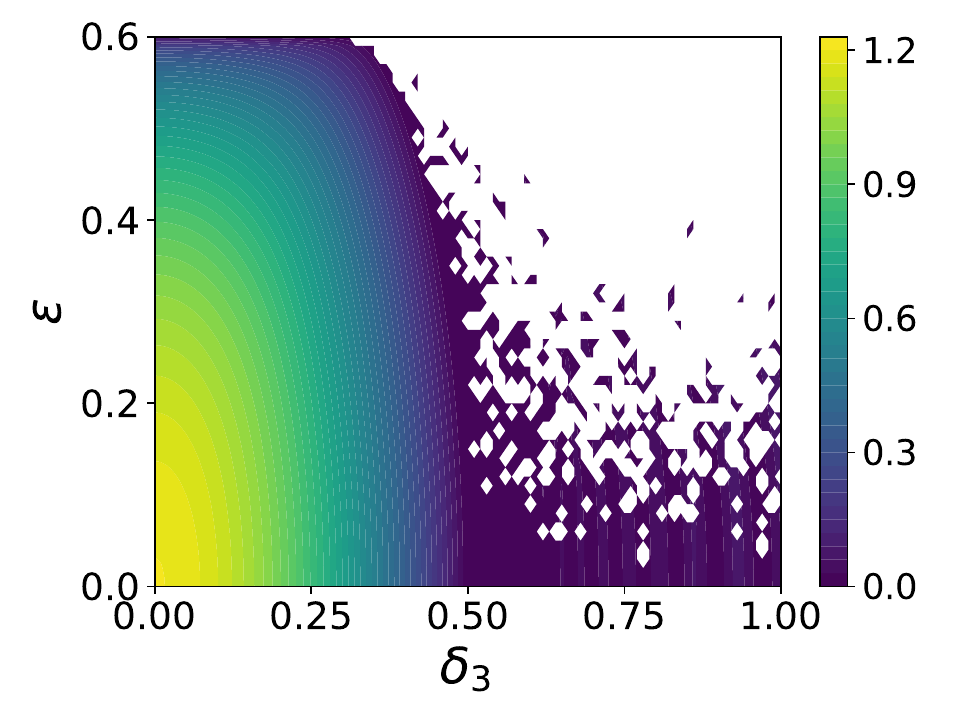}
                \caption{}
                \label{bg_obc}
        \end{subfigure}
        \hfill
        \begin{subfigure}{0.46\linewidth}
                \centering
                \includegraphics[width=\textwidth]{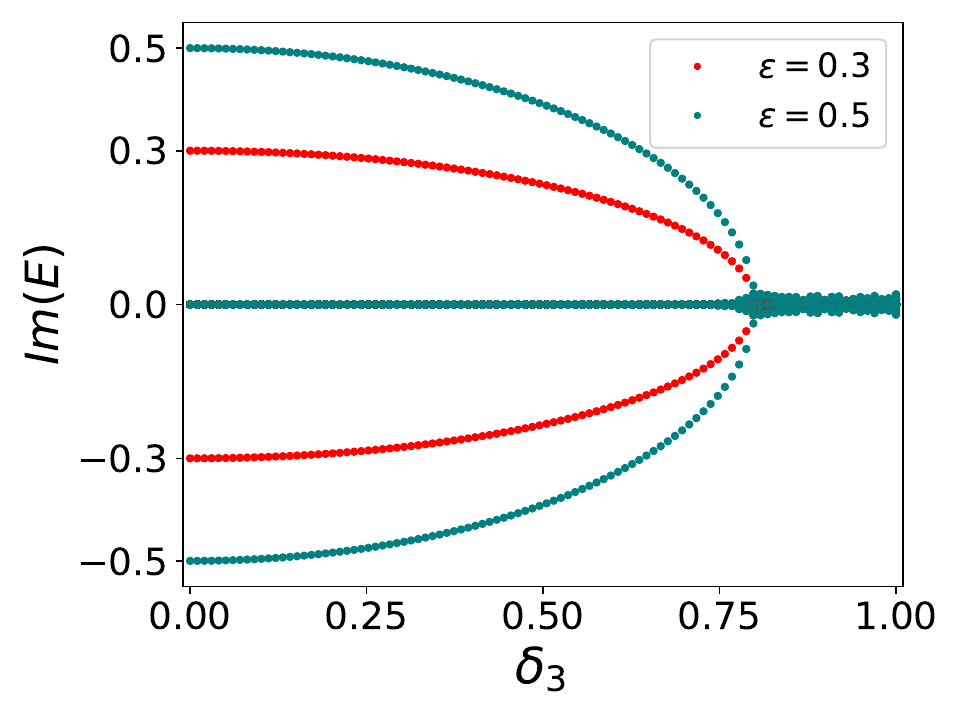}
                \caption{}
                \label{im_obc}
        \end{subfigure}
	\vfill
	\begin{subfigure}{0.46\linewidth}
                \centering
                \includegraphics[width=\textwidth]{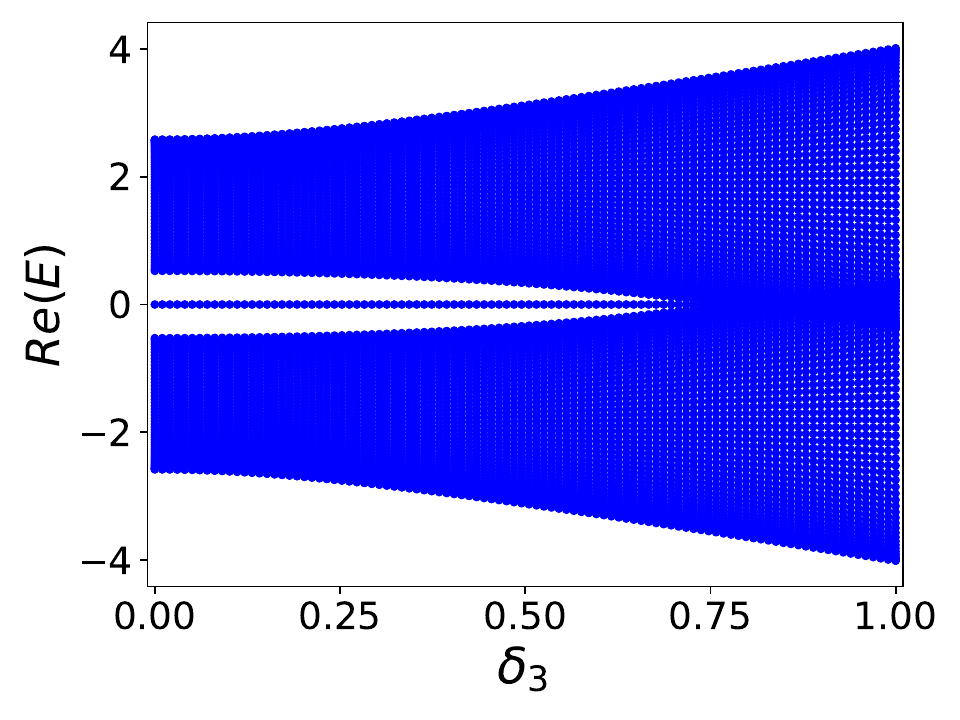}
                \caption{}
                \label{re1_obc}
        \end{subfigure}
	\hfill
	\begin{subfigure}{0.46\linewidth}
                \centering
                \includegraphics[width=\textwidth]{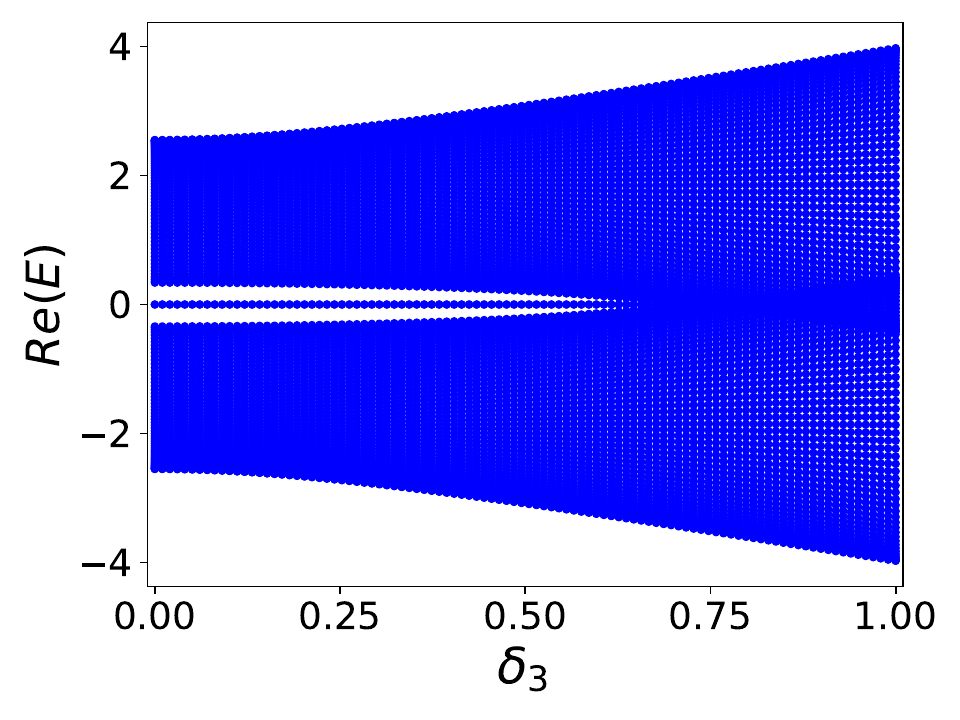}
                \caption{}
                \label{re2_obc}
        \end{subfigure}
	\caption{(Color online)(a) Band gap under OBC in trivial parametric regions and the white portion 
	of this contour plot indicates complex eigenvalues. (b) Imaginary part of the eigenvalues 
	under OBC. (c) and (d) : Real part of the eigenvalues under OBC.
        Parameter Values : $\delta_1 = 1$,$N=120$; Fig(a) :  $\delta_2 = 0.4$;
	Fig(b) : $\delta_2 = 1.6$; Fig(c) : $\delta_2 = 1.6$, $\epsilon = 0.3$
	Fig(d): $\delta_2 = 1.6$, $\epsilon = 0.5$.}
        \label{energy_nh_obc}
\end{figure}

The entirely real energy eigenvalues are obtained in the topologically trivial phase
$\delta_1 > \delta_2$ for certain region in the parameter space.
The contour plot of real band gap in the `$\epsilon-\delta_3$' plane is shown in
Fig.  \ref{bg_obc}. In the trivial phase, complex eigenvalues emerge, and the real band 
gap closes beyond a critical value of $\delta_3$, which depends on the BLG parameter $\epsilon$. 
Comparing Fig.~(\ref{bg_obc}) with Fig.~(\ref{bg1_pbc}), we observe that this critical value 
of $\delta_3$ coincides with the value at which the band gap closes under PBC.
In the topologically nontrivial phase ($\delta_1 < \delta_2$), the real parts of the 
eigenvalues are plotted as a function of $\delta_3$ for fixed values of $\epsilon=0.3$ and $\epsilon = 0.5$ in Figs.
\ref{re1_obc} and \ref{re2_obc}, respectively. The corresponding complex eigenvalues as a 
function of $\delta_3$ for $\epsilon = 0.3$ and $0.5$ are shown in Fig.~(\ref{im_obc}). Both the real 
and imaginary eigenvalues are symmetric  with respect to the $Re(E)=0$ and $Im(E)=0$ axes, respectively. 
This is a manifestation of the particle-hole symmetry in the system.
The imaginary eigenvalues appear in conjugate pairs up to around $\delta_3 \approx .8$ and
vanish beyond this value. 

\begin{figure}
\centering
	\begin{subfigure}{0.48\linewidth}
		\centering
		\includegraphics[width = \textwidth]{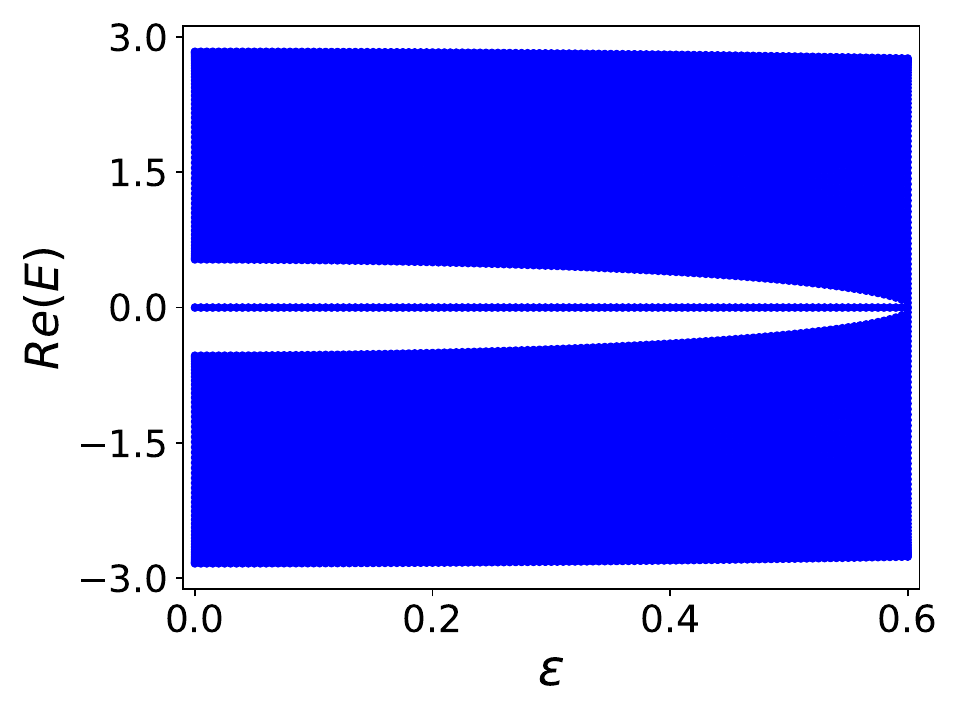}
		\caption{}
		\label{eigval_eps_obc_real}
	\end{subfigure}
	\hfill
	\begin{subfigure}{0.48\linewidth}
                \centering
                \includegraphics[width = \textwidth]{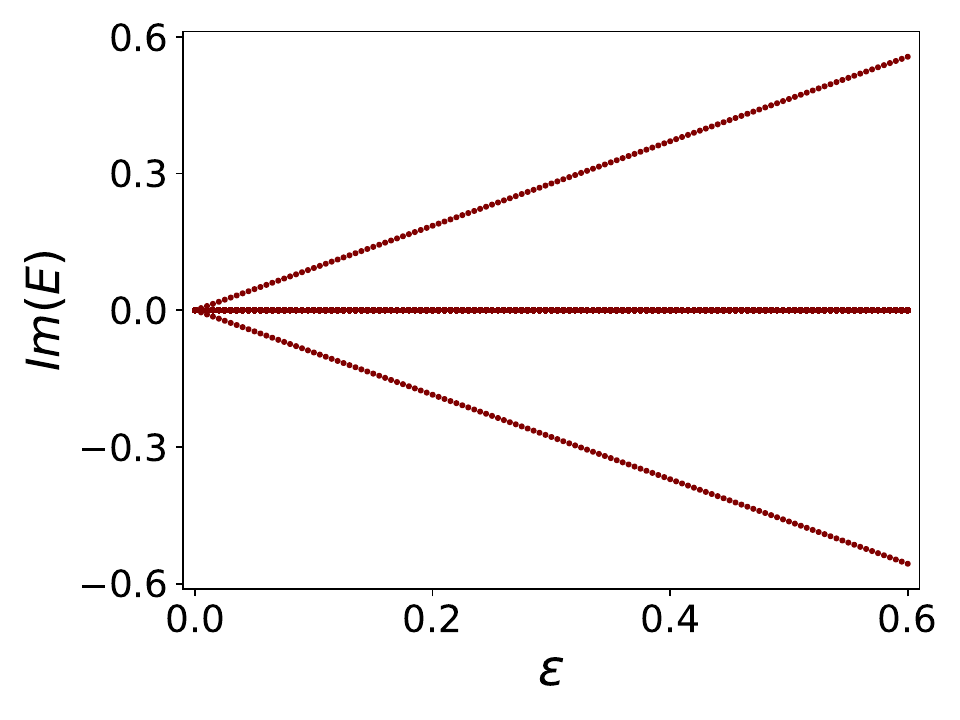}
                \caption{}
                \label{eigval_eps_obc_imag}
	\end{subfigure}
	\caption{(Color online) Real and imaginary part of the eigenvalues (under OBC) with the variation of $\epsilon$.
	Parameter Values : $\delta_1 = 1$, $\delta_2 = 1.6$, $N = 120$, $\delta_3 = 0.3$}
	\label{eigval_eps_obc}
\end{figure}

\begin{figure}
\centering
        \begin{subfigure}{0.48\linewidth}
                \centering
                \includegraphics[width=\textwidth]{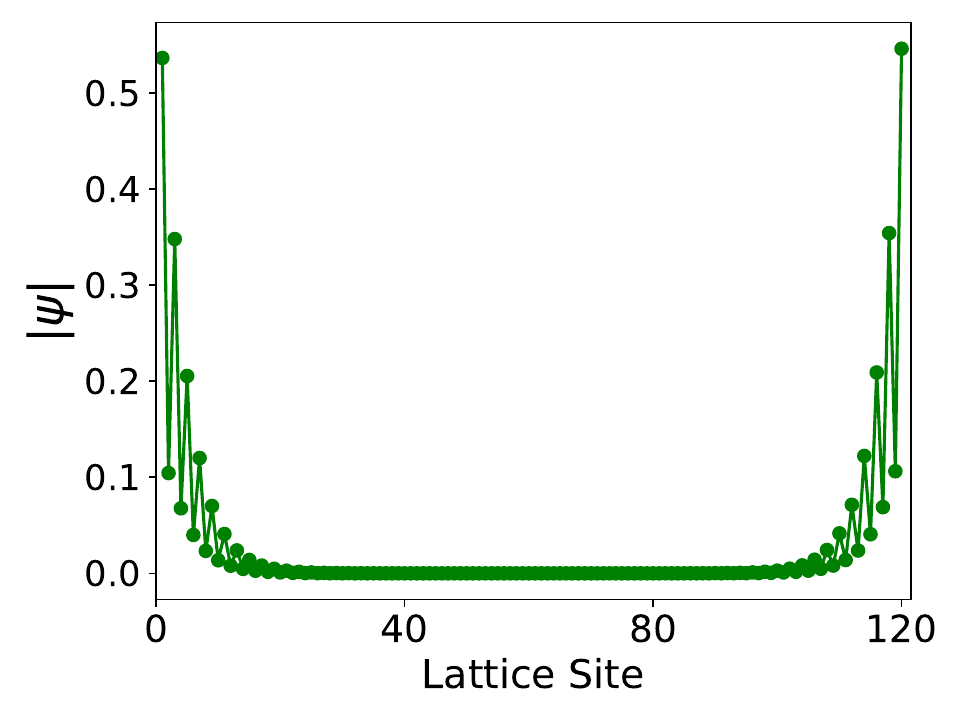}
                \caption{}
                \label{es3}
        \end{subfigure}
	\begin{subfigure}{0.48\linewidth}
                \centering
                \includegraphics[width=\textwidth]{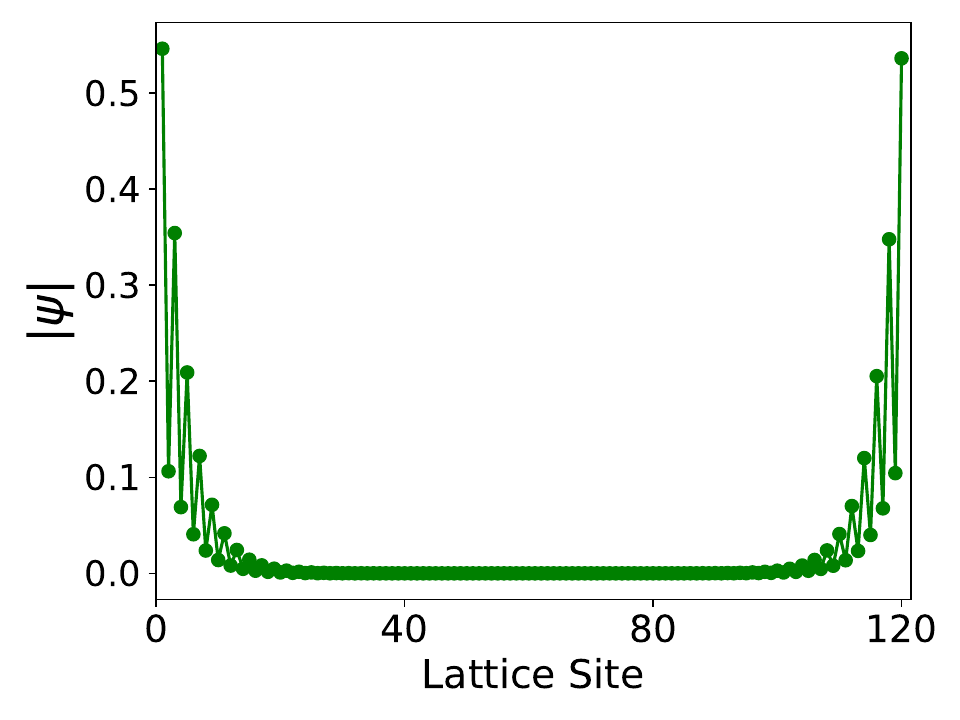}
                \caption{}
                \label{es4}
        \end{subfigure}
        \vfill
        \begin{subfigure}{0.48\linewidth}
                \centering
                \includegraphics[width=\textwidth]{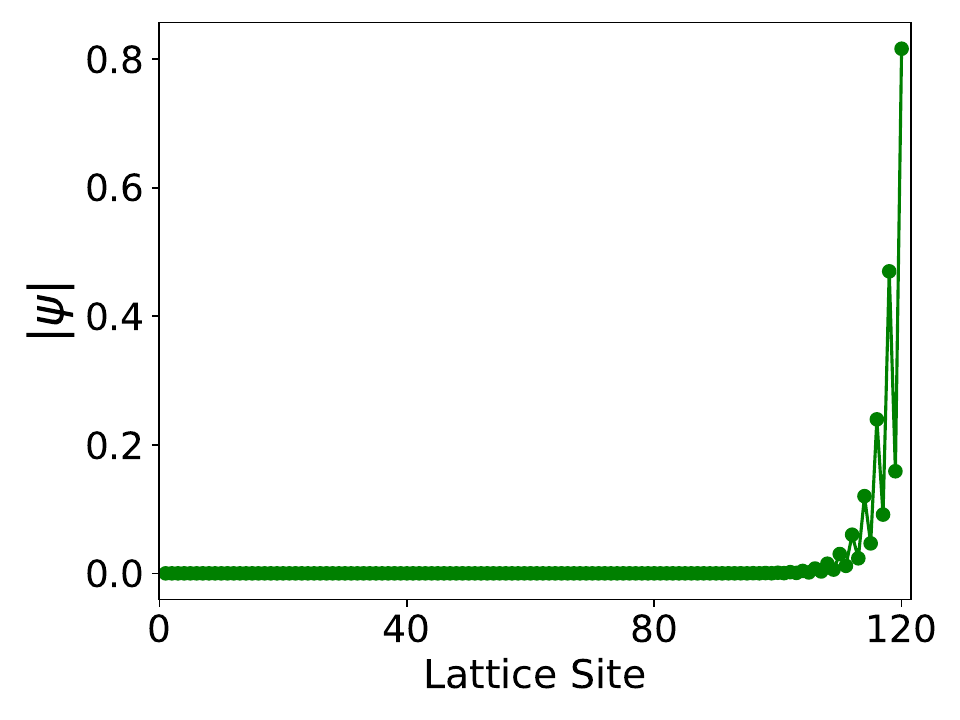}
                \caption{}
                \label{es7}
        \end{subfigure}
 	\hfill
        \begin{subfigure}{0.48\linewidth}
                \centering
                \includegraphics[width=\textwidth]{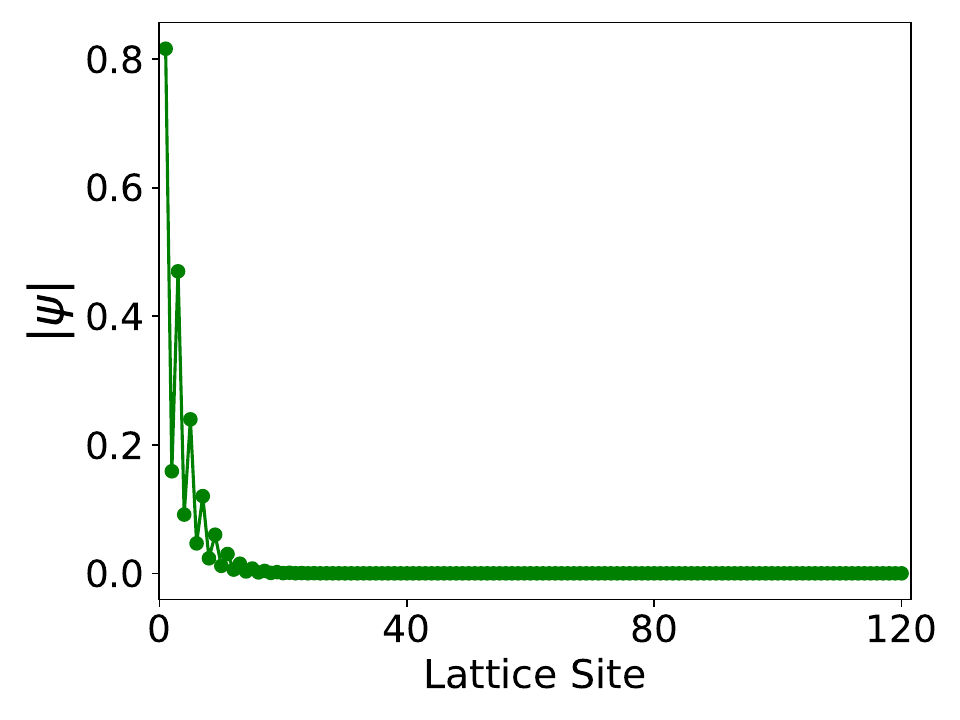}
                \caption{}
                \label{es8}
        \end{subfigure} 
	\caption{(Color online) Plot of edge state wavefunction in the topologically nontrivial
region for $ N = 120, \delta_1 = 1, \delta_2 = 1.6$ and for various combinations
of values of $\epsilon$ and $\delta_3$  \textemdash Figs. (a) and (b) : $\epsilon = 0, \delta_3 = 0.3$;
Figs. (c) and (d): $\epsilon = 0.3,
\delta_3 = 0.3$; Figures in each row correspond to complex-conjugate energy eigenvalues.}
\label{edgestate_obc}
\end{figure}

The real and imaginary parts of the eigenvalues are plotted as functions of $\epsilon$ in 
Figs.~\ref{eigval_eps_obc_real} and \ref{eigval_eps_obc_imag}, respectively. The isolated 
state appearing between the two bands indicates the presence of an edge state. From 
Figs.~(\ref{re1_obc}),(\ref{re2_obc}), and (\ref{eigval_eps_obc_real}), 
we observe that the edgestate persists despite small variations in the parameters $\delta_3$ and $\epsilon$. 
This indicates that the existence of edgestate is robust under the variation of $\delta_3$ and $\epsilon$. 
The appearance of complex eigenvalues is a signature of existence of edge states.
The edge states for $\delta_1 < \delta_2$ and various values of $\delta_3, \epsilon$
are shown in Fig. \ref{edgestate_obc}. The wave-functions are localized at the edges.
The localization of states is symmetric around the edges for $\epsilon=0$
, and are shown in Figs. \ref{es3} and \ref{es4}.
The asymmetric edge states appear for $\epsilon  \neq 0$
and shown in Figs.  \ref{es7} and \ref{es8}. The edge
states vanish around $\delta_3 \approx .8$ for which imaginary eigenvalues disappear.

The edge states appear for $\delta_1 > \delta_2$ only if $\delta_3\neq 0, \epsilon \gtrapprox .7 $ and
the lattice size is of the order of $N \sim 300$ or larger. The size dependence for
the existence of edge states is very typical to non-hermitian systems\cite{Guo2021Prl}, and
the same is observed for $H$ in the region $\delta_1 > \delta_2$. The constraints on
non-vanishing $\delta_3$ and $\epsilon \gtrapprox .7$ for the existence of edge states is new
compared to the the case of $\delta_1 < \delta_2$. The number of edge states is very large for
$\delta_1 > \delta_2, \delta_3 \neq 0, \epsilon >{\vert \delta_1 - \delta_2 \vert}, N \sim 300$.
\begin{figure}
\centering
        \begin{subfigure}{0.48\linewidth}
                \centering
                \includegraphics[width=\textwidth]{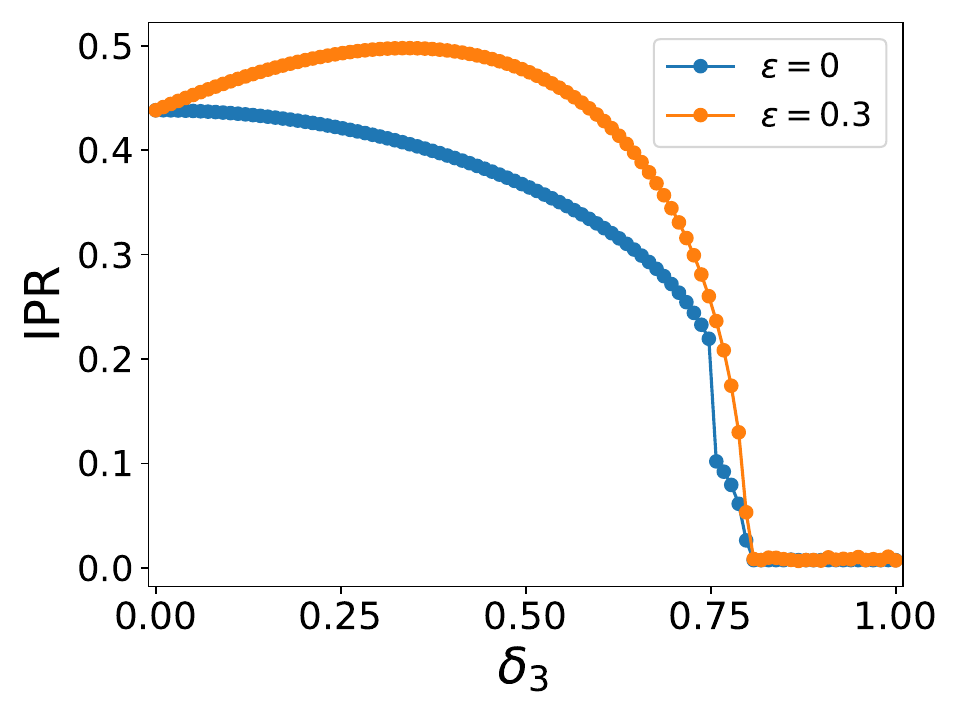}
                \caption{}
                \label{}
       \end{subfigure}
        \hfill
        \begin{subfigure}{0.48\linewidth}
                \centering
                \includegraphics[width=\textwidth]{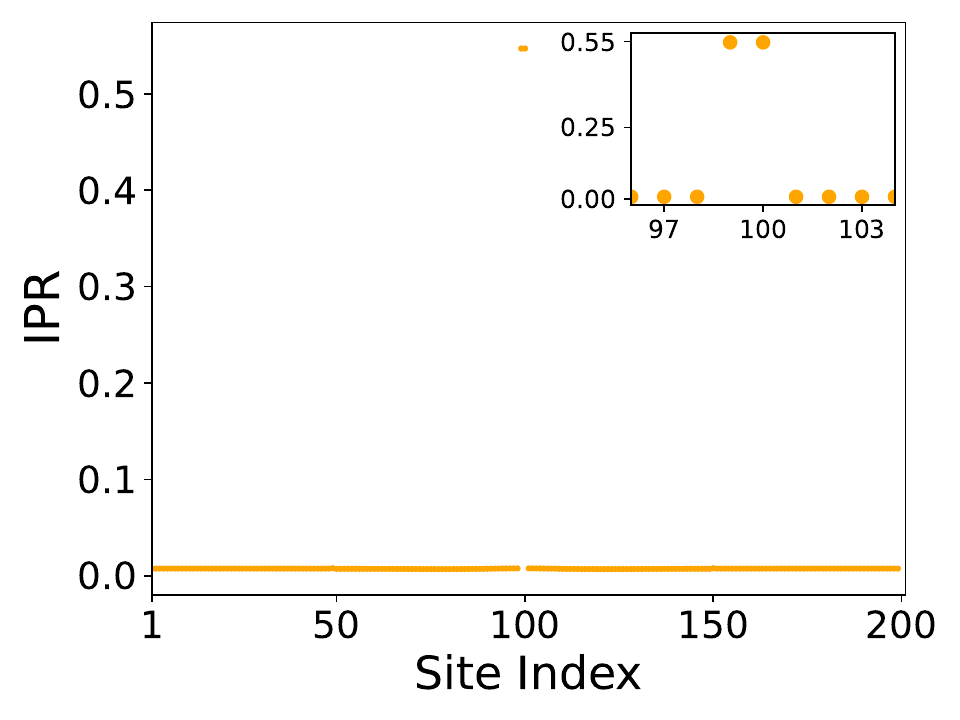}
                \caption{}
                \label{}
        \end{subfigure}
	\caption{(Color online) Fig(a) : Plot of the IPR of the m-th state with the variation of 
	$\delta_3$ (OBC). Parametric values : $\delta_1 = 1, \delta_2 = 1.6
	, N = 200, \phi = 0$. Fig(b) : Plot of the IPR of all the eigenstates(OBC). parametric values : 
	$\delta_1 = 1, \delta_2 = 1.6, \delta_3 = 0.5, \epsilon = 0.5, N = 200, \phi = 0$}
	\label{ipr_obc}
\end{figure}
\begin{figure}
\centering
        \begin{subfigure}{0.48\linewidth}
                \centering
                \includegraphics[width=\textwidth]{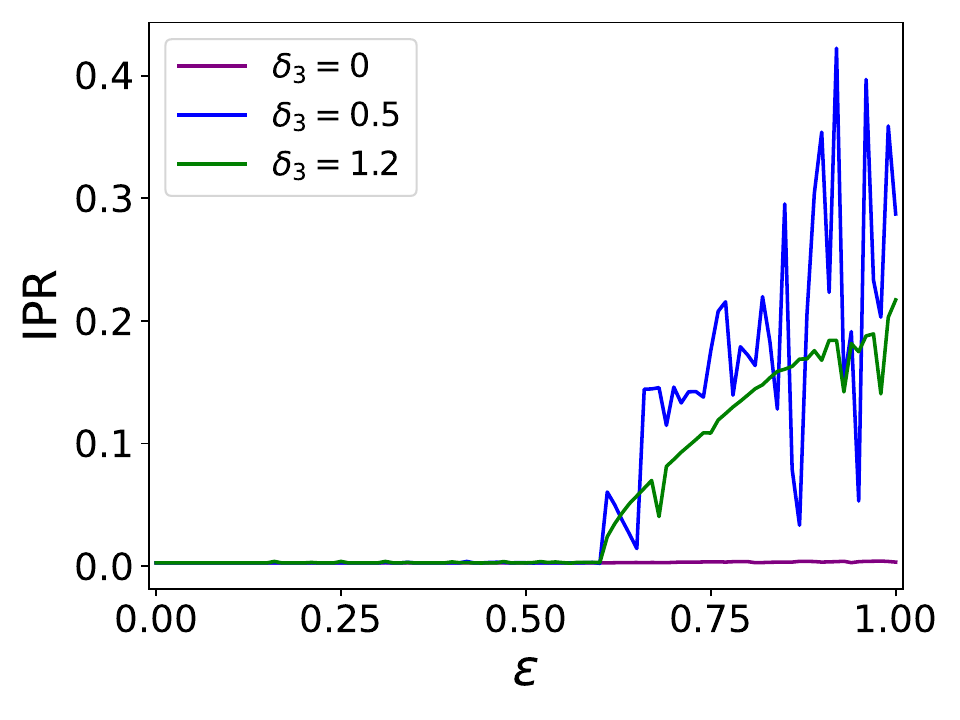}
                \caption{}
                \label{}
       \end{subfigure}
        \hfill
        \begin{subfigure}{0.48\linewidth}
                \centering
                \includegraphics[width=\textwidth]{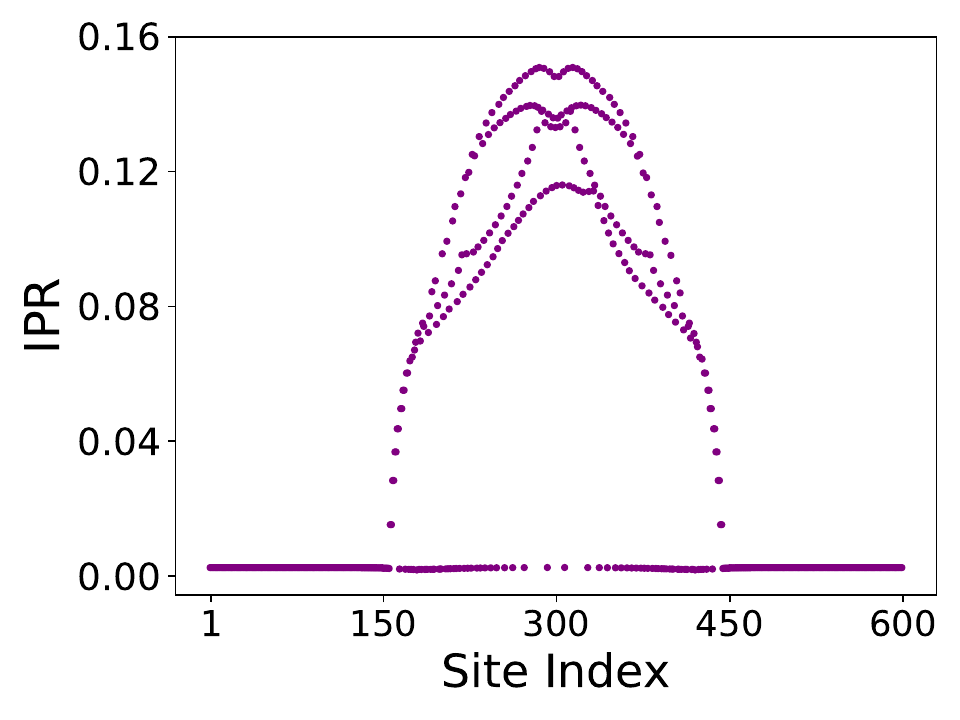}
                \caption{}
                \label{}
        \end{subfigure}
	\caption{(Color online) Fig(a) : Plot of the IPR of the m-th state with the variation of $\epsilon$
	(OBC). Parameter values : $\delta_1 = 1, \delta_2 = 0.4, N = 600, \phi = 0$. Fig(b) : Plot of the 
	IPR of all the eigenstates. Parametric values : $\delta_1 = 1, \delta_2 = 0.4, N = 600, \phi = 0, 
	\delta_3 = 0.5, \epsilon = 0.7$.}
	\label{ipr_ttr}
\end{figure}
The Inverse Participation Ratio(IPR) $I$ is a measure of localization  of states,
and is defined as,
\bea
I = \frac{\sum_{i} \left(\psi_{n,i}^{*} \psi_{n,i}\right)^2}{\left(\sum_{i}\psi_{n,i}^{*} \psi_{n,i}\right)^2},
\eea
\noindent where $\psi_{n,i}$ represents the nth eigenstate in the lattice site `i'. 
If the eigenstate is perfectly localized in a singe site(say j), for that site 
$\psi_{n,j}^{*}\psi_{n,j} \approx 1$ and for other sites($i \neq j$) 
$\psi_{n,i}^{*}\psi_{n,i} \approx 0$ leading to IPR $\approx 1$. If the eigenstate is spread 
across all the sites then the probability at each site is approximately 
$\psi_{n,i}^{*}\psi_{n,i} \approx \frac{1}{N}$ leading to $I \approx N \times \frac{1}{N^2} 
\approx \frac{1}{N}$. In the thermodynamic limit $N \infty$, the IPR stabilizes in a finite value for 
localized state and goes to zero for delocalized state. We show $I$ vs. $\delta_3$ in Fig. \ref{ipr_obc}
for $\delta_1 < \delta_2$. The delocalization of states is seen around $\delta_3 \approx .8$
for both $\epsilon=0$ and $\epsilon=.3$. The $I$ vs. $\epsilon$ for $\delta_3=0, 0.5, 1.2$ plot
in the $\delta_1 < \delta_2$ region is shown in Fig. \ref{ipr_ttr}. The states are delocalized
for $\delta_3=0$ for the whole range of $\epsilon$, while localized for $\delta_3 \neq 0$ and
$\epsilon \gtrapprox 0.6$.

The BBC and NHSE manifest in the Hamiltonian $H$ as follows. 
We have shown that NNN hopping amplitude $\delta_3$ does not play any
role in identification of the topological phases under the PBC. However, the nature
of band spectra crucially depends on $\delta_3$.
The system admits edgestates under OBC in non-trivial parametric regions when the value of the
BLG strength $\epsilon$ is such that the system is in unbroken $\mathcal{PT}$-symmetric regions
i.e. $\epsilon < \vert \delta_1 - \delta_2 \vert$. The edgestates are destroyed beyond a critical
value of NNN strength. In case of PBC the bandgap closes if $\delta_3$ goes beyond  the same critical
value. This is an indication that the system respects BBC in unbroken $\mathcal{PT}$-symmetric regions.
However, it is evident from Fig. \ref{ipr_ttr} that in the broken 
$\mathcal{PT}$-symmetric regions the BBC breaks down and large number of 
eigenstates are localized at the edges even when $\delta_1 > \delta_2$ . This indicates the NHSE. The
interplay between the NNN interaction and the BLG are the reasons of this NHSE. 

\subsection{Interpolating General Boundary Condition}

The bulk Hamiltonian is characterized by four parameters ($\delta_1, \delta_2, \delta_3, \epsilon$)
and the boundary terms contain additional four parameters ($\alpha_L, \alpha_R, \beta_L, \beta_R$),
leading to a eight dimensional parameter space. With such a large parameter-space, even the numerical
investigations of the system become cumbersome. We introduce two new parameters
$-\frac{\pi}{2} \leq \xi \leq \frac{\pi}{2}$ and $0 \leq \delta \leq 1$, and express
the boundary-parameters as, 
\bea
&& \alpha_{R} = \delta \delta_2 \left(1 - \cos \xi \right) \,
\ \alpha_{L} = \delta \delta_2 \sin \xi,\nonumber \\
&&\beta_{R} = i \delta \delta_3 \left (1 - \cos \xi \right) \ ,
\ \beta_{L} = -i \delta \delta_3 \sin \xi,
\eea
\noindent such that the essential features of the system can be described in terms of
six independent parameters ($\delta, \delta_1, \delta_2, \delta_3, \epsilon, \xi)$.
The parameter $\delta$ is a common scale-factor for all the boundary terms, and an increment in $\delta$
for fixed $\xi$ corresponds to the same increment for all the boundary terms. On the other hand, continuous
variation of $\xi$ changes the boundary terms differently such that PBC, OBC, HGBC, AHBC and NGBC appear
as special cases. In particular, different boundary conditions are reproduced as follows: (i)  PBC:
$\xi=\frac{\pi}{2}, \delta=1$, (ii) OBC: $\xi=0$ or $\delta=0$ (iii) AHBC: $\xi=-\frac{\pi}{2}$, (iv) HGBC:
$\xi=\frac{\pi}{2}, \delta \neq 1$, and (v) NGBC: $\xi \neq (0, \pm \frac{\pi}{2})$ and arbitrary
$\delta$. The PBC and OBC, although belong to hermitian boundary boundary condition, are
discussed separately because of their importance. The AHBC has also been mentioned separately
for the same reason. The boundary condition
interpolates from AHBC to PBC by following the sequence AHBC $\rightarrow$ NGBC $\rightarrow$ OBC
$\rightarrow$ NGBC $\rightarrow$ PBC as $\xi$ is varied continuously from $-\frac{\pi}{2}$ to
$\frac{\pi}{2}$ for $\delta=1$. For $\delta \neq 1$, the boundary condition follows the sequence AHBC
$\rightarrow$ NBC $\rightarrow$ OBC $\rightarrow$ NGBC $\rightarrow$ HGBC for continuous variation of $\xi$.

\subsubsection{HGBC: Eigenspectra, Edgestates and Persistent Current}

\begin{figure}
\centering
        \begin{subfigure}{0.48\linewidth}
                \centering
                \includegraphics[width=\textwidth]{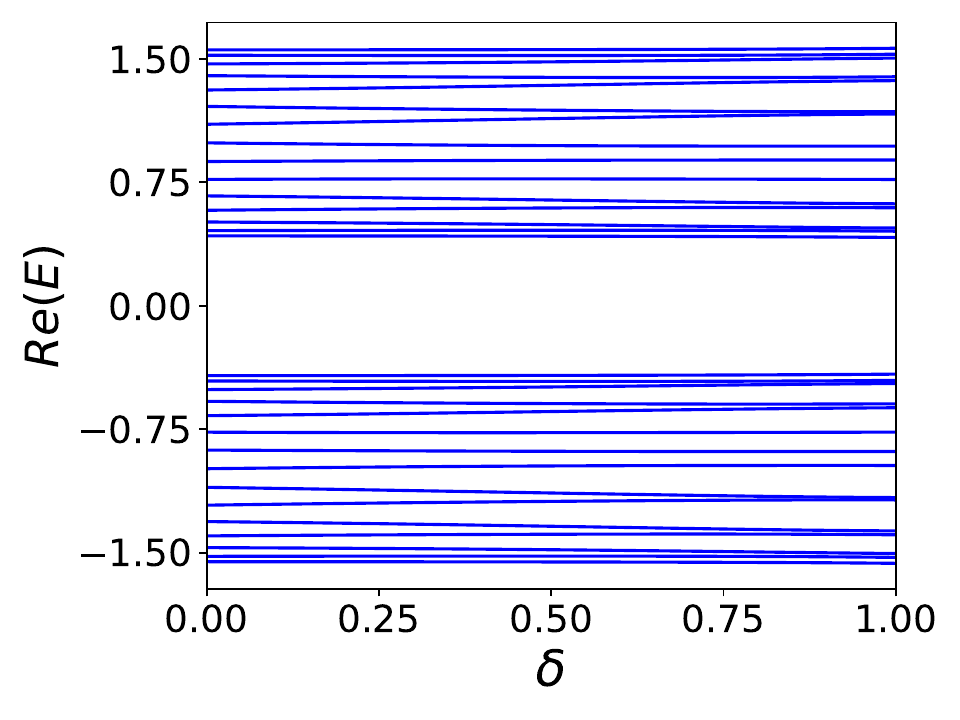}
                \caption{}
                \label{}
        \end{subfigure}
        \hfill
        \begin{subfigure}{0.48\linewidth}
                \centering
                \includegraphics[width=\textwidth]{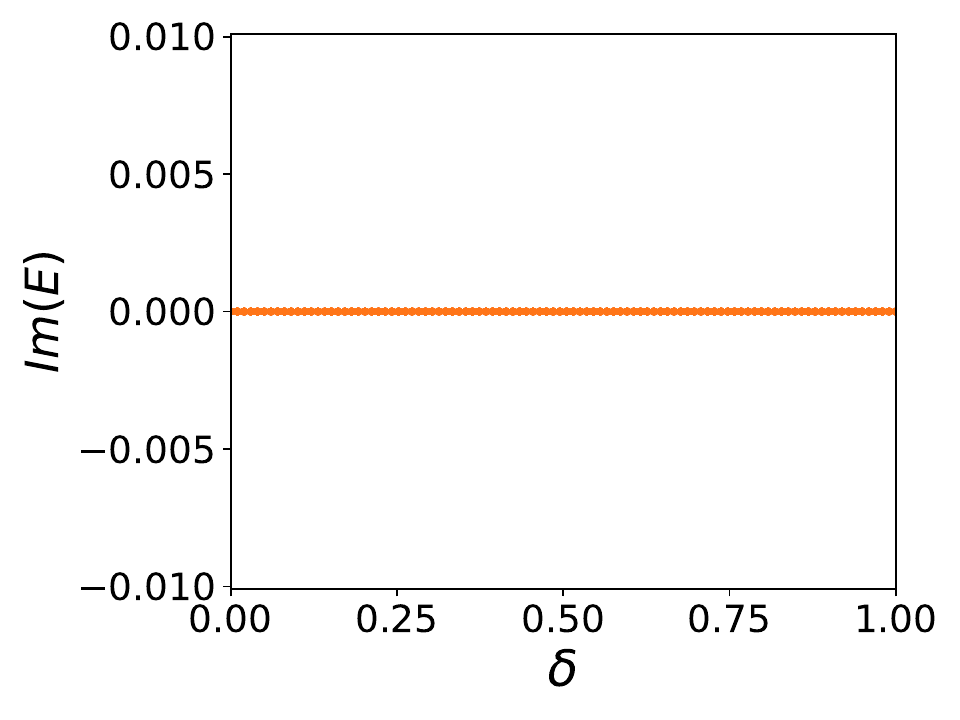}
                \caption{}
                \label{}
        \end{subfigure}
        \vfill
        \begin{subfigure}{0.48\linewidth}
                \centering
                \includegraphics[width=\textwidth]{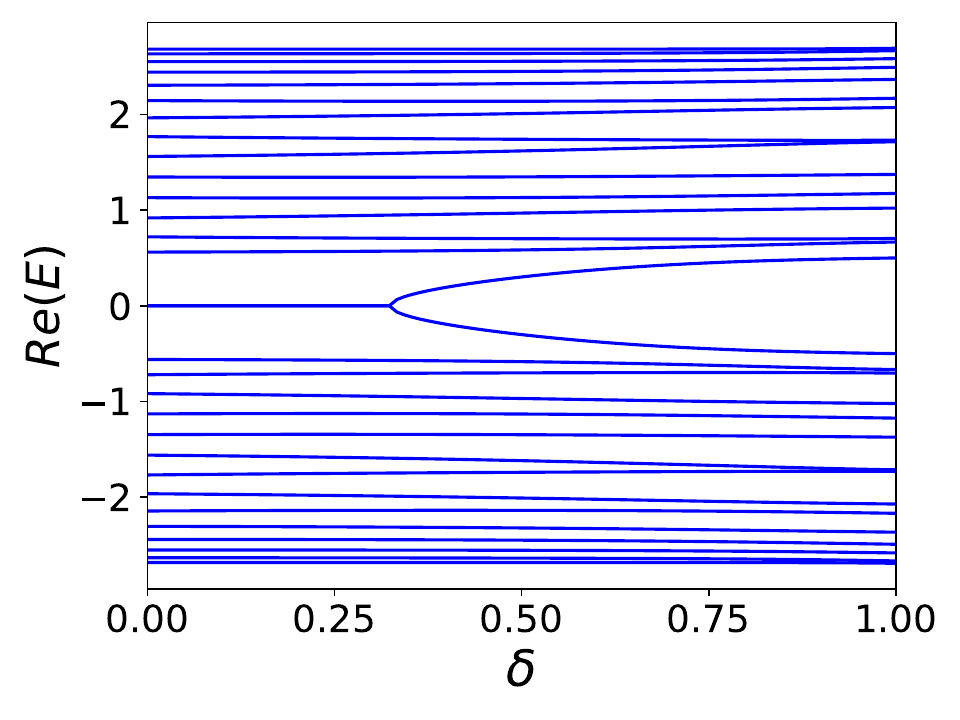}
                \caption{}
                \label{}
        \end{subfigure}
        \hfill
        \begin{subfigure}{0.48\linewidth}
                \centering
                \includegraphics[width=\textwidth]{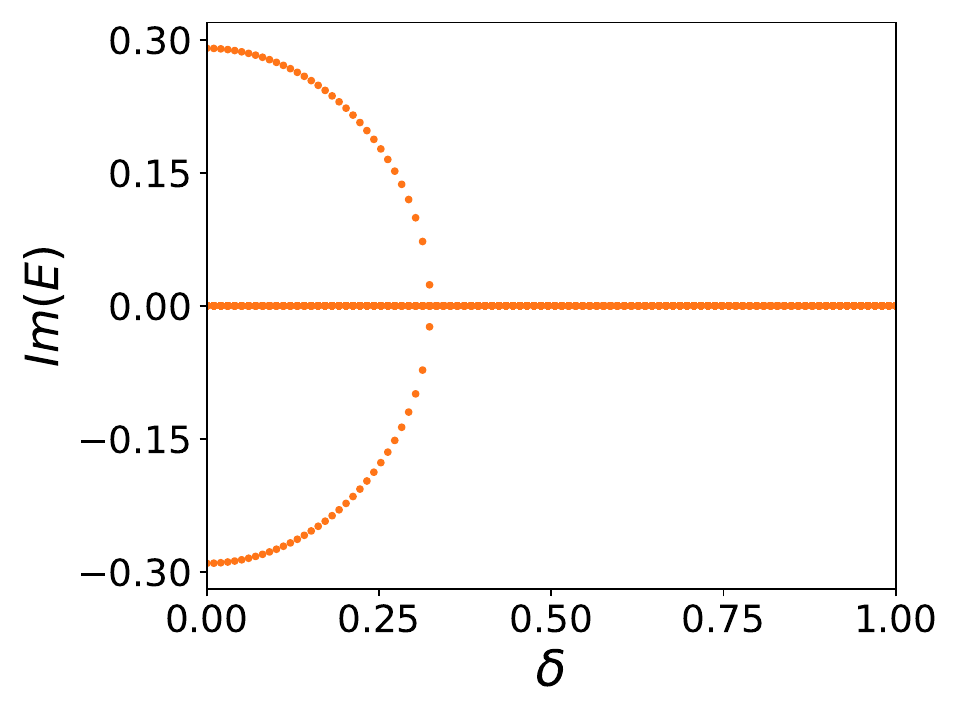}
                \caption{}
                \label{}
        \end{subfigure}
	\caption{(Color online) Plot of eigenvalues under HGBC; The first row corresponds to  
	$\delta_1 > \delta_2$, while  the second row is for $\delta_1 < \delta_2$;
	Parametric Value : $\delta_1 = 1$, $\epsilon = 0.3$, $N = 30$, $\delta_3 = 0.2$, $\xi = \frac{\pi}{2}$,$\phi = 0$; 
	Fig.~a, Fig.~b : $\delta_2 = 0.4$; Fig.~c, Fig.~d : $\delta_1 = 1, \delta_2 = 1.6$.}
        \label{eigval_her_gbc2}
\end{figure}

The balanced loss-gain is the only source of non-hermiticity for HGBC.
The eigenvalues are shown as function of $\delta$ in Fig. \ref{eigval_her_gbc2}. All the
eigenvalues are real for $\delta_1 > \delta_2, \epsilon=.3, \delta_3=.2$ and
$0 \leq \delta\leq 1$. Further, there are no eigenstates in the gap of the two bands.
However, for $\delta_1 < \delta_2$ and all other parameters
remaining the same, pair of imaginary eigenvalues appear up to around $\delta \approx .3$,
and vanish beyond that. The eigenvalues become entirely real for $\delta \gtrapprox .3$.
There is an isolated eigenstate in the gap of the two bands.
\begin{figure}
        \centering
        \includegraphics[scale=0.3]{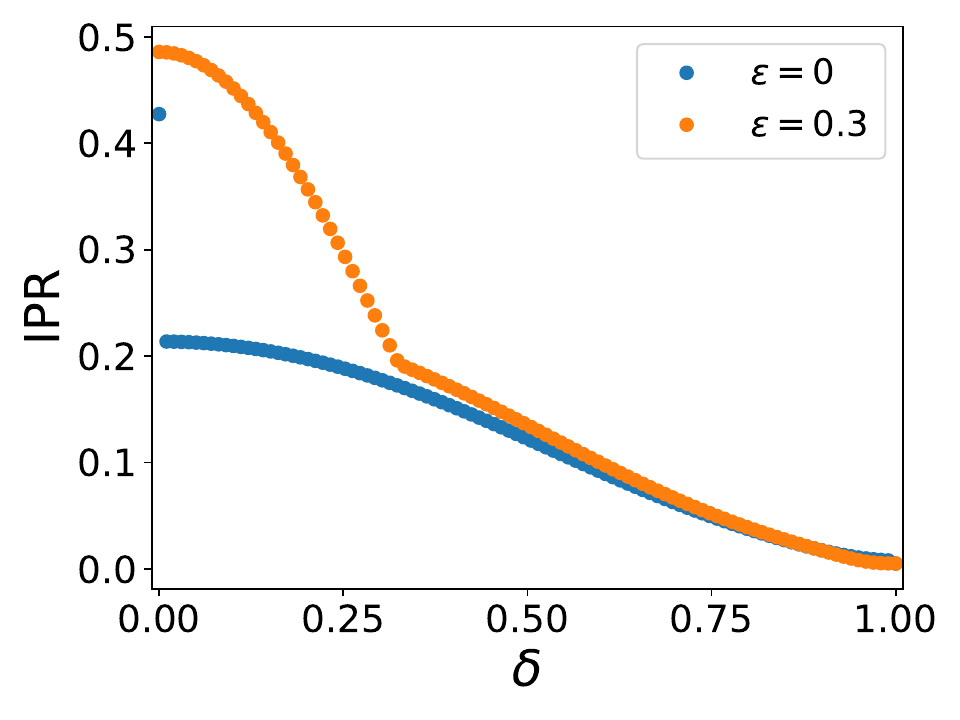}
	\caption{(Color online) Plot of the IPR of the m-th state with the variation of $\delta$ under HGBC;
	Parametric values : $\delta_1 = 1$, $\delta_2 = 1.6$, $\delta_3 = 0.2$,$\xi = \frac{\pi}{2}$,$N = 200$, $\phi = 0$.}
        \label{ipr_hgbc}
\end{figure}
It has been numerically verified by studying the wavefunctions that the edgestate exists for all 
regions of $\delta$ i.e. $0 \leq \delta < 1$ except the PBC. It can also be seen from the IPR plot 
in Fig. \ref{ipr_hgbc}. The eigenvalue of the edgestate is a complex conjugate pair up to 
a critical value of $\delta$. The eigenvalue of the edgestate is real when $\delta$ is greater than the 
critical value. 
\begin{figure}
\centering
        \begin{subfigure}{0.48\linewidth}
                \centering
                \includegraphics[width=\textwidth]{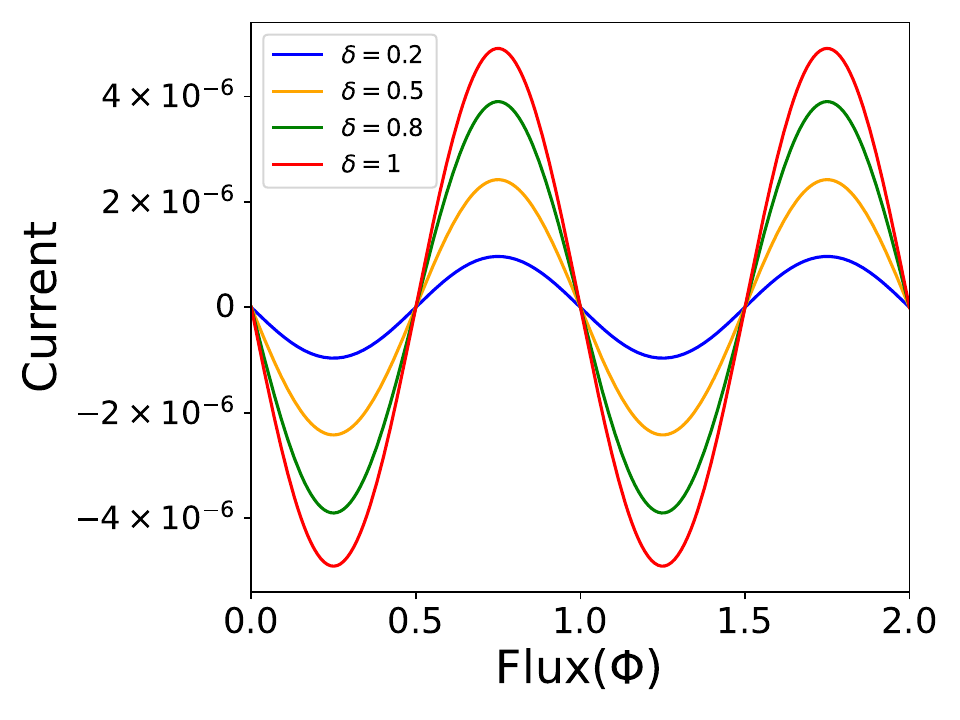}
                \caption{}
                \label{}
        \end{subfigure}
        \hfill
        \begin{subfigure}{0.48\linewidth}
                \centering
                \includegraphics[width=\textwidth]{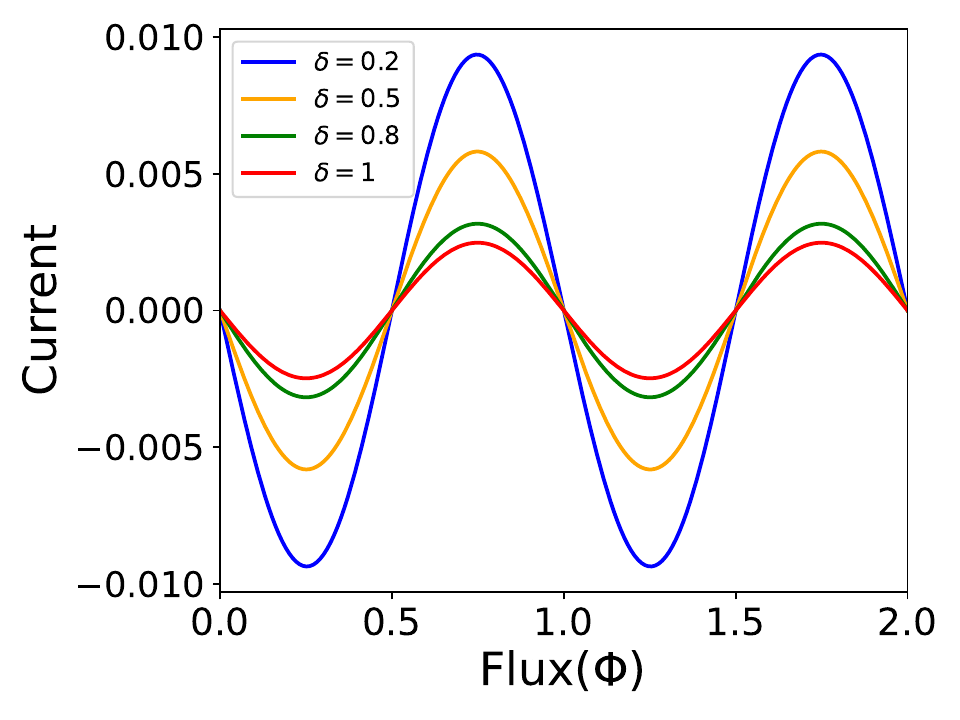}
                \caption{}
                \label{}
        \end{subfigure}
	\caption{(Color online) Plot of the current vs. flux for HGBC;
	Parameter Values : $\epsilon = 0.3, \delta_3 = 0.2, N = 30$,$\xi = \frac{\pi}{2}$;
        Fig(a) : $\delta_1 = 1, \delta_2 = 0.4$ (Topologically trivial region);
        Fig(b) : $\delta_1 = 1, \delta_2 = 1.6$ (Topologically non-trivial region)}
        \label{current_her_gbc}
\end{figure}
The persistent current for the HGBC is shown in the Fig.~(\ref{current_her_gbc}). The 
parameter $\delta$ acts as a modulation parameter and the amplitude persistent current increases 
with the increasing value of $\delta$ in the trivial region. But in the non-trivial region the 
amplitude of the persistent current decreases with the increasing value of $\delta$. 
\subsubsection{NGBC: Eigenspectra, Edgestates and Persistent Current}
\begin{figure}
\centering
        \begin{subfigure}{0.48\linewidth}
                \centering
                \includegraphics[width=\textwidth]{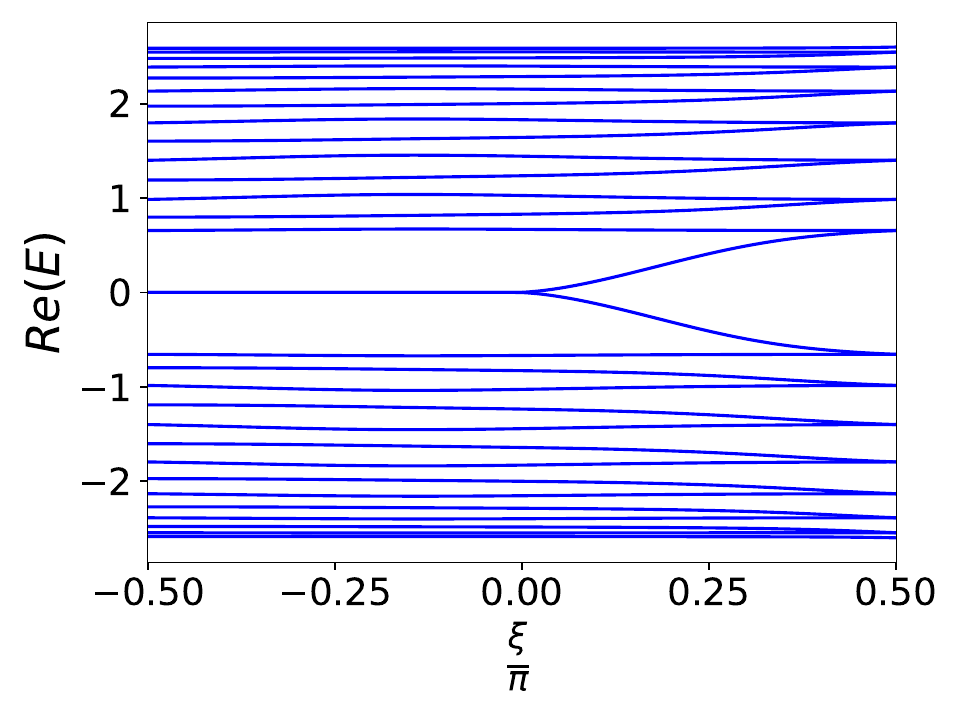}
                \caption{}
                \label{}
        \end{subfigure}
        \hfill
        \begin{subfigure}{0.48\linewidth}
                \centering
                \includegraphics[width=\textwidth]{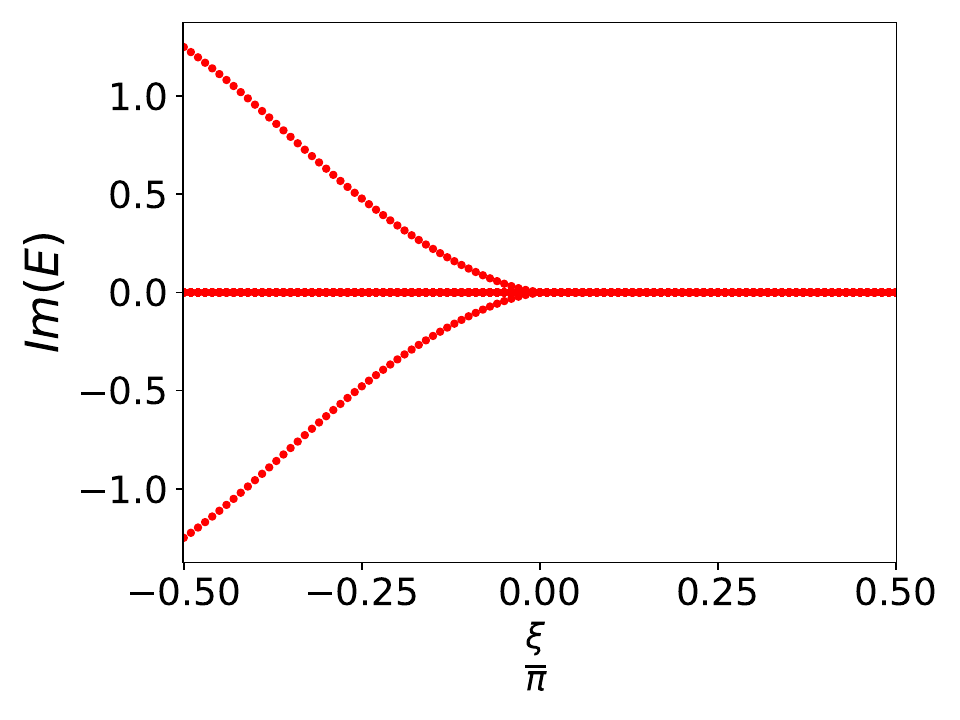}
                \caption{}
                \label{}
        \end{subfigure}
        \vfill
        \begin{subfigure}{0.48\linewidth}
                \centering
                \includegraphics[width=\textwidth]{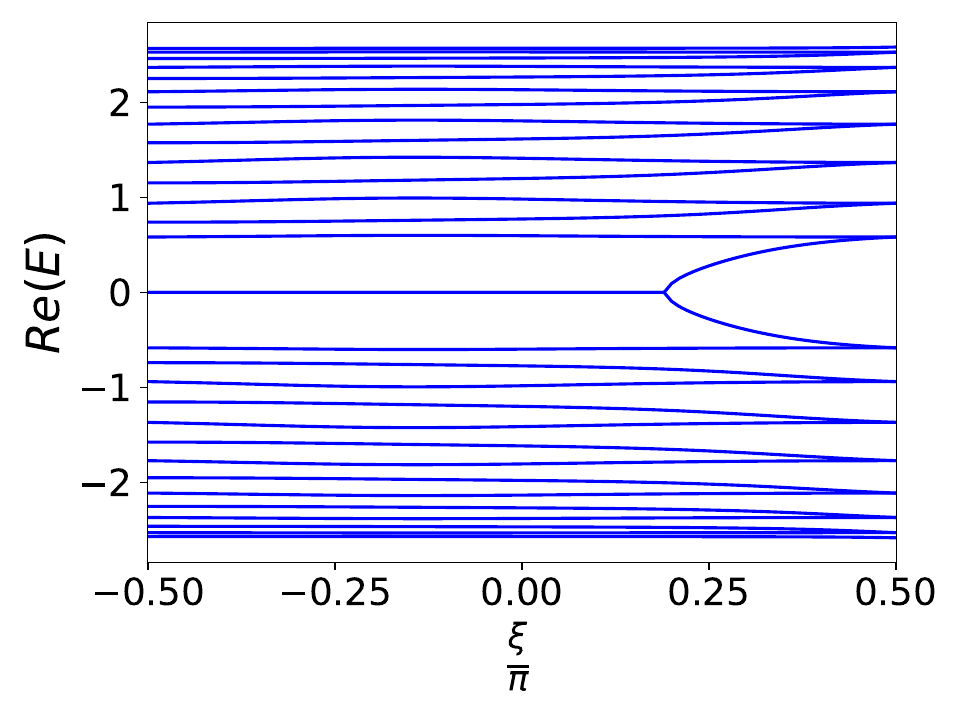}
                \caption{}
                \label{}
        \end{subfigure}
        \hfill
        \begin{subfigure}{0.48\linewidth}
                \centering
                \includegraphics[width=\textwidth]{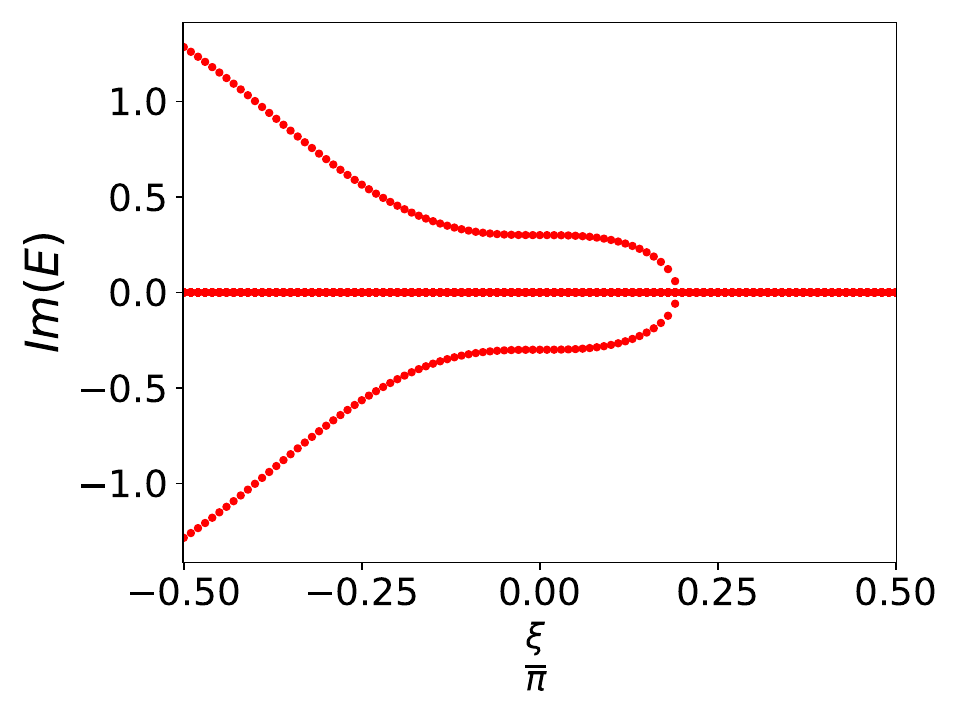}
                \caption{}
                \label{}
        \end{subfigure}
	\caption{(Color online) Plot of eigenvalues  under NGBC for $\delta_1 < \delta_2, \delta_3=0$;
Parametric Values : $\delta_1 = 1$, $\delta_2 = 1.6$, $\phi = 0$, $N = 30$,$\delta = 1$; 
Figs.~a and b : $\epsilon = 0$; 
Figs.~c and d : $\epsilon = 0.3$.}
\label{eigval_gbc_nh1}
\end{figure}
\begin{figure}
\centering
        \begin{subfigure}{0.48\linewidth}
                \centering
                \includegraphics[width=\textwidth]{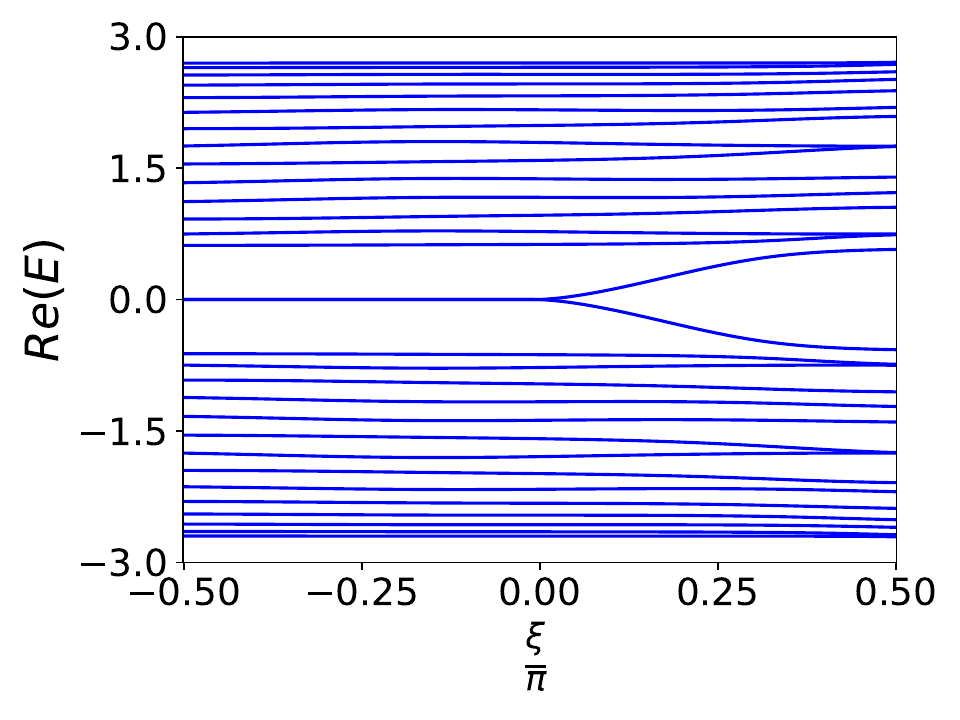}
                \caption{}
                \label{}
        \end{subfigure}
        \hfill
        \begin{subfigure}{0.48\linewidth}
                \centering
                \includegraphics[width=\textwidth]{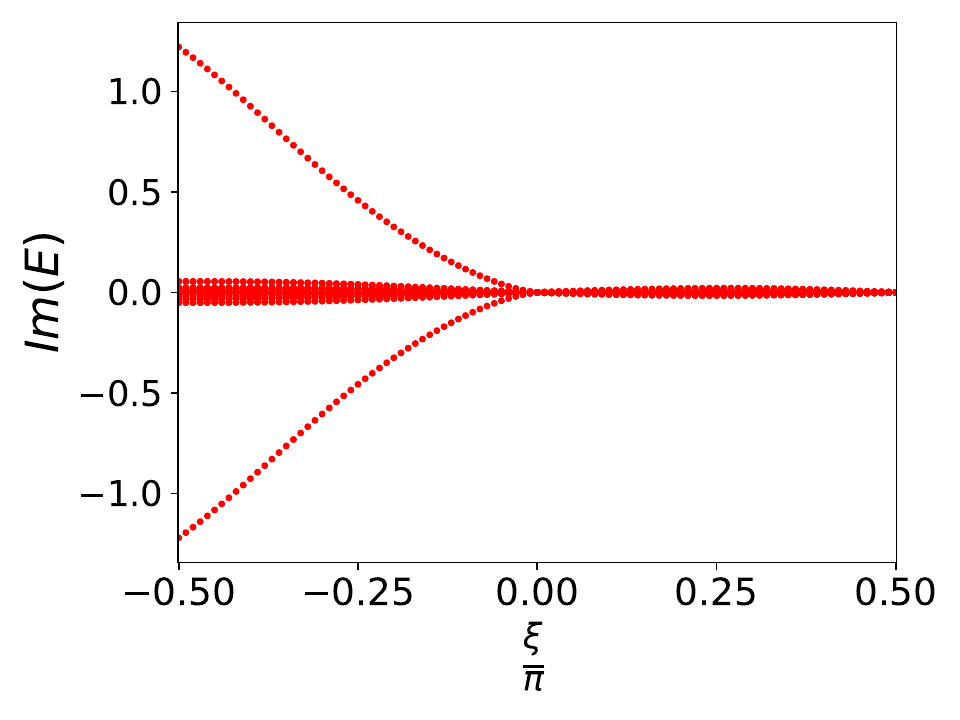}
                \caption{}
                \label{}
        \end{subfigure}
        \begin{subfigure}{0.48\linewidth}
                \centering
                \includegraphics[width=\textwidth]{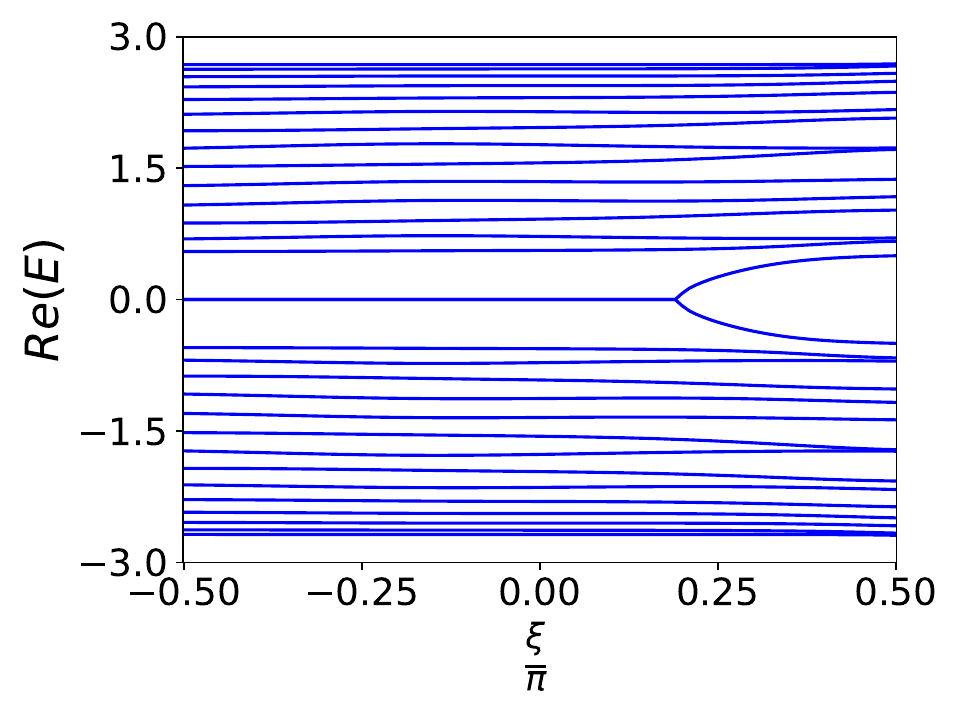}
                \caption{}
                \label{}
        \end{subfigure}
        \hfill
        \begin{subfigure}{0.48\linewidth}       
                \centering
                \includegraphics[width=\textwidth]{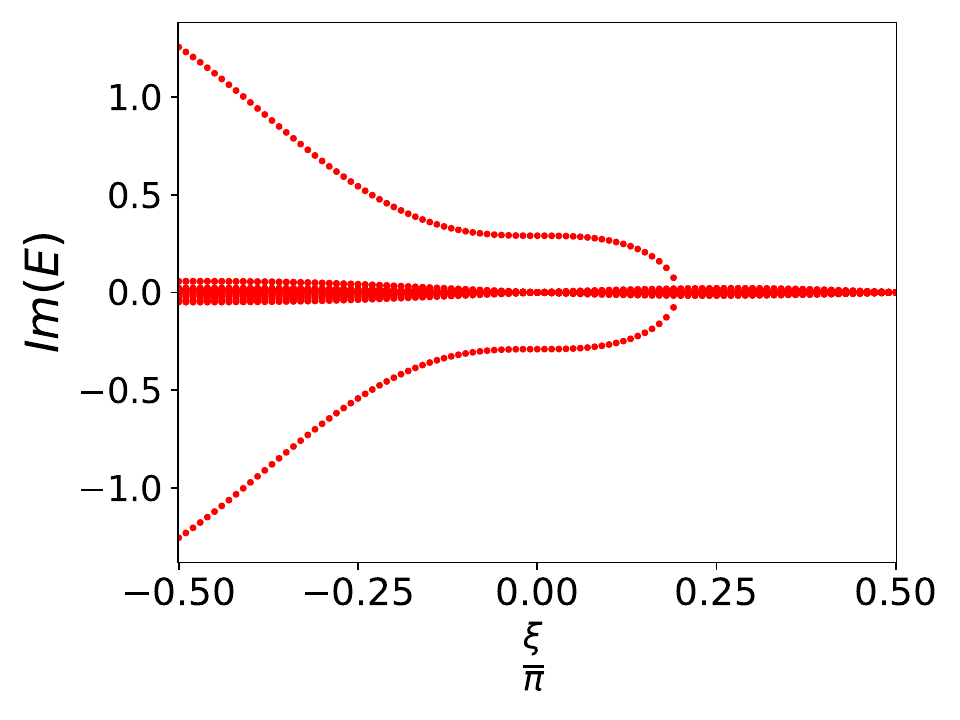}
                \caption{}
                \label{}
        \end{subfigure}
	\caption{(Color online) Plot of eigenvalues  under NGBC for $\delta_1 < \delta_2, \delta_3 \neq 0$;
Parametric Values : $\delta_1 = 1, \delta_2 = 1.6, \delta_3=0.2,
N = 30$,$\delta = 1$,$\phi = 0$; Fig.~a, Fig.~b : $\epsilon = 0$; Fig.~c,Fig.~d : 
$\epsilon = 0.3$.}
\label{eigval_gbc_nh2}
\end{figure}
The spectra of the SSH model under NGBC with and without NNN interaction
is shown  Fig.~\ref{eigval_gbc_nh2} and Fig.~\ref{eigval_gbc_nh1}, respectively. 
\begin{figure}
        \centering
        \includegraphics[scale=0.3]{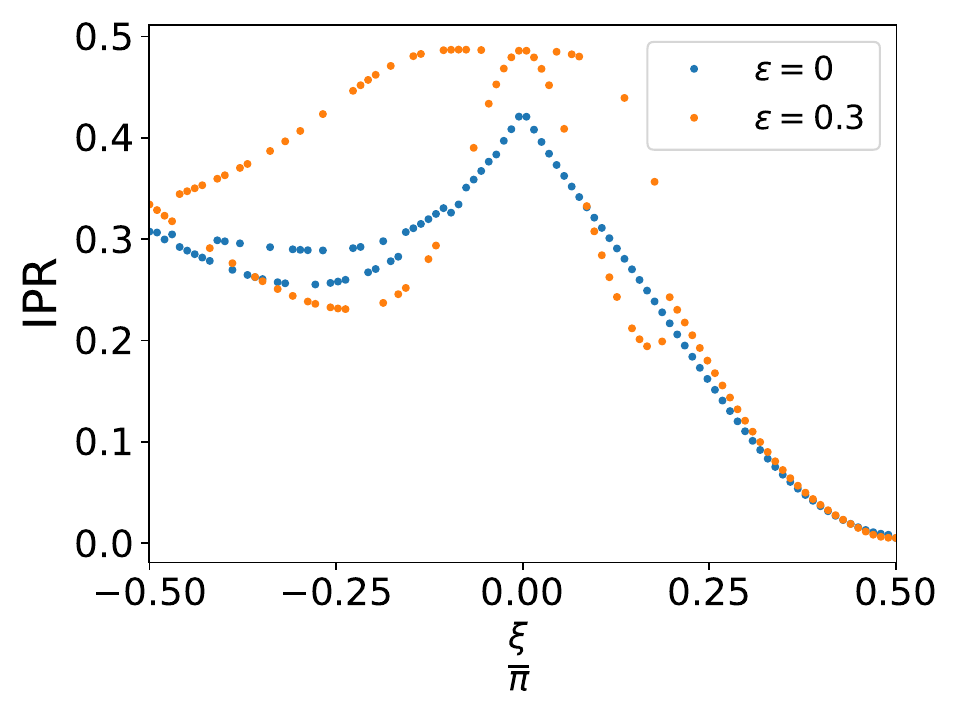}
	\caption{(Color online) Plot of the IPR of the m-th state with the variation of $\xi$(NGBC).
	Parametric values : $\delta_1 = 1$, $\delta_2 = 1.6$, $\delta_3 = 0.2$, $\phi =0$, 
	$N = 200$,$\delta = 1$.}
        \label{ipr_ngbc}
\end{figure}
It is observed that the localized state occurs in the topologically nontrivial phase
except for $0.47\pi < \xi <0.5\pi$, and the states are localized at the
edges. We plot the IPR for two values of $\epsilon$ in Fig. \ref{ipr_ngbc}. 

\begin{figure}
\centering
        \begin{subfigure}{0.48\linewidth}
                \centering
                \includegraphics[width=\textwidth]{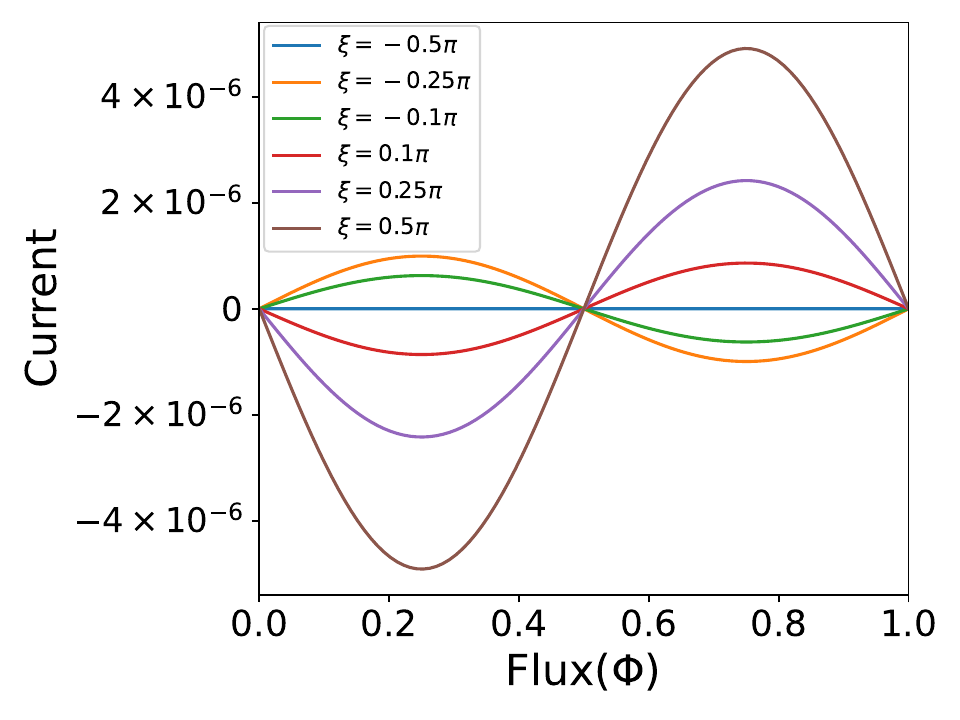}
                \caption{}
                \label{}
        \end{subfigure}
        \hfill
        \begin{subfigure}{0.48\linewidth}
                \centering
                \includegraphics[width=\textwidth]{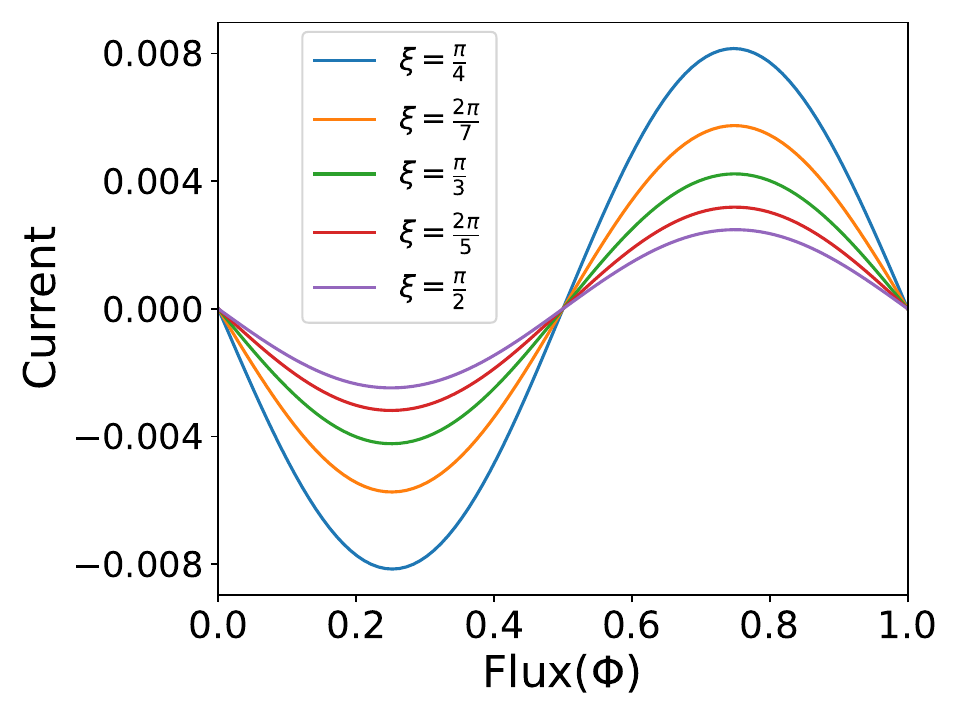}
                \caption{}
                \label{}
        \end{subfigure}
	\caption{(Color online)Plot of the persistent current under NGBC.
	Parametric Values : $\delta_1 = 1, \epsilon = 0.3, N = 30, \delta_3 = 0$,$\delta = 1$,
	Fig.~a : $\delta_2 = 0.4$, Fig.~b : $\delta_2 = 1.6$.}
        \label{current_ngbc}
\end{figure}
The variation of the persistent current as a function of flux under the NGBC is shown in Fig. \ref{current_ngbc}.
The NGBC admits entirely real spectra only within certain regions in the parameters space. The
results are presented for such parametric regions, since the Fermi level for the complex eigenspectra 
is not well defined. The parameter $\xi$ can be used to modulate the amplitude of the persistent current. 
The current increases with increasing $\xi$ in the topologically trivial region, while it decreases with
increasing $\xi$ in the topologically nontrivial region for $0 < \xi \leq \frac{\pi}{2}$.

In the numerical analysis of the GBC we have constrained the BLG parameter $\epsilon < \vert \delta_1 - \delta_2 \vert$. 
We have got the edgestates even when the boundary condition deviates from OBC. But the BBC is respected --- 
the edgestate appear in the parametric regions $\delta_2 > \delta_1$ where the system shows non-trivial 
topological phases under PBC.  
\section{Conclusions \& Discussions}

We have studied an extended SSH model by incorporating NNN
interaction, BLG terms and an external uniform magnetic field into the standard SSH model.
The bulk Hamiltonian is $\mathcal{PT}$-symmetric and the only source of non-hermiticity
is BLG terms. We have studied the system under the GBC \textemdash the boundary Hamiltonian
is $\mathcal{PT}$-symmetric for HGBC, and $\mathcal{PT}$-symmetry with the standard notion is
not preserved under the NGBC.

{\bf PBC}: We have diagonalized the Hamiltonian under the PBC, and obtained analytic expressions for
the eigenvalues and the eigenstates. Although the Hamiltonian is non-hermitian with the
standard notion of hermiticity, it admits entirely real spectra in certain regions
of the parameter space corresponding to unbroken ${\mathcal{PT}}$-symmetry. We have also
shown that the Hamiltonian is pseudo-hermitian, and the equivalent hermitian Hamiltonian
has been constructed. The system admits point as well as line band gaps. Although
the the NNN interaction does not play any role in determining whether the eigenvalues 
are real or complex, the line gap shows closed-loop like structure only when it is non-vanishing.
Further, the introduction of the NNN interaction converts the direct band gap into an indirect band gap.
The band gap reduces as the NNN hopping amplitude increases, and closes the gap beyond a critical value.
The Zak phase receives no contribution from the NNN interaction \textemdash the classification
of topological phases remains the same, namely, the non-trivial region corresponds to
the intercell hopping strength being greater than the intracell hopping strength.
We have also studied the combined effect of NNN interaction and BLG terms on the persistent current,
and shown that the BLG parameter within its allowed range always enhances magnitude of the persistent
current irrespective of the NNN interaction.The persistent current is independent of the
strength of the NNN interaction in the half filled band, while its magnitude increases with
a increase in the strength of the NNN interaction for other fillings of the band gap less than
the half-filling. The combined effect of the NNN interaction and the BLG on the persistent current
has not been studied earlier. 

{\bf OBC}: In the topologically trivial phase $\delta_1 > \delta_2$, the eigenspectra is entirely
real up to a certain value of $\delta_3$ \textemdash the complex energies appear and real band gap
closes beyond this critical value. The topologically non-trivial phase ($\delta_2 > \delta_1$) admits
complex eigenvalues in conjugate pairs, which is a signature of existence of edge states.
We have obtained analytic expressions of the edge states for vanishing NNN interaction
and non-zero BLG terms. The edge states for non-vanishing NNN interaction have
been studied numerically. In the topologically non-trivial phase, the edge states  exists only up to a 
critical value value of the NNN strength and vanishes beyond this critical
value. We have seen numerically that edge states in the topologically trivial phase appear for
non-vanishing NNN interaction only when $\epsilon > {\vert \delta_1 - \delta_2 \vert}$.

{\bf GBC}: We have investigated the extended SSH model with GBC from the viewpoint of exact solvability. 
We have derived exact analytical expressions for the eigenvalues and eigenstates of the Hamiltonian in the absence of NNN 
interactions for a class of GBC of which PBC, APBC and AHBC appear as special cases.
Furthermore, we have theoretically identified the parametric regions that support the existence of edge states
under OBC and AHBC. 

We have analysed the Hamiltonian with GBC numerically for cases where analytical solutions are
absent. 
We have re-parameterized the system parameters such that the essential features are described
in terms of the parameters $-\frac{\pi}{2} \leq \xi \leq \frac{\pi}{2}$ and $0 \leq \delta \leq 1$. The parameter
$\delta$ is a common scale-factor for all the boundary terms, while continuous variation of $\xi$
changes the boundary terms differently such that PBC, OBC, AHBC, HGBC and NGBC appear as special cases.
The system is non-hermitian due to BLG and boundary terms for NGBC. We have studied eigenspectra, edgestates
and persistent Current for the HGBC and NGBC numerically. The system admits entirely real spectra
for HGBC in certain region in the parameter space and exhibits edge states for $\delta \neq 1$
and $\xi=\frac{\pi}{2}$. However, for the case of NGBC, the edge states are observed only for a certain range
of $\xi$.

A few future directions may be outlined as follows:\\
(i) We have shown that the Zak phase receives no contribution from the NNN interaction under the PBC.
The computation of Zak phase under the GBC, and study of its dependence on both NNN interaction and BLG
terms, is worth pursuing.\\ 
(ii) We have analyzed persistent current in the present article for the cases when the entire spectra
is real. The analysis may be extended to the complex spectra following Ref. \cite{Shen2024Prl},
where non-hermitian Fermi-Dirac distribution was introduced along with an analytic expression of
the persistent current.\\
(iii) An equivalent hermitian Hamiltonian may be found for a pseudo-hermitian system endowed
with a positive-definite metric in the Hilbert space. We have constructed equivalent hermitian
Hamiltonian only for the PBC. This may generalized for the GBC.\\
(iv) Recently, investigations on multi-band lattice models are gaining momentum\cite{tkg,Ghuneim2024JPhys}.
The study on non-hermitian oligomer SSH model may provide interesting insights
into the multi-band lattice models.
\section{Acknowledgements}
SG acknowledges the financial support of the DST INSPIRE fellowship of 
Govt.of India(Inspire Code No. IF190276).
\appendix

\section{Consistency condition for the existence of eigenstates under the GBC for $\delta_1 = 0$}
The quartic equation   (\ref{gbc_root}) can be solved analytically leading to four roots
whose analytic expressions for generic values of the parameters prove to be cumbersome for further
analysis. The roots can be expressed in a simple form for $\delta_1 \delta_2=0$ for which
the SSH model reduces to a tight-binding ladder with NNN interaction and subjected to an external
uniform magnetic field. We choose $\delta_1 = 0, \delta_2 \neq 0$, and the results for
$\delta_1 \neq 0, \delta_2 = 0$ can be obtained by simply letting $\delta_2 \rightarrow \delta_1$,
since Eq.  (\ref{gbc_root}) is symmetric under $\delta_1 \leftrightarrow \delta_2$.
With this choice of the parameters and $\tilde{z} = \tilde{z} e^{-i\frac{\pi}{2}}$,
Eq. (\ref{gbc_root}) reduces to the following bi-quadratic form, 
\bea
\left[ \tilde{z}^{2} - \left\{ \frac{E}{\delta_3} + \frac{1}{\delta_3} \sqrt{\delta_2^2 - \epsilon^2} \right\} \tilde{z} + 1 \right] & \times &\nonumber \\ 
\left[ \tilde{z}^{2} - \left\{ \frac{E}{\delta_3} - \frac{1}{\delta_3} \sqrt{\delta_2^2 - \epsilon^2} \right\} \tilde{z} + 1 \right] & =      & 0
\eea
Note that for any solution $\tilde{z}$, $\tilde{z}^{-1}$ is also a solution of the above equation. 
The roots are determined as,
\bea
\tilde{z}_{1,2} & =   &- \frac{1}{2\delta_3} \left(E - \sqrt{\delta_2^2 - \epsilon^2}\right) \nonumber \\ 
	& \pm & \sqrt{\left\{\frac{1}{2\delta_3} \left(E - \sqrt{\delta_2^2 - \epsilon^2}\right)\right\}^{2} - 1} \nonumber \\ 
\tilde{z}_{3,4} & =   &- \frac{1}{2\delta_3} \left(E + \sqrt{\delta_2^2 - \epsilon^2}\right) \nonumber \\
        & \pm & \sqrt{\left\{\frac{1}{2\delta_3} \left(E + \sqrt{\delta_2^2 - \epsilon^2}\right)\right\}^{2} - 1} 
\eea
\noindent which satisfy the relations $\tilde{z}_{1}\tilde{z}_{2} = 1$ and $\tilde{z}_{3}\tilde{z}_{4} = 1$. We parameterize the roots
in terms of $\theta_1, \theta_2$ as, $\tilde{z}_{1} = e^{i\theta_{1}}$, $\tilde{z}_{2} = e^{-i\theta_{1}}$,
$\tilde{z}_{3} = e^{i\theta_{2}}$ and $\tilde{z}_{4} = e^{-i\theta_{2}}$. 
The general expression of the eigenstate in Eq. (\ref{bulk_eqn_gbc}) is 
\bea
\Psi_{n} = \sum_{j=1}^4 \bp c_j A_j \tilde{z}_{j}^{n} \\ 
c_j B_j \tilde{z}_{j}^{n} \ep e^{-i n \left ( \frac{\pi}{2} + 2 \phi \right )}\nonumber
\eea
The boundary equation leads to the matrix equation $ H_{B} C=0$,
where $C$ is a four-component column vector $C= (c_1, c_2, c_3, c_4 )^T$ and $H_{B}$ is a $4\times 4$
matrix with its elements $[H_B]_{ij}, i, j=1, 2, 3, 4$  given by,
\bea
&& [H_{B}]_{1,j} = \left(i\delta_2 B_j -\delta_3 A_j\right), \ [H_{B}]_{2j} = - \delta_3 B_j,\nonumber \\ 
&& [H_{B}]_{3j} = -\delta_3 A_j z_{j}^{m+1}, [H_{B}]_{4j} = - \delta_3 B_j z_{j}^{m+1} -
i\delta_2 z_{j}^{m+1} A_{j}.\nonumber
\eea
The non-trivial solution is obtained for $Det[H_{B}] = 0$ determining $\theta_1$ and $\theta_2$,
and thereby, analytic expressions for the eigenspectra and eigenstates. The resulting equation 
is cumbersome and appears to evade any simple closed-form solutions, and is not pursued in this
article. We present numerical results encompassing all these parametric values.
\section{On trivial solutions of Eq.~(\ref{boundary_eq_con1}) corresponding to $\sin\theta = 0$}

The reason for abandoning the solutions of Eq.~(\ref{boundary_eq_con1}) corresponding to $\sin\theta = 0$
i.e. $\theta=0, \pi$ is that the corresponding wavefunctuions are equal to zero. In order to see this, we
note that for $\alpha_L \alpha_R = \delta_2^2$ the boundary Eqs. $H_{B} \bp c_1 \\ c_2 \ep =0$ 
reduce to,
\bea
\left(1 - \frac{\delta_2}{\alpha_L} z_{1}^m \right) c_1 B_1 + \left(1 - \frac{\delta_2}{\alpha_L} z_{2}^m \right) c_2 B_2
& = & 0, \nonumber \\
\left(1  - \frac{\delta_2}{\alpha_L} z_1^m \right) c_1 A_1 z_1 + \left(1 - \frac{\delta_2}{\alpha_L} z_2^m\right) c_2 A_2 z_2
& = & 0. \nonumber 
\eea
\noindent Substituting $z_j=\tilde{z}_j e^{-2i\phi}, j=1, 2$ and putting $\theta = 0$, the above equations take the form,
\bea
\left(1 - \frac{\delta_2}{\alpha_L} e^{-2im\phi} \right) \left(c_1 B_1 + c_2 B_2 \right) & = & 0, \nonumber \\
\left(1 - \frac{\delta_2}{\alpha_L} e^{-2im\phi} \right) \left(c_1 A_1 + c_2 A_2 \right)   & = & 0. \nonumber 
\eea
\noindent The solutions of the above equations are,\\
(I) $1 - \frac{\delta_2}{\alpha_L} e^{-2im\phi} =0$ or\\
(II) $c_1 B_1 + c_2 B_2 =0=c_1 A_1 + c_2 A_2 $.\\
The only acceptable solution of the first equation is
$$\delta_2=(-1)^l \alpha_L, \phi=\frac{\pi}{2m} l, l \in \mathbb{Z},$$
since all the parameters are real. These solutions correspond to PBC($\nu=1$) for even $l$
or APBC($\nu=-1$) for odd $l$ with specific magnetic flux. Note that the solution $\theta=0$ is already included
in the solutions of the first part of Eq. (\ref{boundary_eq_con1}), i. e. $\cos(m \theta)-\nu \cosh u=0 \Rightarrow
1=\nu \cos(l \pi)=(-1)^l \nu$. The only viable solutions for the generic case correspond to the case-(II)  which
essentially implies that $\Psi=0$. The boundary equations corresponding to $\theta = \pi$ have the form, 
\bea
\left(1 - \frac{\delta_2}{\alpha_L} (-1)^m e^{-2im\phi} \right) \left( c_1 B_1 + c_2 B_2 \right) & = & 0, \nonumber \\
\left(1- \frac{\delta_2}{\alpha_L}(-1)^m e^{-2im\phi} \right) \left(c_1 A_1 + c_2 A_2 \right)   & = & 0. \nonumber  
\eea
\noindent The condition $1 - \frac{\delta_2}{\alpha_L} (-1)^m e^{-2im\phi}=0$ is realized for the following
special case,
$$ \delta_2=(-1)^{l+m} \alpha_L, \phi=\frac{\pi}{2 m} l.$$
Such solutions with specific magnetic flux and corresponding to PBC or APBC with even or odd $l+m$, respectively are already
included in the solutions of the first part of Eq. (\ref{boundary_eq_con1}). The solutions 
$c_1 B_1 + c_2 B_2 =0=c_1 A_1 + c_2 A_2 $ for the generic case leads to $\Psi=0$. Thus, the solutions
corresponding to $\sin \theta=0$ are discarded and Eq. (\ref{boundary_eq_con1}) has required number of solutions.

\section{Derivation of eiegenstates for AHBC}

The expressions of the eigenstates can be obtained by finding $\frac{A_1}{A_2}, \frac{B_1}{B_2}$. 
From the condition $ H_{B} \bp c_1 \\ c_2 \ep = 0$ and using the ratio $\frac{A}{B}$ from Eq. (\ref{gbc_block_eq}), 
we can get the ratio $\frac{A_1}{A_2}, \frac{B_1}{B_2}$. As discussed in the section IV.B, the permissible values 
of $\theta$ are given by $\theta_{s} = \frac{s\pi}{m}$, \ $s = 1,2 \dots (m-1)$ and $\theta_{m} = \pi + \arccos(\alpha)$. 
For simplicity, We consider $\Phi = 0$. In the case of AHBC, all elements in the first row of the matrix $H_{B}$ are zero 
for even $s$, while those in the second row are zero for odd $s$. So, we compute $\frac{A_1}{A_2},
\frac{B_1}{B_2}$ for even $s$ and odd $s$ separately. 

We get the following set of equations for $\theta_{s} = \frac{s\pi}{m}$ where $s$ is even integer. 
\bea
c_2 A_2 & = & - c_1 A_1 \frac{z_1}{z_2} \nonumber \\
c_2 B_2 & = & - c_1 A_1 \frac{\left(E - i\epsilon\right)z_1}{\delta_1 z_2 + \delta_2}  \nonumber \\
c_1 B_1 & = & c_1 A_1 \frac{\left(E - i\epsilon\right)z_1}{\delta_1 z_1 + \delta_2} 
\label{ratio_ahbc_even}
\eea
The following set of equations for $\theta_{s} = \frac{s\pi}{m}$ applies when $s$ is an odd integer.
\bea
c_2 B_2 & = & -c_1 B_1 = -\frac{\left(E - i\epsilon\right)z_1}{\delta_1 z_1 + \delta_2} c_1 A_1 \nonumber \\
c_2 A_2 & = & -\frac{\delta_1 + \delta_2  z_1}{\delta_1 + \delta_2 z_2} c_1 A_1 \nonumber \\
\label{ratio_ahbc_odd}
\eea
Finally, for the $\theta_{m} = \pi + \arccos(\alpha)$, we get
\bea
c_2 A_2 & = & - c_1 A_1 \frac{z_1 + z_1^{m+1}}{z_2 + z_{2}^{m+1}} \nonumber \\
c_2 B_2 & = & - c_1 B_1 \left( \frac{1 - z_1^m}{1 - z_2^m} \right) \nonumber \\
& = & - c_1A_1 \frac{E-i\epsilon}{\delta_1} \frac{1 - z_1^m}{\left(1 - z_2^m
\right)\left( 1 + \frac{\alpha}{z_1}\right)},
\label{ratio_ahbc_edge}
\eea

We get the expression of $\psi_{n,a}$ and $\psi_{n,b}$ for the eigenstate using the Eqs.~(\ref{ratio_ahbc_even}),(\ref{ratio_ahbc_odd}),(\ref{ratio_ahbc_edge}). 
The expression for $\psi_{n,a}$,$\psi_{n,b}$ corresponding to bulk eigenstates at $\theta_{s} = \frac{s\pi}{m}$ for even $s$ is,
\bea
\psi_{n,a} & = & c_1 A_1 \left\{ e^{i\frac{ns\pi}{m}} - e^{-i\frac{(n-2)s\pi}{m}} \right\} \nonumber \\
\psi_{n,b} & = & \frac{c_1 A_1 (E - i\epsilon)}{\delta_1^2} e^{i\frac{s\pi}{m}} \left\{ \frac{e^{i \frac{ns\pi}{m}}}{e^{i\frac{s\pi}{m}} + \alpha} - \frac{e^{-i \frac{ns\pi}{m}}}{e^{-i\frac{s\pi}{m}} + \alpha}\right\} \nonumber
\eea
where $s$ takes the even values in the range $1$ to $(m-1)$. 
The expression of $\psi_{n,a}$,$\psi_{n,b}$ for bulk eigenstates corresponding to the $\theta_{s} = \frac{s\pi}{m}$ for odd $s$ is, 
\bea
\psi_{n,a} & = & c_1 A_1 \left\{ e^{i \frac{ns\pi}{m}} - \frac{1 + \alpha e^{i\frac{s\pi}{m}}}{1 + \alpha e^{-i\frac{s\pi}{m}}} e^{-i\frac{ns\pi}{m}} \right\} \nonumber \\
\psi_{n,b} & = & c_1 A_1 \frac{2i (E - i\epsilon)}{\delta_1 + \delta_2 e^{-i\frac{s\pi}{m}}} \ \sin(\frac{ns\pi}{m}) \nonumber 
\eea
where $s$ takes the odd values in the range $1$ to $m-1$. 
The expression of $\psi_{n,a}$,$\psi_{n,b}$ for bulk eigenstates corresponding to the $\theta_{m} = \pi + \arccos(\alpha)$ is
\bea
\psi_{n,a} & = & \frac{2i c_1 A_1 e^{i\theta_m}}{1 + e^{-im\theta_m}} \left\{ \sin(n-m-1)\theta_m +
\sin(n-1)\theta_m\right\} \nonumber \\
\psi_{n,b} & = & \frac{2i c_1 A_1}{1 + \alpha e^{-i\theta_m}} \frac{E - i\epsilon}{\delta_1\left(1 -
e^{-im\theta_m}\right)} \left\{ \sin(n\theta_m) - \sin(n-m)\theta_m\right\}, \nonumber
\eea

Eigenstate corresponding to the $\theta_{m}$ will give edgestate in the non-trivial region i.e. $\delta_2 > \delta_1$. 
In the non-trivial region, $\alpha > 1$ and the expression of $\theta_{m}$ reduces to 
$\theta_{m}=\pi + i \zeta$ where $\cosh \zeta = \alpha$.  In this limit, $\theta_{m}$ is complex.
In the non-trivial parametric regions, the expression of $\psi_{n,a}$,$\psi_{n,b}$ for the eigenstate 
corresponding to $\theta_{m}$ reduces to
\bea
\psi_{n,a} & = & -(-1)^{n}\frac{2 c_1 A_1 e^{-\zeta}}{1 + (-1)^{m} e^{m\zeta}} \left\{ (-1)^{m} sinh(n-m-1)\zeta \right. \nonumber \\
           & + & \left. sinh(n-1)\zeta\right\} \nonumber \\
\psi_{n,b} & = & -(-1)^{n}\frac{2 c_1 A_1}{1 - \alpha e^{\zeta}} \frac{E - i\epsilon}{\delta_1\left(1 - (-1)^{m} e^{m\zeta}\right)} \left\{ sinh(n\zeta) \right. \nonumber \\
           & - & \left. (-1)^{m} sinh(n-m)\zeta\right\} \nonumber
\eea
The probability density of this eigenstate will be maximum at the edges due to the hyperbolic function. 
So this state is identified as edgestate.

\end{document}